\documentclass[numberedappendix,twocolappendix]{emulateapj}
\usepackage{apjfonts}
\usepackage{color}

\newcommand{\pder}[2]{\frac{\partial #1}{\partial #2}}
\newcommand{\eqref}[1]{Equation (\ref{#1})}

\newcommand{\Ni}{^{56}{\rm Ni}}

\newcommand{\phifa}{\phi_{\rm fa}}
\newcommand{\phiaq}{\phi_{\rm aq}}
\newcommand{\phiqn}{\phi_{\rm qn}}
\newlength{\apjcolwidth}
\shorttitle{SN~Ia Nucleosynthesis}

\begin{document}

\submitted{Submitted to the Astrophysical Journal August 5, 2015, Accepted
April 2, 2016}
\title{A Tracer Method for Computing Type Ia Supernova Yields:\\
Burning Model Calibration,\\
Reconstruction of Thickened Flames,\\ and
Verification for Planar Detonations}

\author{
Dean M. Townsley\altaffilmark{1},
Broxton J. Miles\altaffilmark{1},
F.~X. Timmes\altaffilmark{2,3},
Alan C. Calder\altaffilmark{4,5},
and
Edward F. Brown\altaffilmark{6,3}
}

\altaffiltext{1}{
Department of Physics \& Astronomy,
University of Alabama, Tuscaloosa, AL; Dean.M.Townsley@ua.edu
}
\altaffiltext{2}{
School of Earth and Space Exploration,
Arizona State University, Tempe, AZ
}
\altaffiltext{3}{
The Joint Institute for Nuclear Astrophysics
}
\altaffiltext{4}{
Department of Physics \& Astronomy,
Stony Brook University, Stony Brook, NY
}
\altaffiltext{5}{
Institute for Advanced Computational Sciences,
Stony Brook University, Stony Brook, NY
}
\altaffiltext{6}{
Department of Physics \& Astronomy,
Michigan State University, East Lansing, MI
}
\begin{abstract}

We refine our previously introduced parameterized model for explosive
carbon-oxygen fusion during thermonuclear supernovae (SN~Ia) by adding corrections to
post-processing of recorded Lagrangian fluid element histories to
obtain more accurate isotopic yields.  Deflagration and detonation
products are verified for propagation in a uniform density medium.  A new
method is introduced for reconstructing the temperature-density history
within the artificially thick model deflagration front.  We obtain better
than 5\% consistency between the electron capture computed by the burning
model and yields from post-processing.  For detonations, we compare to a
benchmark calculation of the structure of driven steady-state planar
detonations performed with a large nuclear reaction network and
error-controlled integration.  We verify that, for steady-state planar
detonations down to a density of $5\times 10^6$~g~cm$^{-3}$, our post
processing matches the major abundances in the benchmark solution typically to better
than 10\% for times greater than 0.01 s after the shock front passage.
As a test case to demonstrate the method, presented here with
post-processing for the first time, we perform a two dimensional simulation of a SN~Ia
in the Chandrasekhar-mass deflagration-detonation transition (DDT) scenario.  
We find that reconstruction of deflagration tracks leads to slightly more complete silicon burning than without reconstruction.
The resulting abundance structure of the ejecta is consistent with inferences from spectroscopic studies of observed SNe~Ia.
We confirm the absence of a central region of stable Fe-group material for the multi-dimensional DDT scenario.
Detailed isotopic yields are tabulated and only change modestly when using deflagration reconstruction.

\end{abstract}

\keywords{supernovae: general -- nuclear reactions, nucleosynthesis, abundances -- methods: numerical}

\section{Introduction}

Type Ia Supernovae (SNe~Ia) are at once a pillar of modern cosmology and one
of the persistent puzzles of stellar physics.  These bright stellar
transients are characterized by strong P Cygni features in Si and a lack of
hydrogen or helium in their spectra.  It has generally been accepted that
these events follow from the thermonuclear incineration of a white dwarf (WD)
star producing between 0.3 and 0.9 $M_\odot$ of radioactive $^{56}$Ni, the decay
of which powers the light curve \citep[see][and references
therein]{Filippenko1997,HillebrandtNiemeyer2000,Roepke2006,Calderetal2013}.
The light curves of SNe Ia have the property that the brightness of an event
is correlated with its duration \citep{Phillips1993}.  This relation is the
basis for light curve calibration that allows use of these events as distance
indicators for cosmological studies (see \citealt{Conleyetal2011} for a
contemporary example).  However, their exact stellar origin remains unclear,
even in the face of extensive observational and theoretical study.
Recent early-time observations of the nearby SN~Ia 2011fe are challenging for
a variety of common progenitor scenarios, both single and double degenerate
\citep{Nugentetal2011,Lietal2011,Bloometal2012,Chomiuketal2012}.
Fitting of a wide variety of light curves with a simplified
ejecta model appears to require ejecta masses both at and below the
Chandrasekhar mass \citep{Scalzoetal2014}, however, indicating a variety of
progenitors may be present.

In thermonuclear supernovae, explosive nuclear combustion of a degenerate
carbon oxygen mixture proceeds in one or both of the deflagration and
detonation combustion modes.  In a deflagration, or flame, the reaction front
propagates by thermal conduction
\citep{TimmesWoosley1992,ChamulakBrownTimmes2007}, and is therefore subsonic.  In a
detonation, the reaction front propagates via a shock that moves
supersonically with respect to the fuel \citep{Khokhlov1989,Sharpe1999}.
These two combustion modes have been used to construct a variety of possible
explosion scenarios, either in combination, as in the deflagration-detonation
transition (DDT) scenario \citep{Khokhlov1991DelayedDet}, an example of which is
presented in this work, or singly as in the double-detonation model
\citep{LivneArnett1995,Finketal2010} or the pure-deflagration model
\citep{Finketal2014}.

\begin{figure}
\includegraphics[width=\columnwidth]{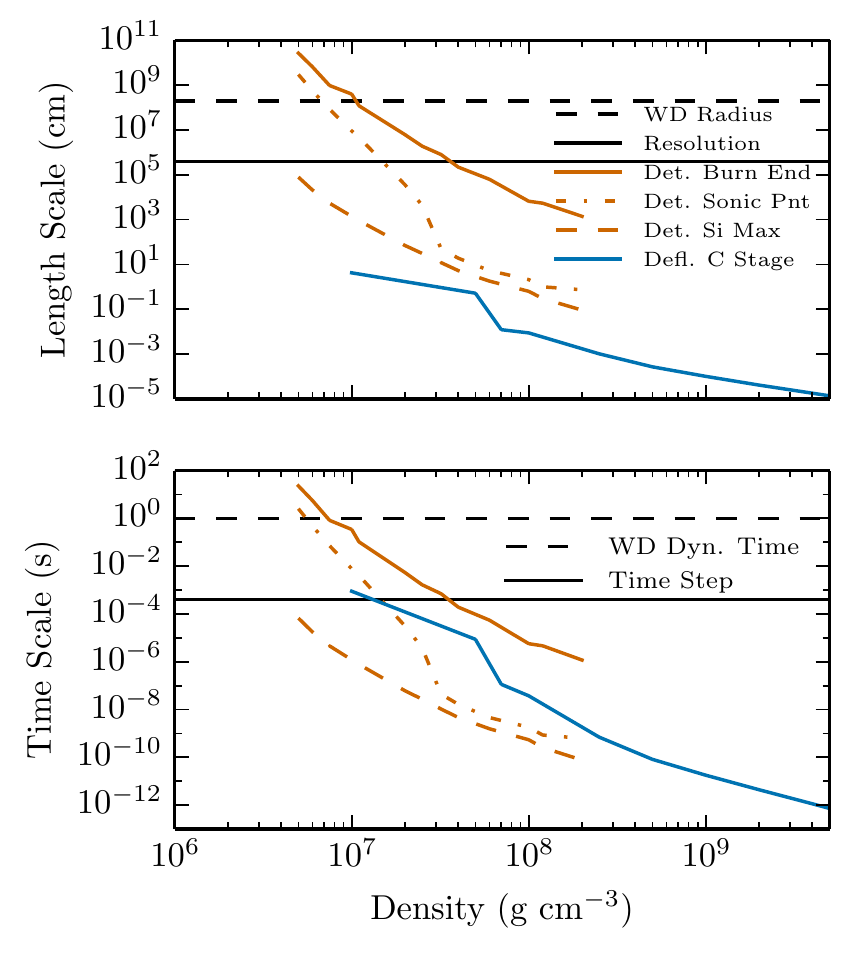}
\caption{\label{fig:ltscales}
Length and time scales of burning processes as a function of fuel density for
a mixture of $^{12}$C:$^{16}$O:$^{22}$Ne in the proportion 50:48:2 compared
to typical simulation resolution (4km, solid black lines) and scales of the
star (dashed black lines).  Stellar scales are taken to be the radius of the
initial star and the dynamical time.  The top panel shows the size scale of
various reaction front features while the bottom shows the self-crossing time
of these features at the propagation speed of the reaction front.  The
behavior of the $^{12}$C flame is shown by the solid blue curve that extends to high density \citep{ChamulakBrownTimmes2007}.
Scales for planar steady-state detonations are shown in red.  We show the
distance from the shock to three points in the detonation structure:  the
peak of $^{28}$Si abundance (dashed), which is also the end of $^{16}$O
consumption; the sonic point (dash-dot), also called the pathological point,
which is also the size of the detonation driving region; and the attainment
of the fully burned NSE state (solid), which is the completion of consumption
of $^{28}$Si.
The detonation driving region transitions
from being just resolved to being larger than the radius of the star between
densities of about $1.5\times 10^7$~g~cm$^{-3}$ and $7\times
10^6$~g~cm$^{-3}$.
} \end{figure}

A major challenge in simulations of SNe~Ia is capturing these burning
processes with confidence and accuracy.  The carbon-oxygen reaction fronts
transition from being unresolved by many orders of magnitude,
to being partially resolved, to finally being larger than the time
and length scales of the star.  Figure \ref{fig:ltscales} shows length and
time scales for detonations (red) and deflagrations (blue,
\citealt{ChamulakBrownTimmes2007}) at various densities.  For the stellar
scales we take the initial WD radius, $R=2\times 10^8$~cm, and the dynamical
time $2\pi\sqrt{R^3/GM}\approx 1$~s where $M$ is the WD mass.
A representative simulation resolution of 4~km is shown, along with the corresponding timestep of approximately the sound crossing time of a cell.
This resolution was found by \citet{Townsleyetal09} to be sufficient to give convergence in 1D with the thickened flame reaction front.
As a result our multi-dimensional (multi-D) simulations are commonly performed between 4 and 1 km to study resolution dependence in a regime in which convergence is demonstrated in 1D.
Several
different stages within a steady-state planar detonation front are indicated,
with distances measured from the shock that initiates the reactions and
propagates the front.  The shortest length scale shown (dashed line) is that
when the $^{28}$Si abundance peaks in time, which also corresponds to the
completion of $^{16}$O consumption.  The next length scale (dot-dashed line)
is the size of the detonation driving region, which is the distance to the
sonic point.  Finally the solid line shows the distance
to completion of burning, reaching the Fe-group element (IGE) dominated nuclear statistical
equilibrium (NSE) state.

The total yields of the explosion are determined by how and when the reaction
fronts stop propagating as well as what portion of the burning occurs within
the reaction front as opposed to what occurs after the reaction front itself
has passed.  The latter can then be influenced by the expansion of the star.  The
fairly thin range of densities, $1.5>\rho_{7}>0.7$, $\rho_7$ being density in
units of $10^7$~g~cm$^{-3}$, in which the detonation driving region
transitions from being unresolved to being larger than the radius of the star
is a manifestation of the difficulty of capturing the reaction dynamics
appropriately.  As the driving time and length scales get large, the
detonation may not be able to attain the planar steady-state structure.
Curvature of the front on scales comparable to the driving length, which will
occur due to the structure of the star, reduce the detonation speed and the
completeness of the burning \citep{Sharpe2001,DunkleySharpeFalle2013}.  The
long reaction times also mean that an ignited detonation may not reach steady
state before the star expands \citep{TownsleyMooreBildsten2012}.

Many recent results on multi-D simulations of SNe~Ia have computed
nucleosynthetic yields by post-processing the density and temperature
recorded by a Lagrangian fluid history during the simulation
\citep[e.g.][]{Travaglioetal2004_snia}.  A large nuclear reaction network is used to
integrate a set of species subject to this $\rho(t)$, $T(t)$ history.  The
burning model used in the simulation is therefore critical, as it determines
these histories.  Recent multi-D work
\citep{Maedaetal2010,Seitenzahletal2010,CiaraldiSchoolmannetal2013,Seitenzahletal2013}
has used the method described in the appendix of \citet{Finketal2010} to set
the energetics of the burning model used in the hydrodynamics.  In this
technique, the results of the post-processing are used to revise the output
of the model of burning and the process is iterated until the yields no
longer change.

In this work we pursue a different route toward construction of our burning
model and post-processing methods which, instead of an iterative bootstrap,
is based on comparison to separate resolved calculations of the deflagration and detonation modes.  The
burning model and post-processing method are then constructed with the goal
that the post-processed results reproduce the results of resolved
calculations of the steady-state structure of the reaction front even though the
actual reaction front is unresolved.  The resolved calculations to which we
want to compare are standard methods \citep[e.g.][]{FickettDavis1979} for the
computation of reaction front structure that can be performed with fairly
complete nuclear reaction networks and using error-controlled time
integration methods to eliminate most computational uncertainty.  Here we
succeed in matching steady-state yields for detonations at high densities and in planar
geometry.  Further development of benchmarks and methods for lower densities
and other geometries in future work will enable confident higher-accuracy
yields for an even larger fraction of the ejecta.

The burning model presented here is the successor to that initially
presented by \citet{Calderetal07} and \citet{Townsleyetal07}, with
tabulations presented by \citet{Seitenzahletal2009_nse}, that has been
used in a number of studies using large multi-D simulations of
SN~Ia
\citep{Jordanetal2008,Meakinetal2009,Jordanetal2012_puls,Jordanetal2012_faildet,Kimetal2013,Longetal2014}.
The capability to treat neutron-enriched fuel was added by
\citet{Townsleyetal09} in order to study how neutron enrichment in the
progenitor might influence the explosion.  The model presented here includes
a change in dynamics to better match iron-group production in detonations and
extends the treatment of initial composition to spatially non-uniform
abundances, allowing more realistic WD progenitors.  This has been used in
work exploring systematic effects of progenitor WD composition and central
density in the DDT scenario,
\citep{Kruegeretal2010,Jacksonetal2010,Kruegeretal2012}, as well as a
study of the turbulence-flame interaction during the deflagration phase
\citep{JacksonTownsleyCalder2014}, and consideration of hybrid C-O-Ne progenitor WDs \citep{Willcoxetal2016}.
Those studies, however,
did not proceed to nucleosynthetic post-processing, which is discussed in detail here for the first
time for our burning model.  The first work utilizing the post-processing for
astrophysical study is an investigation of spectral indicators of progenitor
metallicity \citep{Milesetal2016}.

We present below the structure of our burning model and post-processing
methods, along with particular assumptions currently in use in our SN~Ia
simulations, as well as tests performed so far comparing to calculations of
steady-state deflagrations and detonations.  Our burning model is based on
tabulation of physical quantities and fits of parameters based on resolved
steady-state calculations.  To improve accuracy in post-processing, we explore
supplementing the Lagrangian $\rho$-$T$ history recorded during the
hydrodynamic simulation with a reconstruction of unresolved processes based
on conditions near the reaction front when the fluid element is burned.

In section \ref{sec:burningmodel} we present the structure of our model for
carbon-oxygen burning including the basic variables and the form of their
dynamics.  Following this, we discuss our post processing treatment for
tracks (fluid elements) burned by the deflagration front in section
\ref{sec:def}.  This section is fairly brief since the application 
of a burning model
like that presented here to deflagrations was a major topic of previous work
detailed by \citet{Calderetal07} and \citet{Townsleyetal07}.  Detonations are
discussed in two sections.  Section \ref{sec:dethydro} develops the
error-controlled computation of steady-state detonation structure that we use
as a benchmark, calibrates the timescales in the burning model dynamics based
on this, and compares the resulting dynamics of the burning model in
hydrodynamic tests to the benchmark calculations.  The full method including
track post-processing is then outlined and tested in section
\ref{sec:detpost}.
Finally in section
\ref{sec:nucleosynthesis_computation} we detail how results from full-star
simulations are post-processed, and in section \ref{sec:2dddtyields} we show
the results of applying these methods to compute the yields of a 2D
simulation of the DDT model of SN~Ia, including a consideration of what we
can infer about current
uncertainties.  We summarize conclusions in section \ref{sec:conclusions}.

\section{Improved Parameterized Model for Explosive Carbon-Oxygen Fusion}
\label{sec:burningmodel}

We present here our current parameterized model for the thermonuclear 
burning of carbon and
oxygen fuel.  The model is intended to capture the dynamics of burning for
densities relevant to SNe~Ia for either the deflagration or detonation mode
of combustion.  Conversion of protons to neutrons (neutronization or deleptonization) is
included.  The initial abundance of carbon and neutron-rich elements (e.g.,
$^{22}$Ne) are allowed to vary with position in the WD.
The model is constructed to use a small number of scalars to
track the reaction state and products in order to improve computational
efficiency.
Accurate final-state energy release and electron capture rates are obtained by tabulation.
Abundances of intermediate burning stages are approximated and the formalism can be further refined by adjusting these if necessary.

The process of explosive carbon-oxygen fusion can be roughly divided into 3
stages -- C consumption, O consumption, and conversion of Si-group to
Fe-group material \citep{Khokhlov1989,Khokhlov2000,Calderetal07}.  The main
processes involved in each of these stages are: C destroyed to produce additional O,
Si, Ne, and Mg; O destroyed to produce Si, S, Ar, and Ca, generally in
nuclear quasi statistical equilibrium (QSE, sometimes called NSQE); $\alpha$
particles liberated by photodisintegration are then captured until this
material is converted into Fe-group, eventually reaching full nuclear
statistical equilibrium (NSE).  Due to differences in the rates of the
nuclear processes involved, at densities of interest these stages are
well-separated, in logarithmic time, and sequential.  This structure makes it
possibly to greatly simplify the complex reaction state and dynamics to the
behavior of a model containing just a few reaction progress variables.

\begin{figure}
\plotone{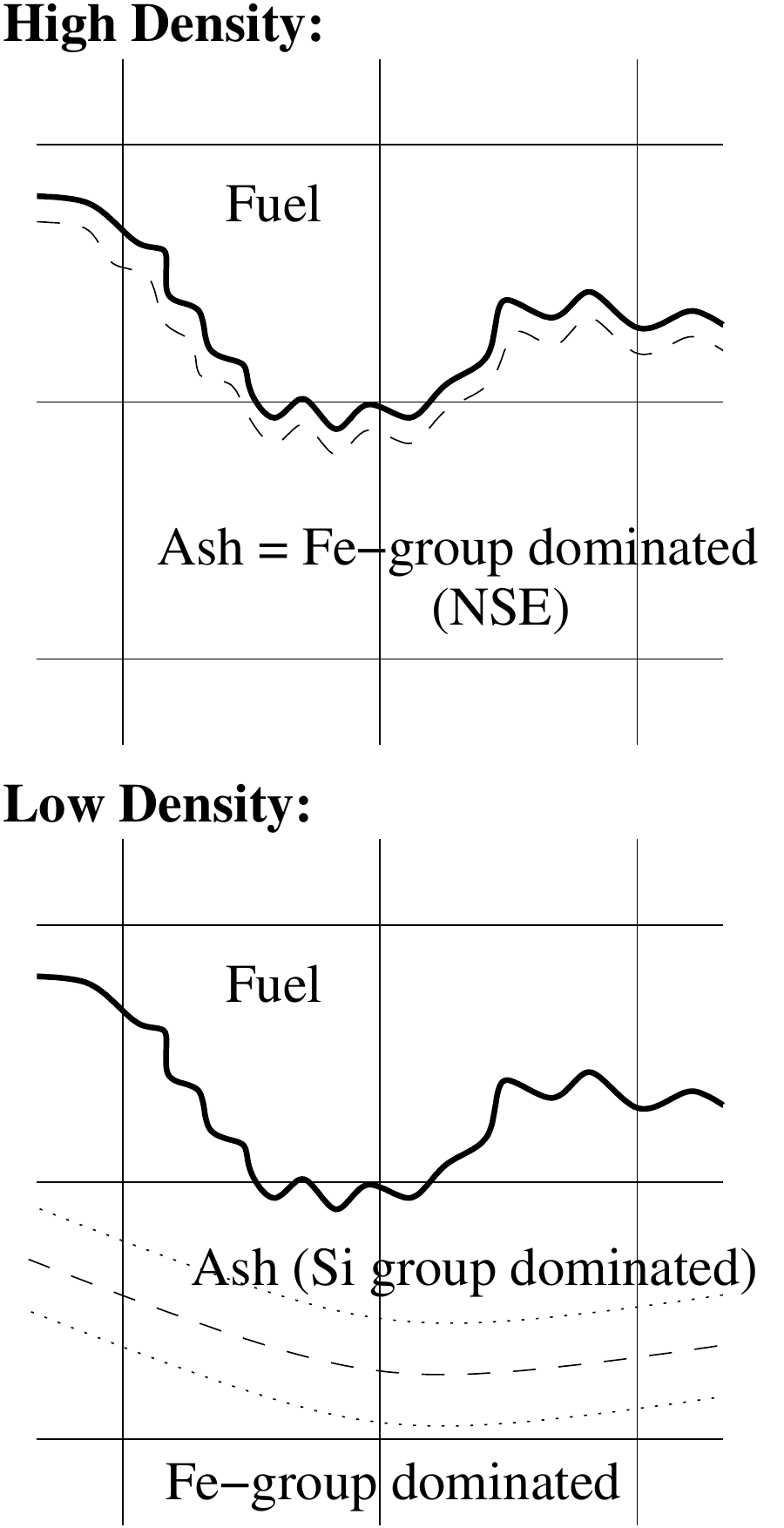}
\caption{\label{fig:subcell_front}
Diagram of structure of thin, multistage reaction fronts embedded in a coarse
computational grid of control volume cells.  At high densities the reaction
front is completely localized -- spatially thin -- such that a mixed computational cell contains
physically separated regions of fuel and ash material.  At lower densities,
some reaction stages remain localized, while others extend over multiple
cells, so that the reactant and product abundances are smoothly varying on
subgrid scales.
} \end{figure}

Individual cells are allowed to contain both unburned and fully burned
material in order to allow modeling of reaction fronts that are much thinner
than the grid scale.  This conceptual structure is shown in Figure
\ref{fig:subcell_front}, where curves are shown to represent two distinct
processes in the overall burn which occur on different timescales.  In this
example we will use O consumption as the shorter-timescale process and Si
consumption as the longer-timescale one.  The curves indicate the contour on
which the O abundance reaches half of its value in the fuel (solid) and where
the Ni abundance reaches half its final value (dashed).  Intermediate Ni
abundances are represented by dotted lines, which would not be distinct at
high densities (the separation between stages is exaggerated at high
density).  At high densities all reaction stages are localized on scales much
smaller than the grid, as indicated by the reaction length scales shown in
Figure \ref{fig:ltscales}, leading to cells which are volumetrically divided
into fuel and ash.  At lower densities, some burning stages become resolvable,
while others remain thin compared to the grid. For resolvable stages, the
actual abundance structure more closely resembles a spatial interpolation of
the coarse grid values.  This is demonstrated in the lower panel of
Figure \ref{fig:subcell_front}.  A structure like this is present for both
detonation and deflagration combustion modes, though in the turbulent
deflagration phase the thin reaction front structure can me much more
irregular than shown in this diagram.

Our burning model is currently implemented in the Flash code, an adaptive-mesh reactive hydrodynamics code with additional physics for astrophysical applications developed at the University of Chicago \citep{Fryxelletal2000,Dubeyetal2009}.
The model is readily adaptable for use with other similar reactive hydrodynamics software and the source code is available as add-on Units for the Flash code\footnote{astronomy.ua.edu/townsley/code}, distributed separately to allow a more liberal license.

\subsection{Definition of Stages and Relation to Fluid Properties}

The first step in abstraction of the fusion processes is defining the
relation of our progress variables to the actual physical properties of the
fluid.  The
transformations taking place via nuclear reactions act most fundamentally on
the abundances in the fluid.  Since we will reduce the burning processes to
just a few stages,
we must define first how these stages are related to the actual abundances.
After this we will proceed to develop reaction kinetics that will reproduce
the effects that the nuclear reactions have on the actual abundances and how
that manifests in the corresponding abstracted stages.

Our basic physics will be phrased in terms of baryon fractions,
denoted by the symbol $X_i$.  These are the fraction of the total
number of baryons which are in the form of the nuclide indicated by the label
$i$.  This is very similar to the traditional definition of mass fractions,
but avoids the ambiguity that rest mass is not conserved as nuclear reactions
take place due to energy release.
Since Baryon number is a conserved quantity, in the absence of sources
the baryon number density, $n_B$, satisfies the continuity equation:
\begin{equation}
\label{eq:barycons}
\pder{(n_B)}{t}  = -\vec\nabla\cdot(n_B\vec v)\ ,
\end{equation}
where $\vec v$ is the fluid velocity.  For reasons of convenience in a
non-relativistic fluid code, we will make the definition
\begin{equation}
\label{eq:rhodef}
\rho \equiv m_u n_B\, [=]\, n_B/N_A\ ,
\end{equation}
where $m_u$ is the atomic mass unit and $N_A$ is Avogadro's
number.  Here $[=]$ is used to denote "is numerically equivalent to in cgs
units."  Our intention is to place distinction between mass and (binding)
energy in the gravitational treatment; we could calculate the mass-energy
density if necessary, but computation of gravity in our simulation will just
use $\rho$ to approximate it.
The baryon fractions $X_i$ then identify directly the number of baryons in species $i$ and therefore, in the absence of reactions, also follow a conservation equation of the form
\begin{equation}
\label{eq:Xcons}
\pder{(Xn_B)}{t}  = -\vec\nabla\cdot(Xn_B\vec v)\ .
\end{equation}
Where $X$ may be $X_i$ or one of the progress variable defined below that will be constructed as linear combinations of the $X_i$.
Taken together
\eqref{eq:Xcons} and \eqref{eq:rhodef} mean that any linear combination of
baryon fractions can be treated as ``mass scalars'' by the advection
infrastructure in conservative fluid dynamics software (e.g.\ Flash; \citealt{Fryxelletal2000}; \citealt{Dubeyetal2009}).

\begin{figure*}
\resizebox{\textwidth}{!}{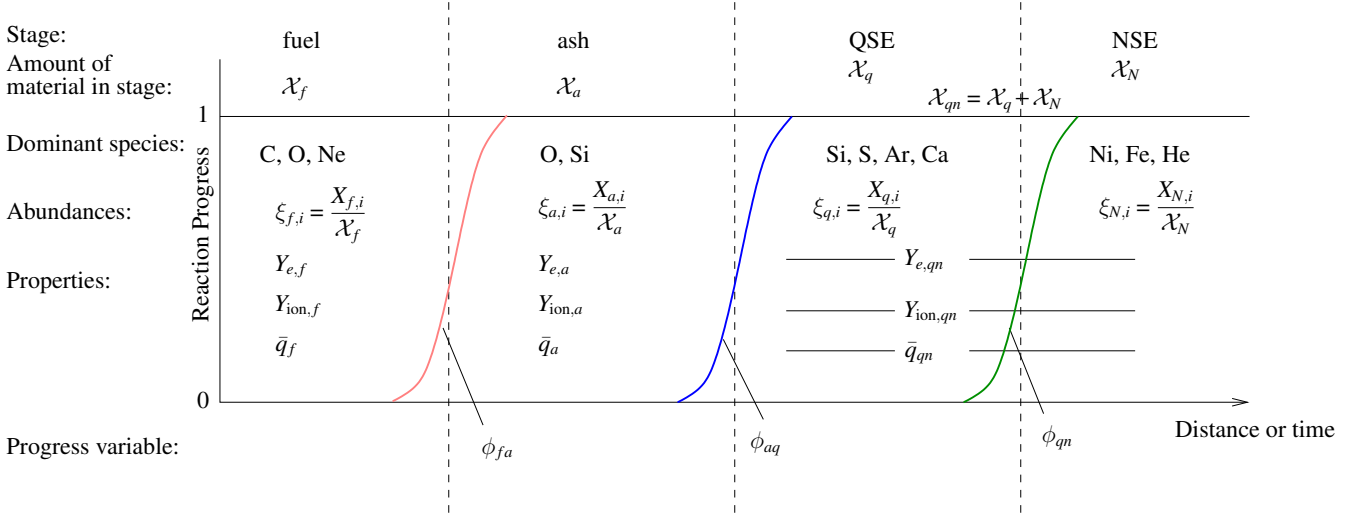}
\caption{\label{fig:stage_diagram}
Diagram of burning stages for C+O burning and associated symbols used here.
The progress of burning of a fluid element is from left to right, with the
horizontal axis indicating distance with respect to the foremost part of the
reaction front or time as a fluid element passes through the reaction front.
See section \ref{sec:space_comparison} for comparison to an actual reaction
front structure.
}
\end{figure*}

In order to start from quantities that satisfy \eqref{eq:barycons},
for purposes of tracking 3 stages of burning (that is three transitions), we
conceive of having four sets of all nuclides, each of which
represents a certain "type" of material:
\[ X_{f,i},\quad X_{a,i}, \quad X_{q,i}, \quad X_{N,i}\ ,\quad\quad
\sum_{\alpha,i}X_{\rm \alpha,i} = 1\ .\]
These denote, respectively, the baryon fractions of individual species
comprising fuel, (intermediate) ash (product of carbon consumption), a quasi-equilibrium (QSE) group, and a
terminal (NSE) group.
These stages and the various symbols used here are laid out in the diagram in
Figure \ref{fig:stage_diagram}.
This means that
any given
baryon has two labels: the type of nucleus in which it resides (e.g.
silicon), and whether we call that material part of, for example,  the ash or
the QSE material.  It is convenient to define 4
``superspecies'' by
\begin{equation}
\mathcal X_\alpha = \sum_i X_{\alpha, i}
\quad \alpha = \{ f, a, q, N\}\ .
\end{equation}
Since each of the burning stages follows in sequence from the earlier ones --
a feature unlike general nuclear species in a reaction network -- it is
convenient to define progress variables such that
\begin{equation}
\mathcal X_f = 1-\phi_{fa}\ ,\ 
\mathcal X_a = \phi_{fa}-\phi_{aq}\ ,\ 
\mathcal X_q = \phi_{aq}-\phi_{qn}\ ,\ 
\mathcal X_N = \phi_{qn}\ ,
\end{equation}
or
\begin{equation}
\phi_{fa} = \mathcal X_a + \mathcal X_q +\mathcal X_N\ ,\quad
\phi_{aq} = \mathcal X_q + \mathcal X_N\ ,\quad
\phi_{qn} = \mathcal X_N\ .
\end{equation}
By virtue of the property $0\le X_{\alpha, i} \le 1$ and thus $0\le\mathcal
X_\alpha \le 1$, we see that $1\ge\phi_{fa} \ge \phi_{aq} \ge \phi_{qn}\ge
0$.  Also, since the $\phi_{\alpha\beta}$ are simply linear combinations of
the $X_{\alpha,i}$ they also satisfy continuity, \eqref{eq:barycons}, in
the absence of sources.

We define a set of specific abundances:
\begin{equation}
\xi_{\alpha,i} \equiv \frac{X_{\alpha, i}}{\mathcal X_\alpha}\,\quad
\alpha = \{f, a, q, N\}\ ,
\end{equation}
so that $\sum_i\xi_{\alpha,i} = 1$.
This is useful because we will, at times, want to specify $X_{\alpha,i}
= \mathcal X_{\alpha} \xi_{\alpha, i}$ by specifying $\xi_{\alpha,i}$.  This
subtle distinction was left unaddressed in our previous revisions of this
burning model \citep{Calderetal07,Townsleyetal07}.
Note that since the $\xi_{\alpha,i}$ are quotients of the $X_{\alpha,i}$,
they are no longer linear combinations.
Nonlinear terms are any that contain products or quotients of fields that are position-dependent.
Linear combinations are required in order for the numerical scheme to be explicitly conservative.
While any general
algebraic combination, including a nonlinear one like a product or power, of
quantities satisfying \eqref{eq:barycons} still satisfies
\eqref{eq:barycons}, once the fields $\rho$, $X$, and $\xi$ are
discretized into values averaged over control volumes, i.e., mesh cells,
conservation of $X$'s no longer implies conservation of $\xi$'s due to
nonlinearities in the advection scheme.  This results from the property that
the average of a quotient is not the quotient of the averages.  Since we will
not compute \eqref{eq:barycons} for both the $X$'s and $\xi$'s, we must make
a choice of which will satisfy explicit conservation to numerical accuracy,
as performed by a conservative advection scheme like that in Flash.  Since
overall energy release is important, we choose to compute conservative
evolution for quantities that are linear combinations of the $X$'s.

In order to evaluate fluid properties and follow nuclear energy release into
the fluid, we must be able to obtain several bulk quantities.  We will
express these in units per baryon, such that obtaining units per cm$^{3}$ is
trivial using the Baryon density $n_B \equiv \rho / m_u$.
The two fluid quantities necessary are:
\begin{equation}
\text{Number of protons per baryon:}\quad
Y_p = Y_e \equiv \sum_{\alpha,i} X_{\alpha,i} \frac{Z_i}{A_i}
\end{equation}
\begin{equation}
\text{Number of ions per baryon:}\quad
Y_{\rm ion} \equiv \sum_{\alpha,i} X_{\alpha,i} \frac{1}{A_i}\ ,
\end{equation}
where, as customary, $Z_i$ is the number of protons and $A_i$ is the number
of protons plus neutrons in nuclide $i$.
Here we have assumed charge neutrality between the number of protons and the
number of non-thermal electrons, and defined $Y_e$ to include only the net
non-thermal electrons. Some $e^+$-$e^-$ pairs are created thermally at high
temperatures and these are accounted for in the EOS \citep{TimmesArnett1999},
and therefore are, in effect, advected with the energy field instead of as
fluid electrons included in our definition of $Y_e$.

For energetic purposes we need to be able to track the rest-mass
of our material rather than the approximation mentioned above.  This is
accomplished by tracking the nuclear binding energy per baryon:
\begin{equation}
\bar q \equiv
 \sum_{\alpha,i} X_{\alpha,i}\frac{Q_i}{A_i} 
= \sum_{\alpha,i} \frac{X_{\alpha,i}}{A_i}\left[
Z_i m_p + N_i m_n -m_i\right]c^2\ ,
\end{equation}
where $m_p$ and $m_n$ are the masses of the (free) proton and neutron
respectively and $m_i$ is the mass of 1 nucleus of nuclide $i$.
Note that $m_i$ is not the atomic mass, which is often given in mass tables and includes electrons and their binding energy.
The average
mass of a baryon in the fluid is
\begin{equation}
\bar m_B=\sum_{\alpha,i} X_{\alpha,i} \frac{m_i}{A_i}=
[m_n + Y_e(m_p-m_n)]c^2 - \bar q\ .
\end{equation}
Thus the actual rest mass density is $\rho_{\rm rest} = \bar m_B n_B = \rho
\bar m_B/m_u$.  Note that because the nuclear binding energy, $\bar q$, is
defined with respect to free protons and neutrons in the same proportion as
the material, calculation of the average rest mass requires both $\bar q$ and
$Y_e$, with the latter specifying the overall relative numbers of protons and
neutrons in the material.

We may now define the group-specific quantities
\begin{equation}
Y_{e,\alpha} = \sum_i \xi_{\alpha,i} \frac{Z_i}{A_i}\ ,\ 
Y_{{\rm ion},\alpha} = \sum_i \xi_{\alpha,i} \frac{1}{A_i}\ ,\ 
\bar q_{\alpha} = \sum_i \xi_{\alpha,i} \frac{Q_i}{A_i}\ .
\end{equation}
so that
\begin{equation}
Y_e = \sum_\alpha \mathcal X_\alpha Y_{e,\alpha}\ ,\ 
 Y_{\rm ion} = \sum_\alpha \mathcal X_\alpha Y_{{\rm ion},\alpha}\ ,\ 
 \bar q = \sum_\alpha \mathcal X_\alpha \bar q_\alpha\ .
\end{equation}
It is again important to note that the group-specific quantities such as
$\bar q_\alpha$ are not linear combinations of the $X_{\alpha,i}$ because the
$\xi_{\alpha,i}$ are quotients of linear combinations of the $X_{\alpha,i}$.
In order to maintain machine-precision advection of the discretized field
$Y_e$, for example, we will need to perform a conservative advection scheme
on the product $\mathcal X_\alpha Y_{e,\alpha}$ instead of separately on
$\mathcal X_\alpha$ and $Y_{e,\alpha}$, since their product will not evolve
conservatively to machine precision.  A similar statement holds for $Y_{\rm
ion}$ and $\bar q$.

We will derive the quantities $Y_{e,f}$, $Y_{{\rm ion},f}$, $\bar q_f$, and
$Y_{e,a}$, $Y_{{\rm ion},a}$, $\bar q_a$ from the initial
state.  Our simulation begins with fuel of known abundances,
\begin{equation}
X_{0,i}(\vec x), \quad \sum_i X_{0,i} = 1\ ,
\end{equation}
which may vary in space as indicated.  These initial abundance will satisfy
\eqref{eq:barycons} throughout our simulation; they will have no sources.
This allows us to, throughout the burning process, know how much of the local
baryons were in what form initially.  From these we define the properties of
the fuel,
\begin{eqnarray}
Y_{e,f}(\{X_{0,j}\}) &=& \sum_j X_{0,j}\frac{Z_j}{A_j}\ ,\nonumber\\
Y_{{\rm ion},f}(\{X_{0,j}\}) &=& \sum_j X_{0,j}\frac{1}{A_j}\ ,\\
\bar q_{f}(\{X_{0,j}\}) &=& \sum_j X_{0,j}\frac{Q_j}{A_j}\ . \nonumber
\end{eqnarray}
Additionally the ashes of the first stage of burning are assumed to be only a
function of the initial composition.
Thus
\[ \xi_{a,i} = \xi_{a,i}(\{X_{0,j}\}) \]
and is therefore also position dependent.  Then
\begin{eqnarray}
Y_{e,a}(\{X_{0,j}\}) &=& \sum_i \xi_{a,i}(\{X_{0,j}\})\frac{Z_i}{A_i}\
,\nonumber\\
Y_{{\rm ion},a}(\{X_{0,j}\}) &=& \sum_i \xi_{a,i}(\{X_{0,j}\})\frac{1}{A_i}\
,\\
\bar q_{a}(\{X_{0,j}\}) &=& \sum_i \xi_{a,i}(\{X_{0,j}\})\frac{Q_i}{A_i}\ .
\nonumber
\end{eqnarray}

As an aside, some concrete examples are useful.  In the
\citet{Townsleyetal07} burning model, the initial abundances were
$\{X_{0,^{12}\rm C}= 0.5,\  X_{0,^{16}\rm O} = 0.5\}$, constant in space, and
with other abundances zero.  Also the ashes of carbon consumption were
specified by $\xi_{a,^{16}\rm O} = X_{0,^{16}\rm O}$ and $\xi_{a,^{24}\rm
Mg}= X_{0,^{12}\rm C}$, with others again zero.  In \citet{Townsleyetal09}
the abundances of the fuel and carbon-consumption ash stages were effectively
modified to add a small amount of $^{22}$Ne, whose abundance was still
uniform in space, so that the initial abundances were $\{X_{0,^{12}\rm C}=
0.5,\  X_{0,^{16}\rm O} = 0.48,\  X_{0,^{22}\rm Ne}=0.02\}$, constant in space,
and carbon-consumption ash abundances were \{$\xi_{a,^{16}\rm O} =
X_{0,^{16}\rm O},\  \xi_{a,^{24}\rm Mg}= X_{0,^{12}\rm C},\  \xi_{a,^{22}\rm
Ne}=X_{0,^{22}\rm Ne}\}$.  In the model at hand we will use two parameters
$X_{0,^{12}\rm C}$ and $X_{0,^{22}\rm Ne}$ that vary in space to define the
initial state, and the $\xi_{a,i}$ are defined as previously.  More detailed
fuel abundances, or those containing additional major constituents such as
$^{20}$Ne or $^{24}$Mg also fit naturally into this scheme.

The fluid properties of the fuel and ashes of just the carbon
burning step depend almost entirely on the initial abundances.
For the equilibrium groups (QSE and NSE), however, all of these
properties change dynamically as the nuclear processing continues at high
temperatures.  The broad rearrangements of abundances which lead to the
variation of properties like $Y_{\rm ion}$ and $\bar q$ in the more processed
ashes are precisely the dynamics that we would like to abstract down to a few
parameters for the sake of computational efficiency.
To this end, we will treat gross properties of the quasi-equilibrium and
equilibrium groups together.  For convenience we define another
superabundance representing the total amount of material in either QSE or
NSE, $\mathcal X_{qn} = \mathcal X_q +\mathcal X_N = \phi_{aq}$.  This allows
us to collect the properties of the equilibrium groups by defining:
\begin{eqnarray}
\delta Y_{e,qn} &\equiv&\mathcal X_{qn} Y_{e,qn}\\
\delta Y_{{\rm ion},qn} &\equiv& \mathcal X_{qn} Y_{{\rm ion},qn}\\
\label{eq:deltaq_def}
\delta \bar q_{qn}&\equiv& \mathcal X_{qn} \bar q_{qn}
\end{eqnarray}
The $\delta$ prepended to the quantities here helps to indicate the somewhat
odd units involved.  For example, $\delta Y_{{\rm ion},qn}$ is the
number of QSE+NSE ions (nuclei) per \emph{fluid} Baryon, whereas $Y_{{\rm
ion},qn}$ itself is the number of QSE+NSE ions per QSE+NSE Baryon.  This
unit convention is the most awkward for $\delta \bar q_{qn}$. 
To restate why this is desirable:  If we had chosen instead to treat
$\bar q_{qn}$ directly, that would cause the total
nuclear energy $\bar q=\mathcal X_f\bar q_f+\mathcal X_a\bar q_a + \mathcal
X_{qn}\bar q_{qn}$, which is a nonlinear combination of $\mathcal X_{qn}$ and
$\bar q_{qn}$, to not be explicitly conserved by the conservative
hydrodynamics scheme.  Applying the conservative hydrodynamics to $\delta
\bar q_{qn}$ maintains conservation of the total nuclear energy.  This form
also makes it straightforward to derive appropriate dynamics, which we do
below.

Using the progress variables, intermediate state definitions, and QSE+NSE
material definitions, the bulk fluid properties can be restated as
\begin{eqnarray}
Y_e &=& [1-\phi_{fa}] Y_{e,f}(\{X_{0,j}\})\nonumber\\
&&\mbox{}+ [\phi_{fa}-\phi_{aq}] Y_{e,a}(\{X_{0,j}\})
  + \delta Y_{e,qn}  \label{eq:yelong}\\
Y_{\rm ion} &=& [1-\phi_{fa}] Y_{{\rm ion},f}(\{X_{0,j}\})\nonumber\\
&&\mbox{} + [\phi_{fa}-\phi_{aq}] Y_{{\rm ion},a}(\{X_{0,j}\})
   + \delta Y_{{\rm ion},qn} \label{eq:yionbreakout}\\
\bar q &=& [1-\phi_{fa}] \bar q_f(\{X_{0,j}\})\nonumber\\
&&\mbox{}+ [\phi_{fa}-\phi_{aq}] \bar q_{a}(\{X_{0,j}\})
   + \delta \bar q_{qn}   \label{eq:qbarbreakout}\ .
\end{eqnarray}
This defines the relationship of our burning model variables to the physical
fluid properties.

\subsection{Posited source terms}
\label{sec:sourceterms}

The previous subsection developed a framework in which the
properties of the parameterized burning stages can be expressed in a way that
can be advected in the absence of sources.  This leaves us to define
dynamical equations (source terms) for the $\phi_{\alpha\beta}$ themselves
and the properties of the equilibrium materials (the $\delta$ prefixed quantities).
By specifying these source terms here, we complete the form of the burning
model.

First a brief note on the form of the source terms which we will posit.
Typically we will write down source terms by specifying the Lagrangian
time derivative
\begin{equation}
\frac{DX}{Dt}=
\pder{X}{t} + \vec v\cdot \nabla X\ .
\end{equation}
In Eulerian form this gives
\begin{equation}
\pder{ (X\rho)}{t} = 
-\vec\nabla\cdot(X\rho \vec v) + \rho \frac{DX}{Dt}\ .
\end{equation}
Thus we are specifying the term in the evolution of the conserved quantity
that is due to transformations rather than just advection.

The evolution of the first stage of burning, $\phi_{fa}$, can be set
from a flame-tracking scheme or via a thermal reaction rate.  This is done
just as it is in \citet{Townsleyetal09},
\begin{equation}
\label{eq:phfadot}
\frac{D\phi_{fa}}{Dt} = \max(0,\dot\phi_{\rm RD})+\dot\phi_{\rm CC}\ ,
\end{equation}
where $\dot\phi_{\rm RD}$ is the reaction due to the
reaction-diffusion (RD) flame propagation calculation, and $\dot\phi_{\rm CC}$ is
thermally activated carbon-carbon fusion \citep{Townsleyetal09}.
The other progress variables then obey
\begin{eqnarray}
\label{eq:phaqdot}
\frac{D\phi_{aq}}{Dt} &=& \frac{\phi_{fa}-\phi_{aq}}{\tau_{NSQE}(T)}\ ,\\
\label{eq:phqndot}
\frac{D\phi_{qn}}{Dt} &=& \frac{(\phi_{aq}-\phi_{qn})^2}{\tau_{NSE}(T)}\ .
\end{eqnarray}
Here the $\tau_{NSQE}$ is the timescale previously determined in
\citet{Calderetal07} for oxygen consumption.  $\phi_{aq}$ reaches completion
at the peak Si abundance, when all oxygen is consumed.  The time and length
scales for completion of this stage were given as the dashed lines in
Figure~\ref{fig:ltscales}.  As can be seen there, this stage is mostly
unresolved in our simulations for steady-state detonations, including at all
densities important for Fe-group production, where the scales for completion
of burning are less than the stellar scales.

In contrast, as shown by the solid lines in Figure~\ref{fig:ltscales}, the
completion of processing of Si- to Fe-group, the $\phi_{qn}$ phase, can occur
on resolved scales for $\rho\lesssim 3\times 10^{7}$~g~cm$^{-3}$.
Additionally, this stage can be left incomplete for $\rho\lesssim
10^{7}$~g~cm$^{-3}$ by the limited length scales in the star and time of
expansion of the star.  Therefore, the dynamics of this phase are very
important for accurate total Fe-group yields.
The dynamics we are now using for $\phi_{qn}$, \eqref{eq:phqndot}, differs
from that used previously
\citep{Townsleyetal09,Jordanetal2008,Meakinetal2009,Jordanetal2012_puls,Jordanetal2012_faildet,Kimetal2013,Longetal2014},
$D\phi_{qn}/Dt=(\phi_{aq}-\phi_{qn})/\tau_{\rm NSE}$.
In the
process of performing the comparisons to benchmark detonation abundance
structures presented in section \ref{sec:space_comparison}, it was found that
the dynamics used previously
led to an approximately exponential relaxation of $\phi_{qn}$ that did not match the time
dependence of the consumption of Si as well as was hoped.  In order to
improve accuracy of our recorded Lagrangian histories, the dynamics applied
to $\phi_{qn}$ was altered to that of \eqref{eq:phqndot}.  This also
necessitates recalibration of the parameter $\tau_{\rm NSE}$, which will be
performed below in section~\ref{sec:taunse_calibration}.

Both of the parameterized timescales above, $\tau_{\rm NSQE}$ and $\tau_{\rm NSE}$, depend
on temperature, $T$, and some of the values used below also depend on
density.
However, there will be significant regions in the
artificial flame reaction front -- where $\phi_{\rm RD}$ is not near 0 or 1
-- that have a cell-averaged temperature and density that is not a good representation of the temperature of most of the fluid in the cell.
These are regions where, as shown in
Figure~\ref{fig:subcell_front}, a cell at the reaction front in reality
consists partly of unburned fuel and partly of fully burned material separated
by a thin front.
In this region our use of an artificially thickened reaction front gives a temperature intermediate between that of the fuel and ash.
The evaluation of the timescales also needs to be
stable as $\phi_{\rm RD}$ changes to obtain reasonable burning
dynamics.  By assuming the rest of the burning will occur at either constant
density or constant pressure, the final burned state, $\rho_f$, $T_f$, and
abundances can be determined based on the current local abundances and
thermal state \citep{Calderetal07}.  The constant pressure prediction
provides a reasonable approximation for the final burning state that will be
reached by the flame, and so the $\rho_f$ and $T_f$ of this final state are used to evaluate $\tau_{\rm
NSQE}$, $\tau_{\rm NSE}$, $\bar q_{\rm NSE}$, $Y_{\rm ion,NSE}$, and $\dot
Y_{e,\rm NSE}$ (see below) in regions were $10^{-6}<\phi_{\rm RD}<0.99$.
Otherwise, in regions away from the artificial flame the local temperature is
used to evaluate $\tau_{\rm NSE}$ and the temperature predicted for an
isochoric evolution is used to evaluate $\tau_{\rm NSQE}$, $\bar q_{\rm
NSE}$, $Y_{\rm ion,NSE}$, and $\dot Y_{e,\rm NSE}$.

Evolution of $Y_{e}$ due to electron capture occurs mainly by conversion of
Fe-group material, that is material that has at some point fully relaxed to
NSE.  At the
densities relevant to our SNIa computation, the timescale for relaxation to
NSE and the timescale for electron capture are well enough separated that
electron capture in material that is only partially fully relaxed to NSE is
not an issue.  However, due to the artificially thickened reaction front in
our SN~Ia simulations, a single cell at high densities will consist of an
artificial mixture of unburned fuel and fully relaxed NSE ash undergoing
electron capture.  To constrain electron capture evolution to relaxed NSE
material, we separate the components of $Y_e$ further into QSE and NSE portions:
\begin{equation}
\delta Y_{e,qn} = \delta Y_{e,q} + \delta Y_{e,N} =
\mathcal X_q Y_{e,q} + \mathcal X_N Y_{e,n}\ .
\end{equation}
For all but the NSE material,
$Y_{e,f} = Y_{e,a} = Y_{e,q} = Y_{e,0}\equiv Y_{e}(\{X_{0,i}\})$.
This simplifies \eqref{eq:yelong} to
\begin{equation}
Y_e = (1-\phi_{qn})Y_{e,0} + \delta Y_{e,n}\ .
\end{equation}
Applying the chain rule to $\delta Y_{e,n}$ gives
\begin{equation}
\frac{D(\delta Y_{e,n})}{Dt} = 
\frac{D\mathcal X_N}{Dt} Y_{e,n} + \mathcal X_N\frac{DY_{e,n}}{Dt}\ .
\end{equation}
The right hand side terms each have a distinct physical interpretation.  The
first is the change due to newly produced material, while the second is due
to the adjustment of the pre-existing material.
New NSE material is created with $Y_{e,0}$ and old NSE material evolves according to the
tabulated $\dot Y_{e,\rm NSE}$, which naturally gets scaled by the fraction of
material currently fully relaxed to NSE, $\mathcal X_N\equiv\phi_{qn}$, so that
\begin{equation}
\label{eq:yedynamics}
\frac{D(\delta Y_{e,n})}{Dt} = 
\frac{D\phi_{qn}}{Dt} Y_{e,0} + \phi_{qn}\dot Y_{e,\rm NSE}\ .
\end{equation}

Next we consider $\delta \bar q_{qn}$.  This represents the average
binding energy of all material involved in incomplete Si burning, whether
currently in QSE or having progressed fully to NSE.  Using the chain rule on
\eqref{eq:deltaq_def} splits this into two contributions,
\begin{eqnarray}
\frac{D(\delta \bar q_{qn})}{Dt} &=& 
\frac{D( \phi_{aq})}{Dt} \bar q_{qn}
+\phi_{aq} \frac{D(\bar q_{qn})}{Dt}\ .
\end{eqnarray}
In earlier versions (\citealt{Townsleyetal09} and prior) of this burning
model, we posited dynamics in which the binding energy relaxed to the NSE
value on the shorter
relaxation timescale, $\tau_{\rm NSQE}$.  However, in verification
comparisons to detonation structures it was found that at low densities this
released energy too quickly and led to under-prediction of the temperature
just behind the unresolved portion of the detonation front.  In order to
improve this behavior, we here introduce a $\delta\bar q_{\rm QSE}$ that
changes as Si-group material is converted to Fe-group, as measured by the
progress variable $\phi_{\rm qn}$,
\begin{equation}
\delta\bar q_{\rm QSE}=(\phi_{aq}-\phi_{qn})\bar q_{\rm QSE0} + \phi_{qn}\bar q_{\rm NSE}\ .
\end{equation}
Relaxation toward this value is assumed to occur via $\alpha$ capture or
photodisintegration, and thus take place on the shorter timescale $\tau_{\rm NSQE}$.
To capture these two timescales we posit the following dynamics,
\begin{eqnarray}
\label{eq:qbardynamics}
\frac{D\delta q_{qn}}{Dt} &=&
\frac{D(\phi_{aq})}{Dt} \bar q_{\rm QSE0} \nonumber\\
&&\mbox{} + \frac{1}{\tau_{\rm NSQE}}\left[
(\phi_{aq}-\phi_{qn})\bar q_{\rm QSE0} + \phi_{qn}\bar q_{NSE}-\delta \bar
q_{qn}\right]\ ,
\end{eqnarray}
where the evolution on the timescale $\tau_{NSE}$ is contained in
$\phi_{qn}$.
Here $\bar q_{\rm QSE0}$ represents the $\bar q$ of the material at the
completion of O consumption, that is the initial QSE state.  This state is less
easily quantified at high densities, as it may contain a significant, and
density-dependent, fraction of $\alpha$ particles, but it will only be
important at low densities when the Si burning is resolved in the simulation.
While, therefore, the most
appropriate value for $\bar q_{\rm QSE0}$ is likely to be density- and
composition-dependent, for simplicity we will use $\bar q_{\rm
QSE0}=q_{^{28}\rm Si}$, which appears mostly sufficient in the verification
tests performed.

The evolution of $Y_{\rm ion}$, or equivalently the ion mean molecular
weight, $\bar A$, poses a similar challenge to $\bar q$.
Each of the QSE and NSE materials will relax the balance between heavies and
$\alpha$/protons/neutrons on approximately the NSQE relaxation time, whereas
the conversion between QSE and NSE occurs more slowly.  We resolve this by
using the scalar that tracks relaxation toward NSE, $\phi_{qn}$, to
appropriately mix approximations of $Y_{\rm ion}$ for the QSE and NSE states
and then set our dynamics to move toward this value on the NSQE timescale.
Working in a way similar to the construction of \eqref{eq:qbardynamics},
\begin{eqnarray}
\label{eq:yiondynamics}
\frac{D(\delta Y_{{\rm ion},qn})}{Dt} &=&
\frac{D\phi_{aq}}{Dt} Y_{\rm ion,QSE0}\\
&&\mbox{} +
\frac{1}{\tau_{NSQE}}\left[
(\phi_{aq}-\phi_{qn})Y_{\rm ion,QSE0} + \phi_{qn}Y_{{\rm ion},NSE}
-\delta Y_{{\rm ion},qn}
\right]\ .
\nonumber
\end{eqnarray}
It is left to obtain a suitable estimate of $Y_{{\rm ion},QSE0}$ for the QSE
state.  We found that at densities $\lesssim 10^7$~g~cm$^{-3}$
that the simple estimate $Y_{\rm ion,QSE0}\equiv Y_{^{28}\rm Si}=1/28$
provides a well-behaved approximation that matches $\bar A$ produced by
benchmark detonation calculations within 10\%
(see section~\ref{sec:space_comparison}).
A somewhat complex approximation was proposed in \citet{Townsleyetal09}, but
it did not yield a better match to $\bar A$ in testing.

The dynamics of our parameterized model for CO burning are contained in
Equations (\ref{eq:phfadot}), (\ref{eq:phaqdot}), (\ref{eq:phqndot}),
(\ref{eq:yedynamics}), (\ref{eq:qbardynamics}), and (\ref{eq:yiondynamics}).
The energy release is computed based on conservation of energy, giving the
energy release rate per mass,
\begin{equation}
\epsilon_{\rm nuc} =
\dot{\bar q} - \phiqn[\dot Y_{e,\rm NSE}N_{\rm A}c^2(m_p+m_e-m_n) +
\epsilon_{\nu,\rm NSE}]\ ,
\end{equation}
where $m_p$, $m_e$, and $m_n$ are the masses of the proton, electron, and
neutron respectively, and $\epsilon_{\nu,\rm NSE}$ is the energy loss
to emission of neutrinos based on the local predicted NSE abundances.
While the burning dynamics has been stated analytically, the resulting
differential equations must now be implemented in a way that is numerically
efficient.  It is possible to exploit some aspects of the separation between
timescales and the strict ordering of the burning stages to make the
integration of these dynamical equations extremely efficient.  This is
discussed in Appendix \ref{sec:burning_implementation}.

\subsection{Calculation of Nucleosynthesis Using Post-Processed Lagrangian
particle Histories}

The burning model presented here is intended to give approximately the right
energy release, as determined by direct computation of steady-state reaction
front structure with large, complete nuclear networks and error-controlled
numerical methods, but with a relatively low computational cost.  In order to
recover detailed abundances, Lagrangian fluid histories are recorded from the
hydrodynamic simulation and post-processed.  Our post-processing is described
in later sections.

The Flash code includes the capability to produce Lagrangian fluid histories
through the use of "tracer" particles \citep{Dubeyetal2012}.  These are particles whose position is
calculated as
\begin{equation}
\label{eq:particles}
\vec x(t) = \vec x_0 + \int_{t_0}^{t} \vec v[\vec x(t'),t']\, dt'
\end{equation}
Where the time-dependent velocity field $\vec v[\vec x,t]$ is simply that
determined by the hydrodynamic evolution.  Generally the number of particles
followed and the distribution of initial positions $\vec x_0$ are chosen to
provide a sampling that is useful for nucleosynthesis
\citep{Seitenzahletal2010}, though here we use a simple weighting in which
each tracer represents an equal mass and initial positions are chosen
randomly to follow the mass distribution.
This random distribution is achieved as follows:
The domain is decomposed into blocks of $8^d$ cells, where $d$ is the dimension, 2 in this case, and we are using blocks that are 8 cells on a side.
The mesh structure in Flash provides an ordering for these blocks, called the Morton ordering \citep{Fryxelletal2000}.
We split the mass of the star into segments based on how much mass is contained in each block, using the same order as the Morton ordering.
A random number between zero and the total mass is then generated for each particle, and the segment in which it falls determines the block in which that particle is initially placed.
A similar procedure is repeated at the block level, using the mass of material in each cell.
Once the cell in which the particle will be placed is chosen, each coordinate of the location of the particle within the cell is chosen randomly and uniformly across each dimension of the cell.
The impact of the finite sampling represented by this distribution on the uncertainty of our results is discussed in Appendix~\ref{sec:samplinguncertainty}.

The Lagrangian tracks are then computed at
the same time as the hydrodynamics.  The method used to perform the
integration of the particle positions
is essentially a second-order Runge-Kutta scheme with the
velocity field sampled at the end of each combined hydrodynamics and energy
source step and linearly interpolated to the particle position.
Note that in the directionally-split hydrodynamics solver, which is used
here, each hydrodynamics step consists of multiple sweeps of the
1D PPM method to allow for multi-D problems
\citep{Fryxelletal2000}.

\section{Deflagration Fronts}
\label{sec:def}

\begin{figure}
\plotone{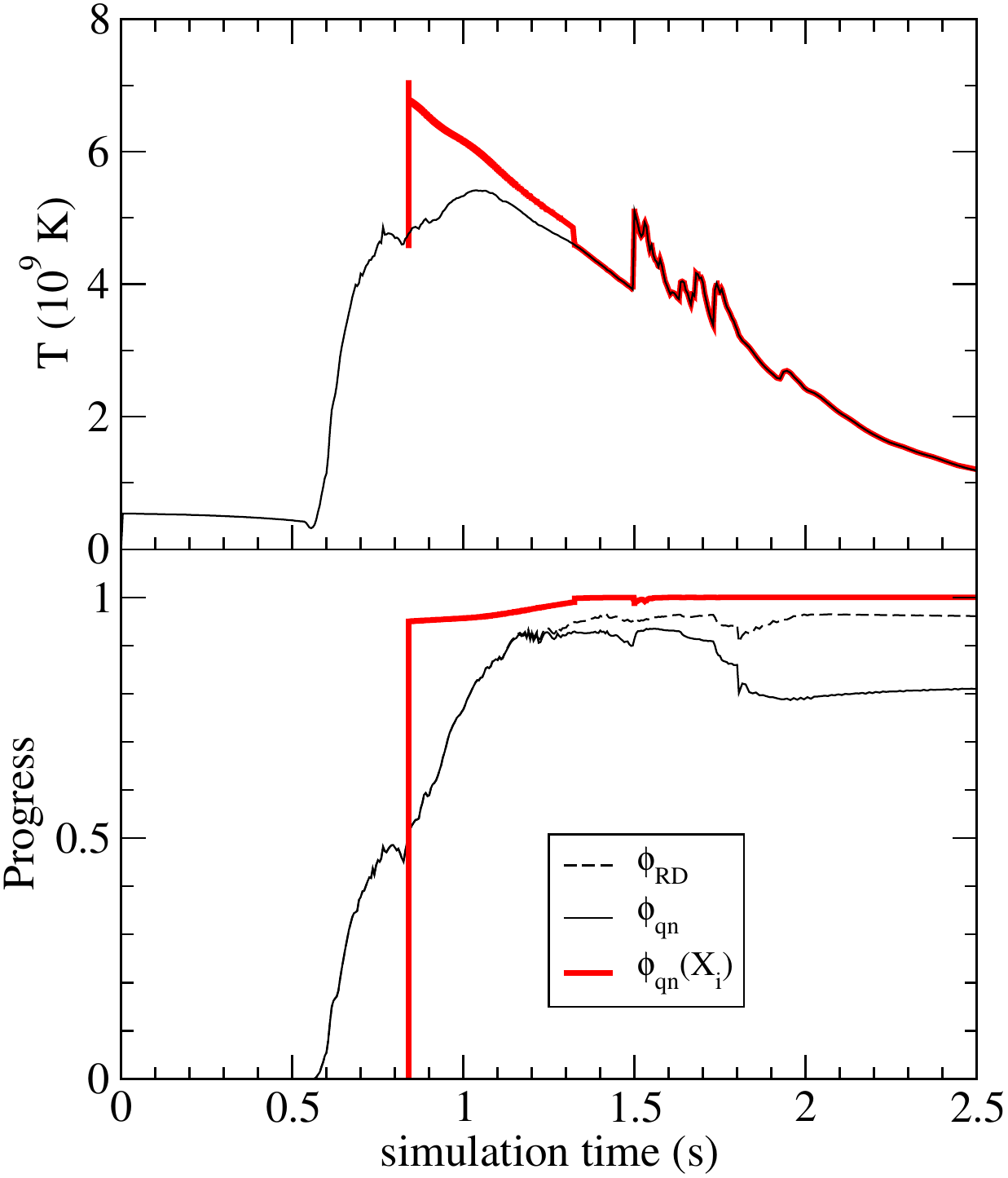}
\caption{\label{fig:flame_post_demo}
Thermal and burning progress histories for fluid burned by a deflagration
front.
Shown is the temperature (upper panel thin solid), reaction-diffusion front
progress variable ($\phi_{\rm RD}$, lower panel dashed) and QSE to NSE
progress variable ($\phi_{qn}$, lower panel thin solid) recorded at the position of the Lagrangian
tracer particle embedded in material ejected at approximately
5\,000~km~s$^{-1}$.  $\phi_{\rm RD}$ and $\phi_{qn}$ are identical up to about
1.2~s, at which time the fluid element reaches low enough temperature and
density that the separation between burning stages begins to become spatially
resolvable.  Also shown (thick red solid lines) are the reconstructed temperature history used in the
post-processing calculation of nucleosynthetic yields
and an analog of $\phi_{\rm qn}$ constructed from the detailed
abundances, $X_i$, computed during post processing (see Equation \ref{eq:phiqnabund}).
}
\end{figure}

Particles representing fluid burned by a deflagration front must be treated
differently from those undergoing detonation because the true burning
structure differs from the effective one used in the simulation.
In some ways treatment
of the particles undergoing deflagration is more straightforward because the
combustion in the hydrodynamical calculation has been made into a spatially resolved
process by coupling it to the RD front as given in \eqref{eq:phfadot}.  The
parameterized dynamics used for the RD front are the same as those discussed
in \citet{Townsleyetal09}, basically causing the 4-zone wide reaction front
to propagate at a specified speed.  However, since the flame is generally
quite subsonic, with Mach number ${\rm Ma} \ll0.01$,  it will typically take
many time-steps, approximately $4/{\rm Ma}$, for a fluid element tracked by a
Lagrangian tracer particle to pass fully through the RD front.  In our
simulations this is several tenths of a second, as can be seen by the
progress variable and temperature histories shown by the solid black lines in
Figure~\ref{fig:flame_post_demo}.  During this time, by
construction \citep[][and section~\ref{sec:sourceterms} above]{Townsleyetal09}, the local temperature is not physical,
but a mixture between burned and unburned states in approximate pressure
equilibrium.  This makes it essential to perform a reconstruction
of the portion of the particle's thermodynamic history during which it is
still inside the RD front, before the fully burned state is reached.

The black line in the upper panel of Figure \ref{fig:flame_post_demo} shows a
typical temperature history for a tracer particle embedded in material
ejected in a DDT SNIa at approximately $5\,000$~km~s$^{-1}$.  The bottom panel
shows the evolution of the progress variable representing relaxation toward
Fe-group, $\phi_{\rm qn}$ (solid black line).  As can be
seen, the transition from unburned to nearly fully burned covers times from
about 0.6~s to 1.2~s, and the slow rise in temperature seen in the upper
panel covers a similar time range.  During this interval, the density and temperature are
not representative of a physical burning process, but are instead the average of the burned and unburned states based on the fraction of the cell burned as indicated by the artificially thickened reaction front (see
Figure~\ref{fig:subcell_front}).
This makes
calculation of, for example, the electron-capture history of this fluid
element based on a direct post-processing of the $\rho(t)$,
$T(t)$ history inappropriate.

We attempt to reconstruct a reasonable approximation to the
temperature-density history that a fluid element would have undergone passing
through a flame of realistic thickness.  The reconstruction of the portion of
the fluid history that elapses while the particle is within the artificially
broad reaction region is obtained by assuming that the pressure jump across
the flame is small, $\lesssim 1\%$ \citep{VladimirovaWeirsRyzhik2006,Calderetal07}.  This will be true as long as the Mach number of the
flame propagation is low, as is the case for our simulations.  Under this
assumption, although the local density and temperature are not representative
of the actual values, the local pressure should be similar to that near the
actual thin flame front to within approximately the Mach number.
In order to use this feature, we perform
self-heating calculations with a pressure history specified from the fluid
histories extracted from the hydrodynamic simulation.  This novel mode of
specified-pressure-history self-heating was added to the TORCH nuclear reaction
network \citep{Timm99}\footnote{Original sources available
from http://cococubed.asu.edu.  Our modifications are available from
http://astronomy.ua.edu/townsley/code}.
The set of 225 nuclides used includes all
those indicated in the discussion of weak reactions in \citet{Calderetal07},
which includes an extension to neutron-rich nuclides near the Fe group
compared to the standard 200 nuclide set used in TORCH.
Weak
cross sections were taken from \citet{Fulleretal1985}, \citet{Odaetal1994},
and \citet{LangankeMartinezPinedo2000}, with newer rates superseding earlier
ones.

Assuming that the
fluid element actually crosses the flame front when the progress variable
passes through $\phi_{RD}=0.5$, the reconstructed temperature history is
shown by the red curve in the upper panel of Figure
\ref{fig:flame_post_demo}.  It is notable that the temperature peak is much
higher and occurs about 0.2 seconds sooner.  The initial condition for the
trajectory is found by performing a short computation at constant pressure
that was raised high enough for the $^{12}$C to begin burning ($2\times
10^9$~K), continuing until the $^{12}$C abundance is 0.1.  The
specified-pressure self-heating follows this.
Once the fluid element exits the artificial reaction front, post-processing
can proceed from there using the recorded temperature-density history.  We
take this point to be when $\phi_{RD}>0.95$ in the recorded history, or when
$P< 10^{22}$~erg~cm$^{-3}$, whichever comes first.  This $P$ corresponds
roughly to when burning of heavier elements will cease, when $\rho\lesssim
10^6$~g~cm$^{-3}$ and $T\lesssim 2\times 10^9$, and it is more convenient to
impose the condition in $P$ than in $\rho$ or $T$ directly.
In Figure
\ref{fig:flame_post_demo}, this transition occurs just after $t=1.3$~s.

The red line in the bottom panel of Figure \ref{fig:flame_post_demo} shows a
progress variable constructed from the full set of species treated in the
post-processing,
\begin{equation}
\label{eq:phiqnabund}
\phi_{qn}(X_i) \equiv
\frac{X_{\text{IGE+LE}}}{X_{\text{IME}}+X_{\text{IGE+LE}}}
\end{equation}
where
\begin{eqnarray}
X_{\text{IME}} &=& \sum_{2 < Z_i \le 22} X_i\ , \\
X_{\text{IGE+LE}} &=& \sum_{Z_i\le 2,~Z_i>22} X_i\ .
\end{eqnarray}
This effective progress variable measures the process that $\phi_{qn}$ is
intended to track, the conversion of Si-group, or generally intermediate
mass elements (IME) to IGE.  In NSE, there can also
be a significant fraction of light elements (LE, protons, neutrons,
$\alpha$'s) that will be present throughout the transition, but will
eventually be captured to form more IGE as the temperature falls.
Here the completeness of processing from IME to IGE is
comparable between the parameterized burning performed in the hydrodynamic
simulation and the post-processed values, with the post-processing giving
complete conversion to IGE and $\phi_{qn}$ indicating more than 95\%
converted to IGE.  The reduction in $\phi_{qn}$ at
late times, starting at approximately 1.8~s, is due to mixing with
surrounding zones in the hydrodynamic simulation as the grid is coarsened
from 4 to 16 or 32~km cells in order to accommodate the expanding ejecta.

\begin{figure}
\plotone{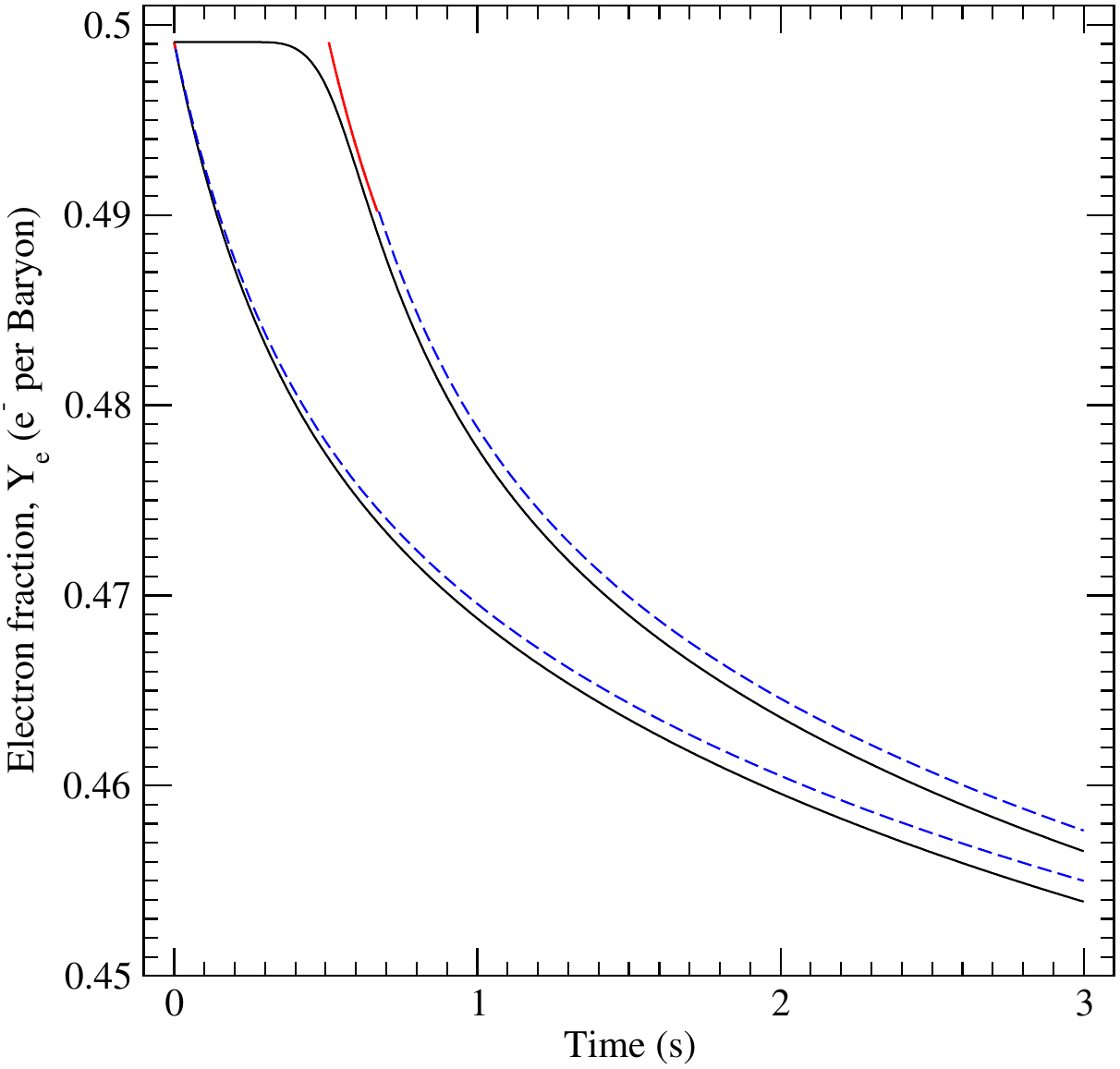}
\caption{\label{fig:ye_comparison}
The time history of $Y_e$ for two fluid elements burned by the artificial
flame starting at a density of $10^9$~g~cm$^{-3}$.  The time history on the
grid computed from the simplified burning model is shown in black, whereas the
history computed in post processing is shown in red (reconstructed portion)
and blue (direct post-processed portion).  Two fluid
elements are treated, one that begins the simulation in the burned state
(lower black), and one that the flame passes through after 0.5~s (upper
black).
}
\end{figure}

In order to verify that the neutronization is captured well by this method,
we turn to 1D simulations in a spatially uniform density medium.
For a low Mach
number flame in these conditions, a constant-pressure self-heating
calculation is a good approximation to the correct fluid history
\citep{VladimirovaWeirsRyzhik2006,Calderetal07,ChamulakBrownTimmes2007}.  A comparison of the $Y_e$ history
obtained from the hydrodynamics and the fluid element history post-processed
as described above is shown in Figure \ref{fig:ye_comparison}.  The histories
of two fluid elements are taken from a simulation in which an artificial flame
is propagated from a hard wall into 50:50 CO fuel at a uniform density of
$10^9$~g~cm$^{-3}$ with a flame speed of $5\times 10^6$~cm~s$^{-1}$.
The first fluid element begins in the burned region.
Its initial state is determined by the Rankine-Hugoniot jump conditions
satisfied across the flame front as used to set the initial condition of the
simulation.  The second fluid element is taken from a position that the
flame passes at about 0.5~s.  The reconstructed portion of the
post-processing is shown in Figure \ref{fig:ye_comparison} by the solid red
portion of the curves and the direct density-temperature history
post-processed portion is shown by the blue dashed lines.  The black curves
show the $Y_e$ according to the burning model, \eqref{eq:yelong}, at the
fluid element position in the hydrodynamic simulation.  The agreement is
fairly good, with the change in $Y_e$ from 0.5 matching within a few percent
for both the initially burned case and the case passing through the
reconstructed portion at 3~s, a few times longer than expected exposure in an
explosion simulation.
This provides confirmation that scaling $\dot Y_e$ with $\phi_{qn}$
in the burning model provides a reasonable behavior even with a thickened
reaction front.  The difference between the $Y_e$ time history given by the
simulation and the post-processing appears consistent with the use of a larger
nuclide set to compute the neutronization rate tables used in the burning
model \citep{Seitenzahletal2009_nse}.  Using a larger nuclear network for
post-processing would improve this difference at some cost to efficiency.

Ideally the $Y_e$ histories of the two fluid elements would just be shifted
by a time delay based on when their burning began.  However, the flame
propagation in physical space is slowing somewhat due to the loss of pressure
due to neutronization of the earlier burned material.  This causes the
later burned fluid element to be at a slightly higher density at a given time
interval after burning began.
The first several tenths of a second of evolution match well in both cases,
demonstrating that the post-flame state is consistent with the
Rankine-Hugoniot calculation as expected.

\section{Detonation Hydrodynamics}
\label{sec:dethydro}

In this section we demonstrate the detonation structure we wish to reproduce
and we test the burning model in hydrodynamic simulations in comparison to
this benchmark.
Although it was developed initially for deflagrations in carbon-oxygen mixtures,
the reaction structure of detonations is similar enough
\citep{Khokhlov1983,Khokhlov1989} that the 3-stage model can also be applied
to them.  In the simplest form, this just involves identifying the first
stage, $^{12}$C consumption, with the rate of the actual $^{12}{\rm
C}+^{12}{\rm C}$ reaction, and then following the later burning stages.  This
was done, for example, in \citet{Meakinetal2009}, and we will do something
similar here, with some adjustments for improved accuracy.

As can be inferred from the length scales shown in Figure \ref{fig:ltscales},
the actual burning structure is not resolved in full-star simulations.
Therefore, somewhat like in the case of the deflagration, the dynamics which
lead to the reaction front propagation are not the same in the simulation as
in reality.  The physics is similar; the energy release determines the
strength, and therefore speed, of the detonation shock.  However, the
acoustic structure in the simulation is not the same as the physical detonation structure.  Reactions
must be suppressed in the numerically unresolved shock in order to prevent
numerical diffusion from dominating the propagation of the reaction front
\citep{FryxMuelArne89}.  This creates an artificial separation of a few
zones between the shock and the reaction zone.  In addition, the reactions
may run to near completion within the single zone in which reactions are
re-enabled downstream of the shock.  We show in Appendix \ref{app:ppmdet}
that the widely-used technique of disabling reactions in the zones adjacent
to the shock reproduces the steady state detonation speed and the resolved
portions of the reaction structure.

Here we present the error-controlled calculation of the 1D
structure of planar detonations that we will use as our benchmark for both
the burning model in hydrodynamics and the Lagrangian post-processing.  After
introducing this benchmark, the remainder of this section will focus on how,
in comparison, the burning model acts in hydrodynamic simulations.
Post-processing will be discussed in section \ref{sec:detpost}.  As already
mentioned in the presentation of the burning model in section
\ref{sec:burningmodel}, simply treating $\phi_{fa}$ according to the C
reaction rate and then proceeding as discussed in \citet{Townsleyetal07}
turned out in testing to not reproduce partially-resolved detonation
temperature and abundance structures at intermediate densities,
$10^{6}$-$10^{7}$~g~cm$^{-3}$.  The successful comparison to benchmarks shown
in this section is the result of making the required adjustments to the
burning model timescales discussed in section \ref{sec:taunse_calibration}.

\subsection{Verification Benchmark: The ZND structure}
\label{sec:znd}

In order to evaluate the realism of our simplified model of burning,
it is necessary to define an authoritative reference with which
it will be compared.  Since, as one might expect,
no direct experimental validation of nuclear
detonations in stellar matter are available, we instead turn to a
hierarchical approach to validation~\citep{calder_2002_aa}. 
Following this practice, our interest is in verifying that
burning characteristics of our models are similar enough to those computed
with methods in which we have more confidence.
A typical benchmark in a hierarchical verification like this
would be a direct numerical simulation (DNS) of similar phenomenon with more
detailed, and typically separately verified, treatments of physical
processes.  Another source of benchmarks is particular configurations or
steady states that can be computed more easily, for example in lower
dimension, or in more detail and with better numerical error control.

As one of the two combustion modes in SN~Ia explosions, the
predicted outcome of C-O detonations have been discussed in some detail
previously in the astrophysical literature.  \citet{Khokhlov1989} presented an
overview of the microscopic structure of steady-state planar C-O and He
detonations at a variety of densities.  Further work by \citet{Sharpe1999}
extended calculations of the structure of the planar steady-state structure
and products beyond the sonic point in the detonation wave, allowing the
completion of burning to be computed at a wider range of densities.
\citet{Sharpe2001} followed this up with computations of detonation speeds
and structure for non-planar, i.e. curved, detonation fronts in steady state,
still in one dimension.  \citet{Gamezoetal1999} and \citet{Timmesetal2000}
investigated the multi-D structure of C-fueled detonations with
high resolution reactive hydrodynamics for cases important for SNe~Ia.
Recently, \citet{DominguezKhokhlov2011} performed a high-resolution
investigation into the stability of C-fueled detonations in 1 spatial
dimension at low densities, $\lesssim 10^6$~g~cm$^{-3}$.

We are interested here in an inherently transient phenomena as the
detonation traverses different densities within the star.
As a result, the ideal benchmark
is simulations of the reactive Euler equations which include all relevant
nuclides (and therefore all relevant reactions) and in which all important
length scales are resolved.
The component models of such a DNS have been separately validated in many
contexts, and their limitations are fairly well understood.
Unfortunately, a DNS is challenging for the nuclear processes under
consideration here.  In order to fully capture the reaction kinetics, it is
necessary to include hundreds of species.  The more severe limitation,
however, as demonstrated in Figure \ref{fig:ltscales}, is the large
separation of time and length scales between the final reaction
stages -- those which process Si-group to Fe-group elements or perform
electron captures -- and the reactions which drive the burning front forward,
fusion of carbon.  At the densities of most interest, where the
nucleosynthetic processing to Fe-group is incomplete due to the finite size
of the star, a few $\times 10^6$~g~cm$^{-3}$, these length scales are
$10^9$~cm and $0.1$~cm respectively.

In this work, we will compare our results with those obtained from the
well-known Zel'dovich, von Neumann and During (ZND) model of detonations
\citep{Zeldovich1940,vonNeumann1942,vonNeumann1963,Doring1943,FickettDavis1979}.
This model predicts both the detonation velocities and final products as well
as the detailed 1D thermal and compositional structure in space for
steady state denotations.  It can also be computed with error-controlled
methods with a large reaction network including all relevant reactions.
Matching these detailed
structures during burning is crucial for our application.  The $\Ni$ yield
of the supernova will be determined by the burning processes that lead to
these structures.  Therefore, if our burning model, including particle post
processing steps, can accurately reproduce the abundance profiles predicted
by the ZND model, it increases our confidence in the yields that it predicts
in more general cases.

The ZND equations describe the detonation structure between the
detonation shock front and the sonic point.  Beyond the sonic point, where the
following flow is moving away from the detonation front at the local sound
speed, disturbances cannot move upstream to change the detonation flow.  The
portion of a propagating steady-state detonation between the shock and the
sonic point is a static (i.e. time
invariant) structure which propagates in space at the detonation speed.  The
flow beyond the sonic point is typically not static, and its form depends
on the boundary condition of the following flow.  \citet{Sharpe1999} computes
the flow beyond the sonic point for asymptotically free propagation, but we
do not undertake that here.

Before moving further in the discussion, it is useful to state the ZND
equations in the form in which we will use them, for plane-parallel,
steady-state detonations \citep{FickettDavis1979,Khokhlov1989}.
In the frame of the detonation front,
\begin{eqnarray}
v&=& \frac{\rho_0D}{\rho}\ ,\\
\label{eq:det_dddt}
\dot \rho &=& \frac{\Sigma}{v^2-c_s^2}\ ,\\
\dot T &=& \left(\pder{T}{P}\right)_{\rho,Y_i}
 \Bigg\{ \left[v^2-\left(\pder{P}{\rho}\right)_{T,Y_i}\right]\dot \rho
 \nonumber\\
 &&\quad\quad\quad\quad\quad\quad\quad\mbox{}
 -\sum_j\left(\pder{P}{Y_j}\right)_{\rho,T,Y_{i\ne j}}\dot Y_j\Bigg\}
\ .
\end{eqnarray}
Here dot indicates an ordinary time derivative, $d/dt$,
$v$ is the flow velocity (with respect to the detonation front), $\rho_0$ is the unburned density, $D$ is the
detonation speed, $c_s$ is the frozen (evaluated with constant $Y_i$)
adiabatic sound speed, $P(\rho,T,Y_i)$ is the pressure.  To be consistent
with the above conventions, $Y_i$ is the number of nuclei of nuclide $i$ per
fluid baryon.  Thus $Y_i = X_i/A_i$, where $A_i$ is the mass number of
nuclide $i$.  The $\dot Y_i$ are given by the nuclear reactions.
The energy release function, in the absence of weak
interactions, is
\begin{equation}
\Sigma = \left(\pder{P}{\mathcal E}\right)_{\rho,Y_i}
\left[ \sum_i B_i\dot Y_i - \sum_j \left(\pder{\mathcal
E}{Y_j}\right)_{P,\rho,Y_{i\ne j}} \dot Y_j\right]\ ,
\end{equation}
where $\mathcal E$ is the internal energy.
The integration of these equations is begun just behind the leading shock,
whose properties are related to those of the fresh fuel by the detonation
speed $D$ and the usual shock conservation equations.

\begin{figure}
\plotone{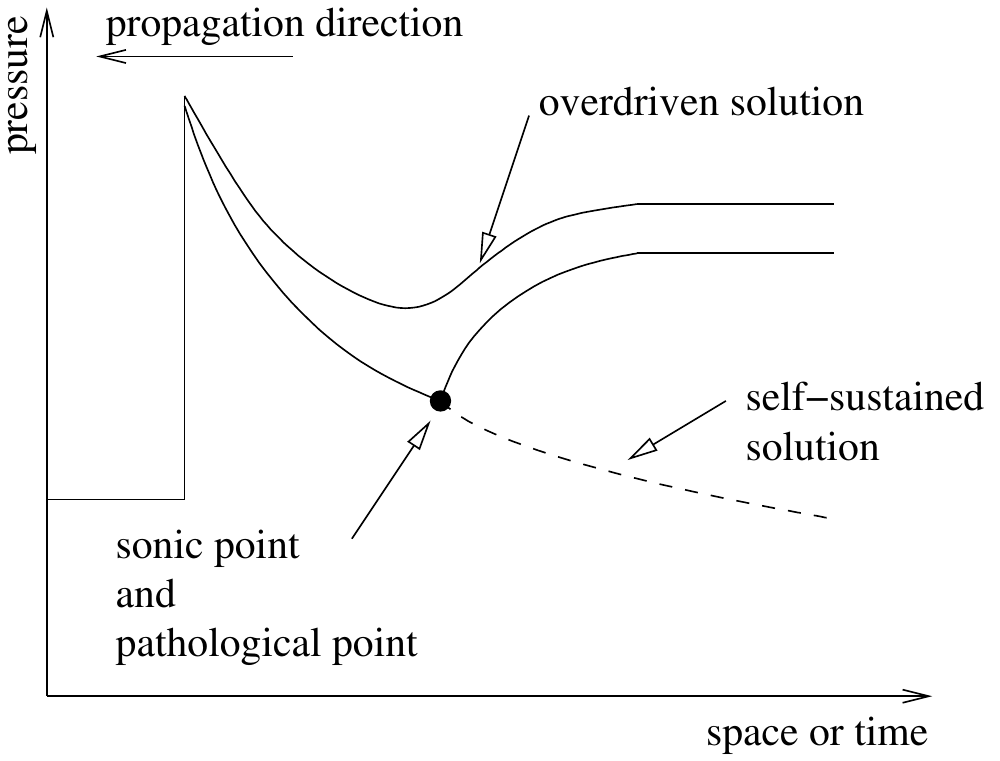}
\caption{ \label{fig:det_diagram}
Detonation pressure structure in space and time for a steady-state detonation
in 1 dimension.
The lower solid and dashed lines that pass through the pathological point represent possible solutions for the cases in
which the sonic point is reached before completion of burning.
These detonations are termed "pathological" or "eigenvalue" detonations.
}
\end{figure}

A diagram of the form of typical solutions are shown in Figure
\ref{fig:det_diagram}.
\eqref{eq:det_dddt} is singular at the sonic point, where $v=c_s$,
unless $\Sigma$ is also zero there.  There is a large class of
solutions for which $D$ is high enough that the entire following flow is
subsonic.  That is, $\Sigma$, and therefore $\dot\rho$, changes sign from
negative to positive before $v$ increases to $c_s$, thus avoiding an
encounter with this singularity.
This type of solution has a higher pressure in the
final state than in the reaction zone, and is called "overdriven" or
"supported" since it is effectively being pushed from behind by an
overpressure.
In this
case the full flow, including $D$ itself, has an inherent dependence on this boundary condition.
As the pressure in the final state, or at the "piston" following the
detonation, is decreased, $D$ also drops, and eventually a sonic point
will appear.  For detonations with lower pressures in the following flow, the
steady portion of the detonation flow then becomes an eigenvalue problem
such that $\Sigma=0$ at the sonic point.

In simplified reaction systems, $\Sigma=0$ at the sonic point because that is
the point at which fuel consumption completes.  This is called a
Chapman-Jouget detonation \citep{FickettDavis1979}, and its speed can be
computed from just the energy release and the EOS, without a need for the
full ZND equations \citep{Khokhlov1989,Gamezoetal1999}.
For reaction systems with complex or reversible reactions or changes in mean
molecular weight, the heat release function $\Sigma$ may not reach
or cross zero at a unique level of progress toward the fully burned state.
That is, $\Sigma=0$ may be attained before burning is "complete" and a static
final state reached.  In this case, the sonic point, and thus the end of the
static portion of the detonation profile, also occurs before a stable final
state is reached.  Such a detonation is termed "pathological" or "eigenvalue" and the sonic
point, where the singularity appears in the ZND equations, and where,
therefore $\Sigma=0$, is called the pathological point.  This is, in fact, the
more common case, and eigenvalue detonation structures in this case
represented a major advancement manifest by the ZND model
\citep{FickettDavis1979}.

The ZND integration can be continued after passing through the pathological
point, but there is more than one way to exit this point \citep{Sharpe1999}.
Figure \ref{fig:det_diagram} shows a diagrammatic representation of the
relation of the pressure profile in overdriven and self-sustained, or
unsupported, detonation.
The lowest overdriven detonation which can be fully integrated using just the
ZND equations without traversing a singularity is that which passes just above the pathological point.
While it is possible with special methods to traverse the pathological point
and obtain the self-sustained solution
\citep{Sharpe1999,MooreTownsleyBildsten2013}, we do not undertake
this here due to our large set of species and complex reactions.  This seems
prudent because even some of the profiles obtained by \citet{Sharpe1999} using
this method show clear indications of having further zero-crossings of
$\Sigma$
beyond the pathological point.  How these would manifest in the detonation
structure is unclear from this level of analysis.

It is now possible to choose a well-defined verification benchmark
problem whose solution can be calculated with both the ZND model with a
fairly complete reaction set and a 1D hydrodynamic simulation with our
simplified burning model.  We choose our benchmark to be the
slightly-overdriven state found by tuning $D$ to be a small
amount above the eigenvalue that leads to the pathological point.
This configuration
can be replicated in a 1D hydrodynamic calculation by manipulation
of the boundary conditions in the following flow to have the appropriate
pressure in the fully burned state.
The static portion of the benchmark structure, between the shock and the
sonic point, can be also used as a reference solution for self-sustained
detonations once they reach steady-state.


\subsection{Calibration of Timescale for Si Consumption}
\label{sec:taunse_calibration}

In order to make use of the simplified dynamics for the transition from the
QSE to the NSE state, given by \eqref{eq:phqndot}, we must calibrate
the timescale $\tau_{\rm NSE}$.
In \citet{Calderetal07}, $\tau_{\rm NSE}$ was calibrated by computing the
consumption timescale in isochoric self-heating as a function of temperature
and then using a fit to that timescale for $\tau_{\rm NSE}$.  Here we will
compare the time evolution of Si group element abundances for our benchmark
detonation, computed using the ZND equations, directly to the behavior
posited in our burning model by \eqref{eq:phqndot}.

\begin{figure}
\plotone{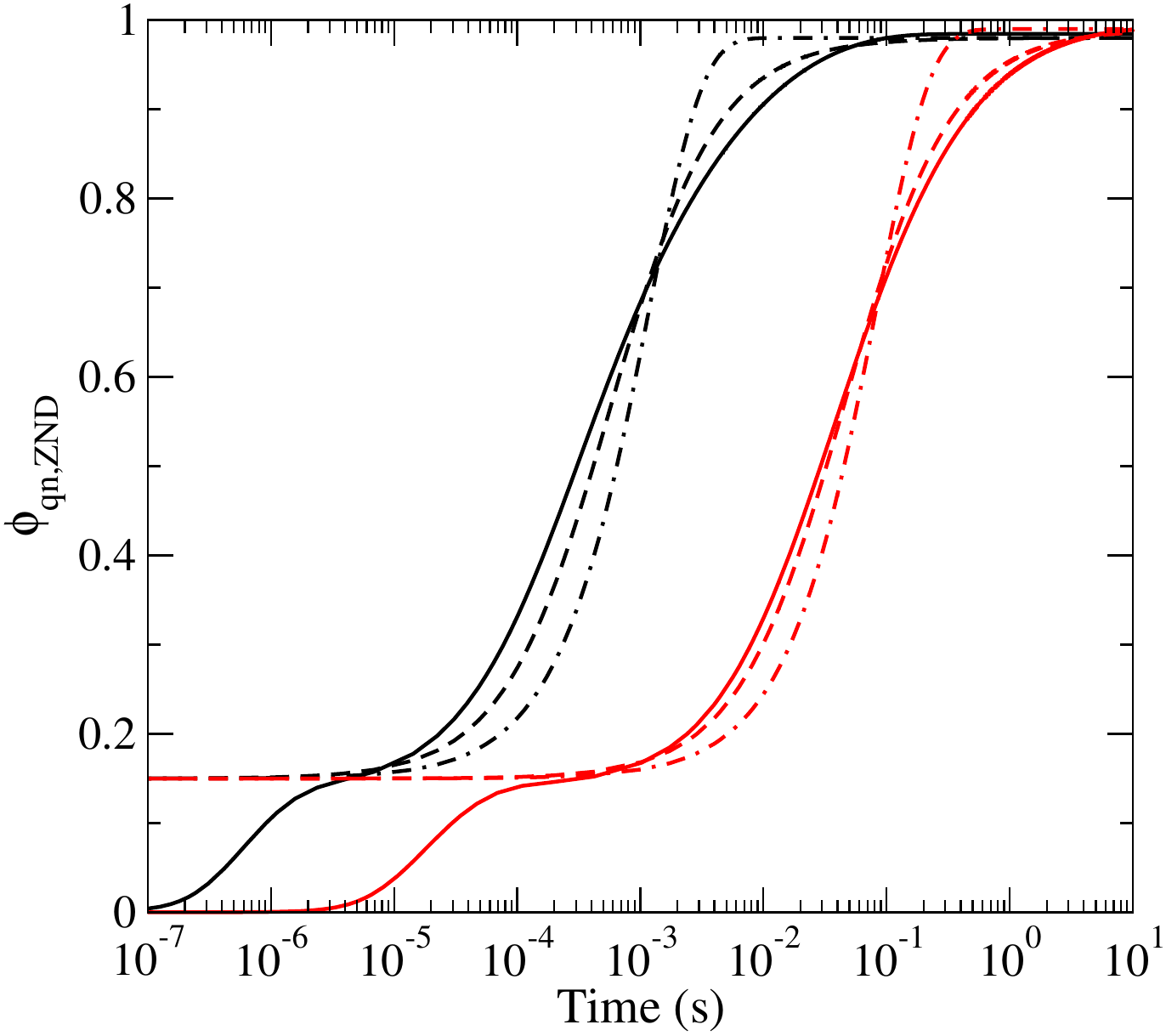}
\caption{
\label{fig:taunse}
Effective progress variable for the conversion of Si- to Fe-group material,
or relaxation toward full NSE (see text for definition).  Driven
detonations computed by a ZND integration are shown at two pre-shock
densities of $10^7$~g~cm$^{-3}$ (black, shorter timescale), and
$0.5\times10^7$~g~cm$^{-3}$ (red, longer timescale).  Also shown are fits
to the dynamics in the current burning model (dashed lines) and
exponential relaxation (dot-dashed lines).
}
\end{figure}

In order to make comparisons we use $\phi_{qn,\rm ZND}\equiv \phi_{qn}(X_i)$
based on \eqref{eq:phiqnabund}, where the $X_{i}$ are the abundances from the
ZND calculation computed with a large network for a driven solution.  In
Figure \ref{fig:taunse},
$\phi_{qn,\rm ZND}$ is shown for two densities spanning the range of interest,
0.5 and $1\times10^7$~g~cm$^{-3}$.  From Figure \ref{fig:ltscales} we see
that at these densities the synthesis of
IGE from IME will occur as a partially or mostly resolved process on the grid during the
explosion of the star, and will largely determine the IGE yield of the
explosion.  Expansion times for the star are in the range of a few tenths
of a second and the hydrodynamic timestep is around
$10^{-4}$~s for typical simulation resolutions of a few km.

As will be shown in section \ref{sec:space_comparison}, the early rise to
$\phi_{qn,\rm ZND}\approx0.15$ in both curves is due to IGE+LE
produced during the oxygen consumption stage.  Therefore we will proceed by
fitting the latter part of the curve only to get a better characterization
of the transition timescale.  If necessary, the inclusion of some IGE in
the intermediate state, $\xi_{q,i}$ in Figure \ref{fig:stage_diagram}, could
be introduced in converting the progress variables to abundances.  However
since we use post-processed yields for our final abundances this is not
necessary.

In the C-O burning process, the stages are well enough separated in time that
oxygen consumption, which is complete about the same time the Si abundance
peaks, completes before the transition from Si- to Fe-group
proceeds very far.  This can be seen clearly in Figure \ref{fig:ltscales} as
the 5 orders of magnitude separating the time of maximum Si abundance (dashed
red line) and the completion of burning (solid red line).  We therefore
assume $\phi_{aq}=1$ and
for a characteristic value of $\tau_{\rm
NSE}$ we may analytically integrate \eqref{eq:phqndot} to obtain
\begin{equation}
\phi_{qn}(t) = \phi_{qn,\rm final} - \frac{1}{1/(\phi_{qn,\rm
final}-\phi_{qn,0}) + t/\tau_{\rm NSE}}\ .
\end{equation}
Here $\phi_{qn,0}$ and $\phi_{qn,\rm final}$ are taken from $\phi_{qn,\rm
ZND}$ at the end of oxygen consumption and in the final state respectively.
In this case they are about 0.15 and 0.99.  $\phi_{qn,0}$ might be different
if we performed this calibration with different initial abundances.
This
form can now be fit to the curves shown in Figure \ref{fig:taunse} using a
non-linear least squares fit.  We use a fitting region
$0.15\le\phi_{qn}\le0.85$, to capture the major portion of the evolution.
The resulting fits are shown by the dashed lines.
The fit timescale is not sensitive to choice of $\phi_{qn,0}$; a 5\%
variation in $\phi_{qn,0}$ only changes the fit $\tau_{\rm NSE}$ by 1\%.
The maximum error in the fits occurs when $\phi_{qn}\approx 0.4$, and is
about 0.06 and 0.03 for the higher and lower density shown in Figure
\ref{fig:taunse} respectively.  We will discuss below in section
\ref{sec:space_comparison} how well the resulting burning model performance in
hydrodynamics compares to the detonation benchmark, and extend this
comparison to abundances in post-processing, compared to those in the
benchmark, in section \ref{sec:particle_verif}.

This fitting
procedure has been repeated at several densities,
between 0.3 and 10 $\times 10^7$~g~cm$^{-3}$.  At each of these
densities the conversion of Si- to Fe-group takes place at a declining temperature.
The decline during this burning stage is, however, much less than the variation from one density to
another.  By evaluating the temperature when the relaxation is
approximately half complete, we can construct and fit a relation between
$\tau_{\rm NSE}$ and $T$.  We obtain
\begin{equation}
\tau_{\rm NSE}(T) = \exp(201.0/T_9 -46.77)\ .
\end{equation}

The $\tau_{\rm NSE}$ timescale found here is not directly comparable to previous work because we have
used different burning dynamics.  However, a similar fit can be performed
with the exponential decay form that results from the simpler dynamics
previously posited, $D\phi_{qn}/Dt = (\phi_{aq}-\phi_{qn})/\tau_{\rm NSE}$
\citep{Townsleyetal07}.  This is shown by the dot-dashed lines in Figure
\ref{fig:taunse} when fit to
the same region indicated above.  This form does not appear to provide a good reproduction
of the late-time behavior of $\phi_{qn}$.  Also the timescales obtained for
the exponential fit are approximately a factor of 10 to 20 shorter than
those given for $\tau_{\rm NSE}$ in \citet{Calderetal07}.  This is
understandable because the definition used in that work measured a timescale
to reach a fairly complete burning stage, whereas we have fit an
exponential form directly.

\subsection{Comparison of parameterized burning hydrodynamics against
200-nuclide ZND Structure}
\label{sec:space_comparison}

The verification that we are attempting to perform involves demonstrating
that the abundance structure produced by post-processing particle tracks from
the hydrodynamics which utilizes the parameterized burning matches the ZND
structure for a steady state detonation.  That comparison will be done
in section~\ref{sec:particle_verif}, but first it is useful to compare
the intermediate result obtained from the parameterized burning model in
the hydrodynamics simulation alone.  This will provide a check on the
realism of spatial thermodynamic structure without the added complication
of the integration of the Lagrangian tracks, and also give some diagnostics
concerning whether the parameters within the burning model are behaving as
expected.

\begin{figure*}
\epsscale{0.5}
\plotone{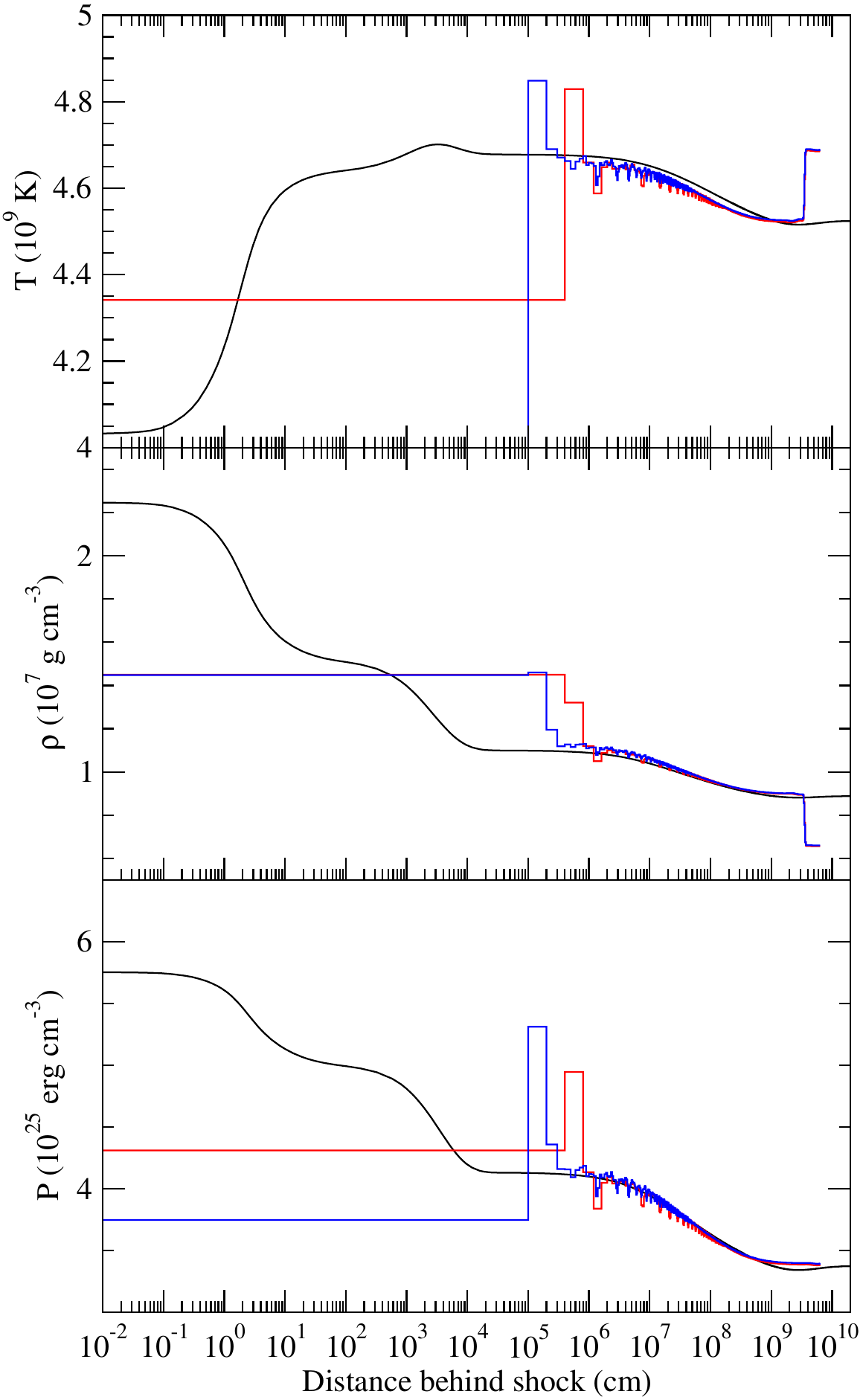}
\plotone{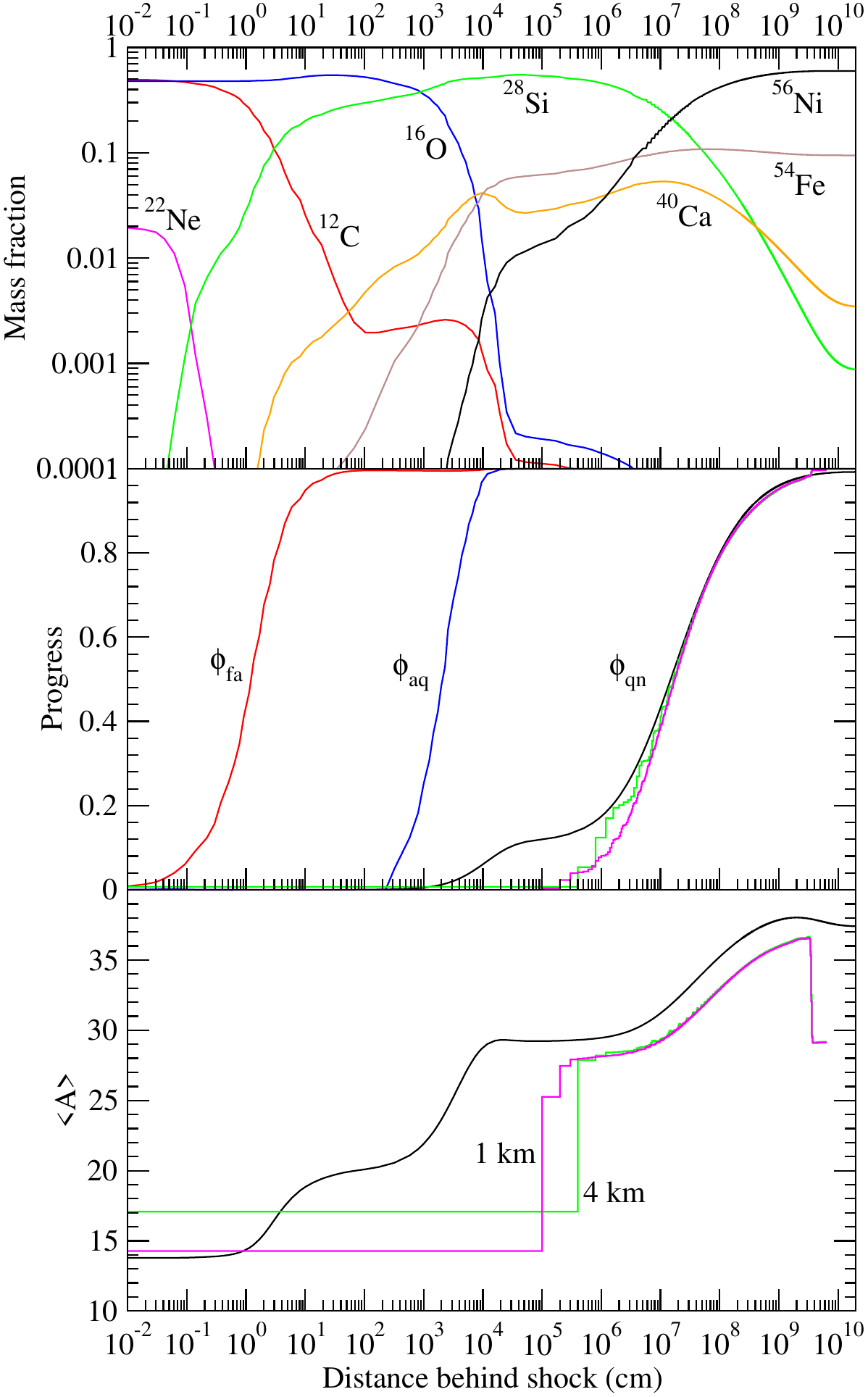}
\epsscale{1.0}
\caption{\label{fig:space_compare_5e6}
Detonation thermal and composition structure displayed by our parameterized
model for C-O burning in 1D hydrodynamic simulations compared to our
detonation benchmark of the steady-state ZND solution of the
equivalent detonation.  This case is at a density of
$\rho=5\times10^6$~g~cm$^{-3}$ and an initial composition of 50\% $^{12}$C,
48\% $^{16}$O and 2\% $^{22}$Ne.
\emph{Left}:
Thermal structure of simulations at spatial resolutions of 4~km (red) and
1~km (blue) compared to the benchmark steady-state ZND solution (black).
\emph{Right}:
Top: The composition structure of the benchmark ZND calculation computed with a
200-nuclide network. Middle: Effective progress variables derived from
the abundances in the benchmark (red, \eqref{eq:phifaabund}; blue,
\eqref{eq:phiaqabund}; black, \eqref{eq:phiqnabund}), compared to the
$\phi_{qn}$ progress variable obtained in the hydrodynamic simulations at
4~km (green) and 1~km (magenta) resolution.
Bottom: Average number of nucleons per nucleus, $\bar A = 1/Y_{\rm ion}$,
derived from the full abundances in the benchmark (black) and obtained from
the progress variables, \eqref{eq:yionbreakout}, in 1D hydrodynamic
simulations at 4~km (green) and 1~km (magenta).
}
\end{figure*}

\begin{figure*}
\epsscale{0.5}
\plotone{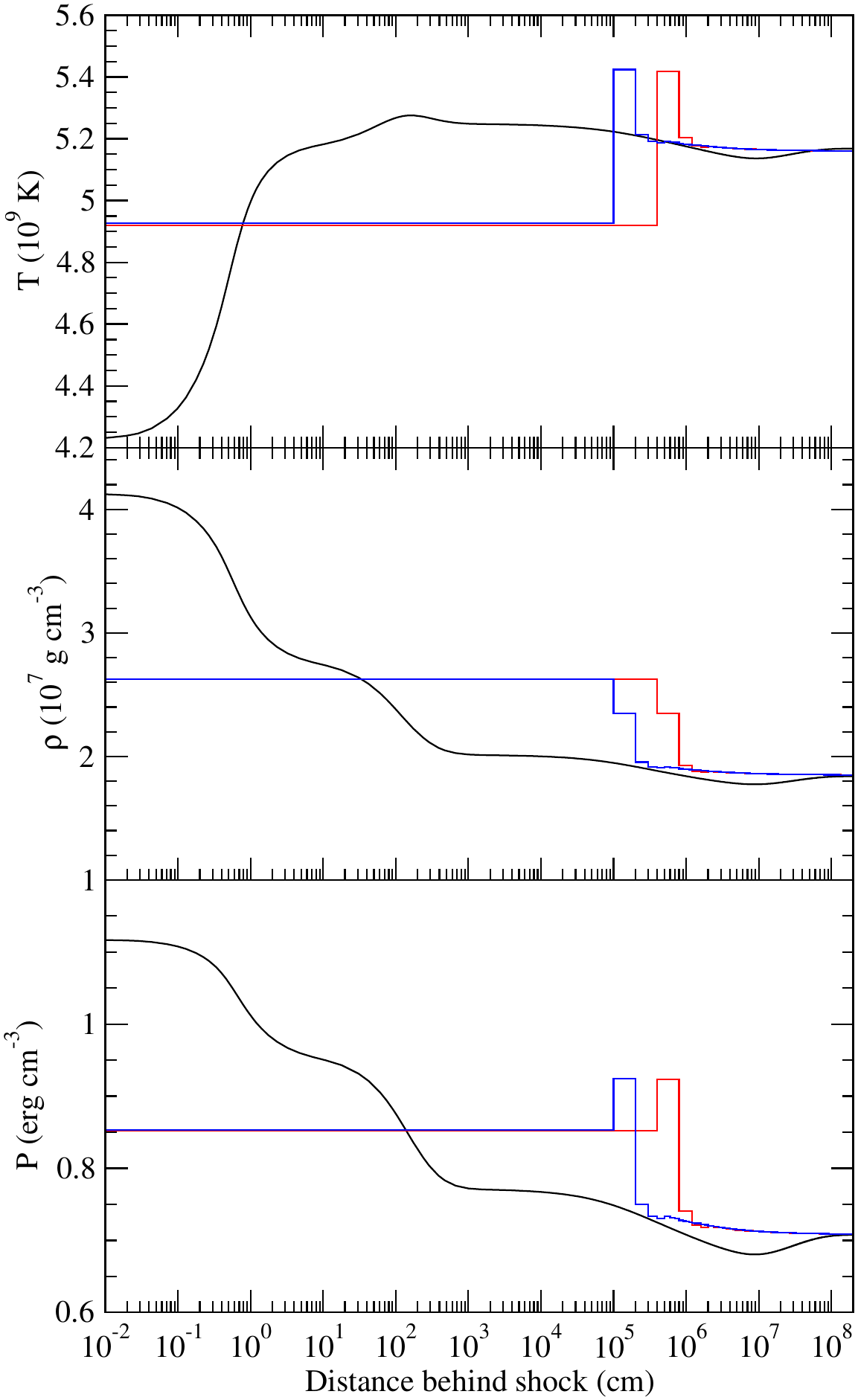}
\plotone{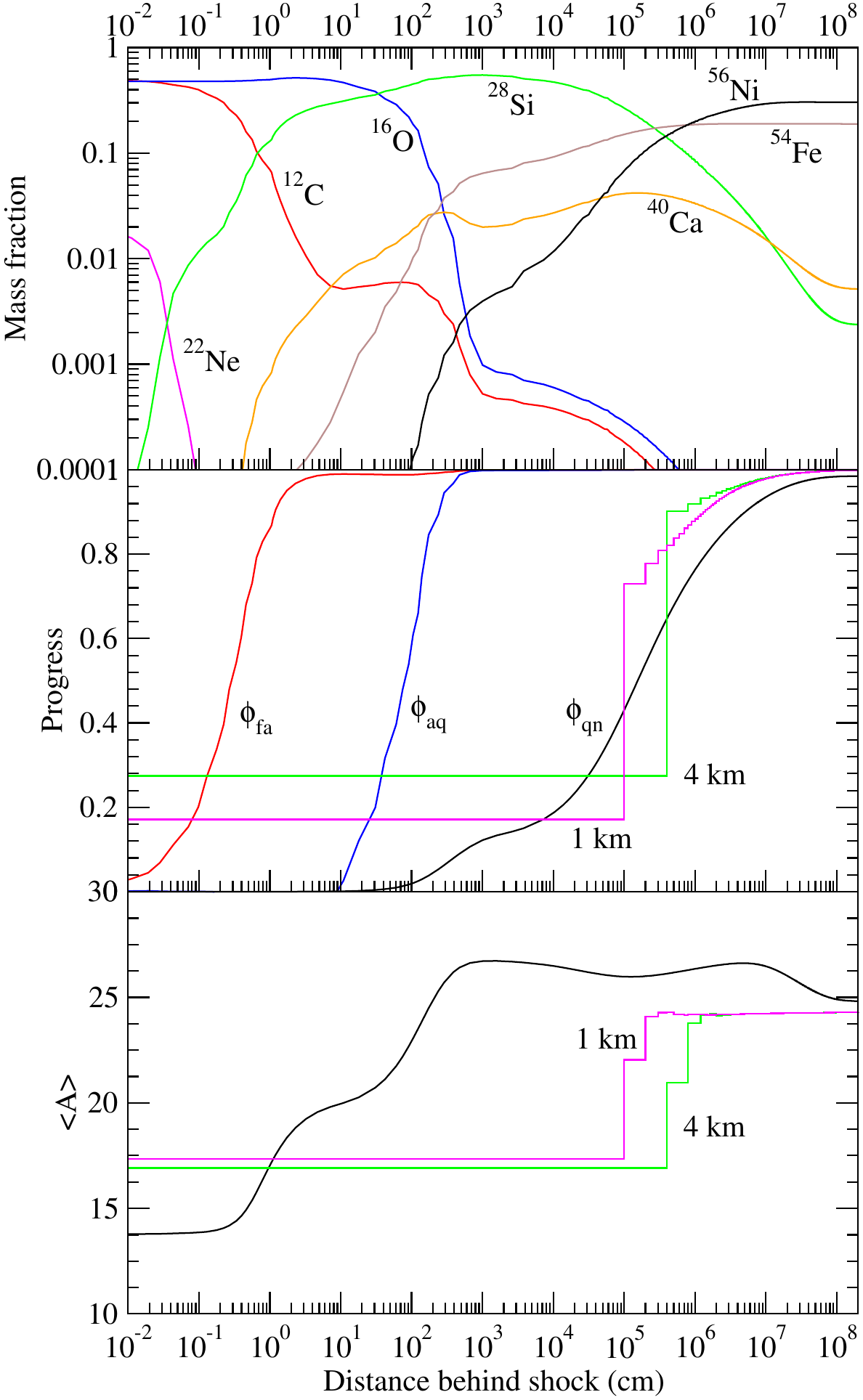}
\epsscale{1.0}
\caption{\label{fig:space_compare_1e7}
Similar to Figure \ref{fig:space_compare_5e6} but for $\rho=10^7$~g~cm$^{-3}$.
}
\end{figure*}

Our benchmark is, as described in section \ref{sec:znd}, the ZND solution for
a steady-state, planar, slightly overdriven detonation in 1 dimension.  This
solution is shown as the reference curves in Figures
\ref{fig:space_compare_5e6} and \ref{fig:space_compare_1e7}, with
thermodynamic quantities, $T$, $\rho$, $P$, in the left panel (black), and
abundances in the top right panel.
The initial condition for
the 1D hydrodynamic simulations is material at spatially
constant density and temperature away from the ignition point.  We consider
cases here with this background temperature set to $4\times 10^8$~K.
The domain extends from $x=0$ to $65,536$~km in order to allow the detonation
to approach steady state.  Two resolutions, 4~km and 1~km, similar to the
resolution of production supernova simulations \citep{Townsleyetal09}, are used to
confirm insensitivity to resolution.  We will refrain from using the term
convergence here, reserving it for circumstances in which gradients are
resolved.  The boundary condition on the opposite end of the domain from the
ignition is reflecting, but has no
impact on the simulation due to the supersonic nature of the detonation and
since the simulation is stopped before the front reaches it.  The
left boundary, at $x=0$, is a zero-gradient boundary.  The initial
perturbation is made in both temperature and velocity.  Along with inflow
from the zero-gradient boundary, the latter will serve to support the
detonation from behind.  Both temperature and velocity are placed as linear
gradients decreasing from a maximum at $x=0$ to the background values of
$T=4\times 10^8$~K and velocity of zero over a size we will call the size of
the ignition region.  The velocity is tuned by hand until the pressure far
behind the detonation front and near the $x=0$ boundary matches the late-time
pressure found for the slightly overdriven ZND solution.  Ignition region
sizes were 1024~km and 128~km for $10^7$ and $5\times 10^6$~g~cm$^{-3}$
respectively.

As above, we will focus on densities at which the transition from Si-group
burning products to Fe-group burning products is fully or partially resolved
on the spatial grid.
At a density of $5\times 10^6$~g~cm$^{-3}$, as indicated by Figure
\ref{fig:ltscales}, nearly the entire Si- to Fe-group
transition is resolved at 4~km resolution for the steady state detonation.
The spatial structure obtained from the ZND calculation and the
hydrodynamics, which uses the parameterized burning, is shown in
Figure~\ref{fig:space_compare_5e6}.  The thermodynamic quantities, $T$,
$\rho$, and $P$, are shown in the left panel.  The hydrodynamic result is at
an evolution time of 5.45 seconds, when nearly the entire domain has been
consumed.  The zero point for the distance behind the shock in the
hydrodynamic simulations is taken as the last zone in which the shock
detection considers the cell inside a shock,
thereby suppressing the reactions in that zone.
See Appendix \ref{app:ppmdet} for more on this suppression.
In steady
state, the shock region in which the reactions are suppressed is
a well-localized region of approximately 4-5 zones.  As a result of this, the
first point from the hydrodynamic simulations, indicated with stair-stepped
lines, is at 4~km and 1~km for simulations of those respective resolutions.
The top right panel shows the spatial abundance structure of a selection of
nuclides for the steady-state detonation from the 200-nuclide ZND
calculation.
From the ZND abundance and thermal structures shown in Figure
\ref{fig:space_compare_5e6} we see that both the $^{12}$C and $^{16}$O consumption
stages are entirely unresolved because they take place on length scales of
approximately 1~cm and several $\times 10^3$~cm respectively.
In the span of less than a single zone, the burning reaches the Si-rich QSE.

The values of $P$, $T$, and $\rho$ at the point chosen as zero distance behind the shock in the simulation
are not quite the same as the post-shock values
expected based on the detonation speed.
This is presumably the result of numerical
mixing in the vicinity of the under-resolved shock and burning front.
The post-shock density is about 35\% lower than the peak value predicted by
the ZND calculation and the pressure, rather than peaking at the shock, peaks
in the first zone in which burning is allowed at a value about 10\% lower
than expected.  The $T$ peak, which also occurs
in the first zone in which reactions are allowed, is about 3\% higher than
the peak in the benchmark.  This transient is also larger in time and space
than the true burning structure due to the resolution, but the thermal state
appears to relax back toward a good approximation of the QSE state very
quickly, within 2 zones.  After this and a small undershoot, the hydrodynamic
solution is a very good match, with 3\% in $P$ and $\rho$, and within about
1\% in $T$, all the way out to the pathological point.  There is noise on a
similar level, but more so in $P$ and $T$ than $\rho$.  An artifact of the
initial ignition is evident at the end of the hydrodynamic curves for $T$ and
$\rho$.  We also find very good consistency between resolutions after the
first few zones behind the shock, matching within a percent, with noise in
each case slightly larger than that.  The hydrodynamic result is probably not
completely relaxed to the steady-state overdriven solution, since there is no
pressure minimum.  However, the pathological point occurs quite close to the
end of the domain even for this large domain and the pressure minimum is
expected to be fairly shallow.

A comparison of some of the parameters in the burning model are shown
in the lower right two panels in Figure~\ref{fig:space_compare_5e6}.  In
order to make a comparison of the progress variables we have defined some
effective progress variables for the 200-nuclide set.  In
addition to \eqref{eq:phiqnabund} above, we define
\begin{eqnarray}
\label{eq:phifaabund}
\phi_{fa}(X_i) &=& 1 - \frac{X_{^{12}\rm C}}{X_{^{12}\rm C,0}}\ ,\\
\label{eq:phiaqabund}
\phi_{aq}(X_i) &=& 1 - \frac{X_{^{18}\rm O}}{X_{^{18}\rm O,0}}\ .
\end{eqnarray}
The spatial structure of both $\phi_{fa}$ and $\phi_{aq}$ are unresolved at
this density and these resolutions.  Thus they are both 1 in the first zone
behind the shock-detection suppression of burning because our data dumps always
follow a reaction sub-step in our operator split time evolution.  For this
reason the $\phi_{fa}$ and $\phi_{aq}$ from the hydrodynamic simulations are
not shown.

We find a good match between the evolution of
$\phi_{qn}$ and the effective equivalent defined for the 200-nuclide set.
The largest discrepancy is due to the production of some Fe-group material
with Si-group in the benchmark.  After $\phi_{qn}\gtrsim0.3$ the discrepancy
is less than $0.05$, and after $\phi_{qn}\gtrsim 0.5$ it is less than $0.02$.
This indicates that our temperature-dependent fits of the timescales for this
evolution, described in section \ref{sec:taunse_calibration}, are acting satisfactorily.  The bottom right panel of
Figure~\ref{fig:space_compare_5e6} shows how the mean ion molecular weight $\bar A$
compares to the equivalent quantity from the parameterized burning $1/Y_{\rm
ion}$.  This quantity is systematically about 4\% low, probably due to our
choice of $\tilde Y_{\rm ion,QSE}= 1/28$ as an estimate of the $Y_{\rm ion}$
of the QSE state.  The QSE state is not pure $^{28}$Si, and therefore this estimate
is slightly off and
creates a systematic offset in the consecutive evolution toward $Y_{\rm
ion,nse}$.  The difference observed in the test may also be magnified by the hydrodynamic simulation having
not quite reached the steady overdriven state.  In either case the
discrepancy in $\bar A$ only leads to less than 1\% discrepancy in $T$, as
found above, so this level of agreement appears sufficient for producing
accurate thermodynamic histories for particle post-processing.

As a second case, shown in Figure~\ref{fig:space_compare_1e7}, we perform a
similar calculation at an ambient density of $10^7$~g~cm$^{-3}$.  At this
density, more than half of the transition from the Si-group dominated QSE to
the Fe-group dominated NSE is unresolved on a 4~km grid.  This is
according to the profile of $\phi_{qn}$ predicted by the 200-nuclide ZND
calculation, shown in the middle right panel of
Figure~\ref{fig:space_compare_1e7}.
We see a region, similar to that in the first case, of about 2 zones in which $T$ is about 3\% higher
than the expected peak and $P$ and $\rho$ are intermediate between the
expected post-shock values and the QSE values, after which all of these
relax to within 3\% of the benchmark values.
The largest source of discrepancy is due to the lack of the expected minimum
near the pathological point at a distance of $10^7$~cm behind the shock.
Instead, the hydrodynamic solution monotonically relaxes to the state given
by the boundary condition pressure.  However, even with this discrepancy the
maximum difference between the benchmark and hydrodynamic result is about 5\%
in $P$ and $\rho$ and less than 2\% in $T$.  As before the two resolutions
match very well, within 0.5\%.

In terms of progress variables, during the partially-resolved transition from
Si- to Fe-group, the progress variable for this process, $\phi_{qn}$, is about
0.1 higher than the benchmark predicts at a given distance behind the shock.
This seems like a reasonable indication of the uncertainty of the progress
variable's reproduction of the real process for partially-resolved cases like
this one.  The $\bar A$ determined in the final state by the burning model in
the hydrodynamics is only about 2\% lower than the benchmark.  However, as
for the thermal profiles, the non-monotonic behavior near the pathological
point is not captured.

The main difference from the benchmark in this case is due to the lack of a
clear pathological point in the hydrodynamic result.  It is unclear if this
is due to the limited resolution, a deficiency in the burning model, or
insufficient time to relax to the steady state.  In any case, the discrepancy
in the thermal quantities used for post processing is, at maximum, a fairly
modest 5\% in $\rho$ and 2\% in $T$.  We will accept this as the approximate
uncertainty in the thermal histories produced by the burning model, and
proceed to investigate the abundances produced in post-processing directly
below.  At higher densities than about $10^7$~g~cm$^{-3}$, as can be seen
from the length scale for completion of Si- to Fe-group conversion, the
conversion will be nearly complete on scales smaller than the resolution.  The
burning model shows good reproduction of the final state, within a few
percent, so that denser cases should also have similar good accuracy.

\section{Verification of Lagrangian particle nucleosynthesis against ZND
solution}
\label{sec:detpost}
\label{sec:particle_verif}

While it is important that the progress variables provide a good reproduction
of the detonation structure, in the end the yields will be computed by
post-processing Lagrangian tracer particle histories.  In this section
we compare computed Lagrangian track yields to the steady-state ZND solutions
that we are using as a benchmark.
Detonation yields are computed by a direct integration of the $\rho(t)$,
$T(t)$ history recorded by the Lagrangian tracer particle from the
hydrodynamic simulation, using them to set the reaction rates in the nuclear
reaction network.

\begin{figure*}
\plotone{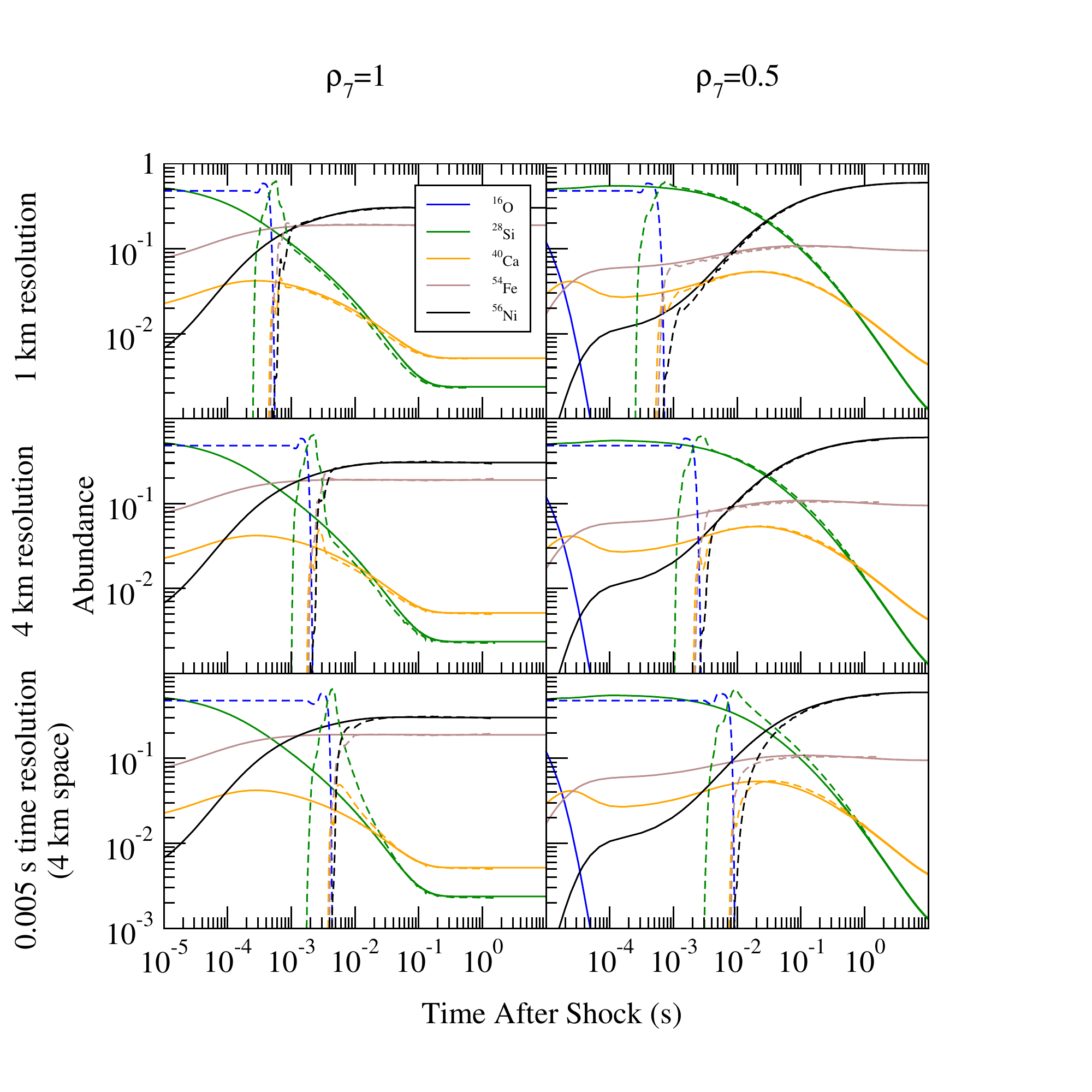}
\caption{\label{fig:particle_multi}
Abundance histories computed from post-processed Lagrangian histories from
hydrodynamic simulations (dashed lines) compared to benchmark steady-state
detonation structures computed from the ZND equations (solid lines).
Shown as mass fractions are the major abundances after C consumption,
$^{16}$O (blue), $^{28}$Si (green), and $^{56}$Ni (black), as well as
$^{40}$Ca (orange) and $^{54}$Fe (brown).
Two densities are shown, $10^7$~g~cm~$^{-3}$ (left column) and $5\times
10^6$~g~cm$^{-3}$ (right column).
Each of these is shown
from a simulation with 1~km (top row) and 4~km (middle row) spatial
resolutions with histories recorded at full time resolution, and at 4~km
spatial resolution with history recorded at a reduced time resolution of
$0.005$~s (bottom row).
}
\end{figure*}

The results of the integration of the reactions over the Lagrangian history
are compared with benchmark calculations in Figure \ref{fig:particle_multi}
for the same two densities, $10^7$ and $5\times 10^6$~g~cm$^{-3}$ (left and
right columns), and two spatial resolutions, 1~km and 4~km (top and middle
row), as used in Section~\ref{sec:space_comparison}.  We also consider a case
with a reduced time resolution for the recording of the Lagrangian history
(bottom row).  Comparison can now be made directly with actual abundances.
We show the major abundances for stages beginning at oxygen consumption,
$^{16}$O, $^{28}$Si, and $^{56}$Ni, as well as the major neutron-rich nuclide
produced before other Fe-group material, $^{54}$Fe \citep{Bravoetal2010}, and
the spectroscopically important $^{40}$Ca.  Each plot shows two curves for
each nuclide: the benchmark solution (solid lines) computed using the ZND
equations and the post-processing of the $\rho(t)$, $T(t)$ history
(dashed lines).

In order to compare structures we must choose a zero time during the
Lagrangian history.  Zero time for the benchmark ZND integration corresponds
to the downstream side of the shock.  We have chosen the zero time for the
Lagrangian history to be at the first timestep that reaches 1\% above the
ambient temperature.  This makes the entire reaction region  visible on these
plots because the C and O consumption timescales in the benchmark are
shorter than the timestep in all cases.  The abundances in the first part of
the reaction region are unrealistic, as expected.  Notably at $\rho_7=1$ the
$^{28}$Si abundance during the first few steps overshoots what should be present.
However, the abundances appear to recover quickly to fairly accurate values
within 0.01~s in all cases with full time resolution in the history.  For the
coarsened time resolution history shown in the bottom row, the recovery
toward the correct solution is slower, taking until nearly 0.1~s at
$\rho_7=0.5$.  This is comparable to the expansion timescale of this material
during the supernova and indicates that a time history at the same time
resolution as the hydrodynamics is required at this density.

In comparison to the benchmark solution we find excellent agreement after
0.01~s. The worst case is $^{28}$Si at $\rho_7=1$, 4~km resolution, off by
less than 0.005, about 20\% of the abundance at that time.  More typical
discrepancies are those near where $^{56}$Ni and $^{28}$Si are similar
abundance for $\rho_7=0.5$, which are between 5 and 10\%.
This comparison verifies directly, for the first
time in the computation of thermonuclear supernovae, that a hydrodynamic
calculation with post processing correctly reproduces detonation yields
computed with an error-controlled integration of the ZND model.  Thus the
dynamics in our parameterized burning model is able to give sufficiently
accurate thermodynamic structures for post-processing abundance
calculations accurate to between 5 and 10\% for steady-state planar detonations down
to $\rho_7=0.5$.  This includes densities at which the detonation structure
is partially resolved.  The driving region extends
to near the plateau of the $^{56}$Ni abundance, as can be inferred from the
location of the density and temperature minima near the pathological point in
the ZND integrations shown in Figures \ref{fig:space_compare_5e6} and
\ref{fig:space_compare_1e7}.


\section{Computation of Complete Nucleosynthesis}
\label{sec:nucleosynthesis_computation}

The previous sections have outlined methods for treating fluid elements
within the star processed by either the detonation or deflagration modes of
burning.  In order to obtain yields for an actual computation of a DDT SNIa,
it is necessary to perform both of these methods on the fluid
element histories of a single simulation.  This involves sorting and
classifying histories to be treated with the two different methods and
treating cases that may overlap.  Also some aspects of the implementation of
energy release in the hydrodynamics must be modified to allow both types of
reactions.  Here we discuss these and other details of the unified
post-processing.

\subsection{Track Classification}

We begin by discussing how a Lagrangian history recorded from the
hydrodynamics, hereafter called a ``track'' is classified as being processed
either in a deflagration or detonation.  This determines how the first
portion of the post-processing is performed, which may involve reconstruction
and replacement of an unresolved portion of the time history.

The recorded values of $\phi_{\rm RD}(t)$ and $\phi_{\rm fa}(t)$ for a track
are scanned starting from the beginning of the time history.
In searching for a detonation, the first few points are ignored
after which we search for a sudden increase in $\phi_{\rm fa}$ to near unity.
This indicates a detonation.  The actual parameters used are for $\phi_{\rm
fa}>0.9$ and $\phi_{\rm fa}$ having increased by more than 0.2 since the 4th
previously
recorded $\phi_{\rm fa}$.  If the track is determined to be a detonation, it
is subject to a direct post-processing of its $\rho(t)$, $T(t)$ history.

During the search for a possible detonation feature, if $\phi_{\rm RD}$
exceeds 0.5 before a detonation is detected, the track is classified as a
deflagration.  For a deflagration, the post-processing begins from the point
in the time history at which $\phi_{\rm RD}=0.5$ and proceeds initially
with a reconstruction as discussed in
section \ref{sec:def}.  This calculation is changed to a direct
post-processing at the time any of 3 conditions are met: $\phi_{\rm
fa}>0.95$, $P<10^{22}$~erg~cm$^{-3}$, or $\phi_{\rm fa}-\phi_{\rm RD}>0.1$.
The latter condition is in addition to those mentioned in section
\ref{sec:def}, and most likely indicates that a fluid element passing through
the artificially thickened flame front has been struck by a detonation shock.
These borderline cases are some of the most challenging for for obtaining accurate yields.
Several such examples are discussed along with others in Appendix \ref{sec:example_tracks}.

A track which does not meet either of the above criteria for detonation or
deflagration will be assumed to have not been processed by either the
deflagration and detonation and will be post-processed directly based on the
$T$, $\rho$ history recorded.

\subsection{Mixed Burning Modes in Hydrodynamics}

As implied above, a fluid element with $\phi_{\rm RD}<0.5$ will not be
considered to have been burned by the deflagration for the purposes of
post-processing.  This also has implications for the hydrodynamic
implementation of the energy release: a detonation must be able to propagate
into regions where $0<\phi_{\rm RD}<0.5$, i.e. regions that have been
partially burned by the RD front that is propagating the deflagration.  This
presents a challenge because the temperatures in these regions are not
physical and therefore can't be used directly to compute a reaction rate like
the $\dot \phi_{\rm CC}$ appearing in \eqref{eq:phfadot}.  In order to
allow detonations to propagate fully into the artificially broad deflagration
reaction front, this issue is treated directly in energy release in the
hydrodynamics rather than in post-processing.

Typically $\dot\phi_{\rm CC}$ is suppressed when $\phi_{RD}$ is larger
than some small threshold.  In order to allow thermal burning in these
regions without it getting out of control, two measures are taken.  First,
$\dot\phi_{\rm CC}$ is only re-enabled in the proximity of non-flame-related
burning.  Carbon reaction unrelated to the deflagration is taken to be
present if $\phi_{\rm fa} -\phi_{\rm RD} > \delta_b$, where $\delta_b$ is a
threshold calibrated based on trials.  $\delta_b=0.1$ has been found to be
suitable in 2D and $\delta_b=0.3$ in 3D.  For a given cell in the Eulerian
hydrodynamics, proximity of thermally activated burning is established if
this condition is satisfied in neighboring cells within one width of the RD
front away, typically 4 cells.  This allows $\dot\phi_{\rm CC}$ to activate
when a detonation arrives at the RD front.

The second control measure attempts to estimate the temperature of the fuel in the absence of the deflagration rather than use the local $T$ directly in the computation of $\dot\phi_{\rm CC}$.
Recall that the
zones in which $0<\phi_{\rm RD}<1$ should be thought of as being regions of
mixed burned and unburned material separated by a thin surface which is the
propagating physical flame, each in approximate pressure equilibrium with the
other.  We would like to estimate the temperature of the unburned material.
This is done by removing the energy which corresponds to the current amount
of material burned and then computing the $T$ that corresponds to the energy
leftover at the local $\rho$.  This is a very rough calculation, but is only
meant to be an estimate.  The resulting temperature is then used to
calculation $\dot\phi_{\rm CC}$.

\subsection{Initial Abundances}

In order to perform post-processing with a large network, it is necessary to
specify a full set of initial abundances.  These initial abundances must
reflect the previous processing of the material in the star by earlier phases
of evolution including the burning phases of the progenitor star and the
core convection that precedes the ignition of the deflagration.  Our initial
abundances are parameterized by three parameters: the $^{12}$C abundance at
ignition, the metallicity of the progenitor, and the $Y_e$ of the material,
parameterized in the hydrodynamics by the mass fraction of $^{22}$Ne in the
fuel.  The value of each of these for a given track is determined based on
location of the tracer particle within the progenitor WD at the beginning of
the simulation.  Note that $Y_e$ is not the same parameter as metallicity due
to the additional electron captures that occur during the pre-explosion core
convection phase.

Given these parameters, the initial abundances are constructed from 4
components:  (1) $^{12}$C of the specified mass fraction. (2) Metallicity
given by scaled solar abundances of all elements heavier than $^{4}$He
\citep{anders.grevesse:abundances} except with the abundances of C, N, and O
added together to give the abundance of $^{22}$Ne used for the initial
abundances \citep{timmes.brown.ea:variations}.  (3) Ashes from the convective
phase made up of equal parts $^{20}$Ne, $^{16}$O, $^{13}$C, and $^{23}$Ne
\citep{PiroBildsten2007,Chamulaketal2008}.  (4) The remainder is taken to be $^{16}$O.  The
contribution associated with the metallicity is assumed to be uniform
throughout the star and any additional depletion of $Y_e$ in the interior
convection zone is matched with the necessary amount of simmering ashes.

\section{Results: 2D DDT Yields}
\label{sec:2dddtyields}

Our model of SNe~Ia using 2D simulations with a DDT is intended to reproduce the large-scale abundance distribution observed in the ejecta of normal SNe~Ia.
The most accessible observational characterizations are the abundance tomography studies \citep{Stehleetal2005,Mazzalietal2008}, though these do require some information about the ejecta as input, and therefore are not free of assumptions.
Reproduction of abundance structure inferred from spectra is one of the metrics by with the original W7 model \citep{NomotoThielemannYokoi1984} and the 1D DDT models \citep{HoeflichKhokhlovWheeler1995} are found to succeed.
Here we will compare our yields to these tomographic reconstructions and the essential aspects of successful theoretical models.

\begin{figure}
\includegraphics[width=\apjcolwidth]{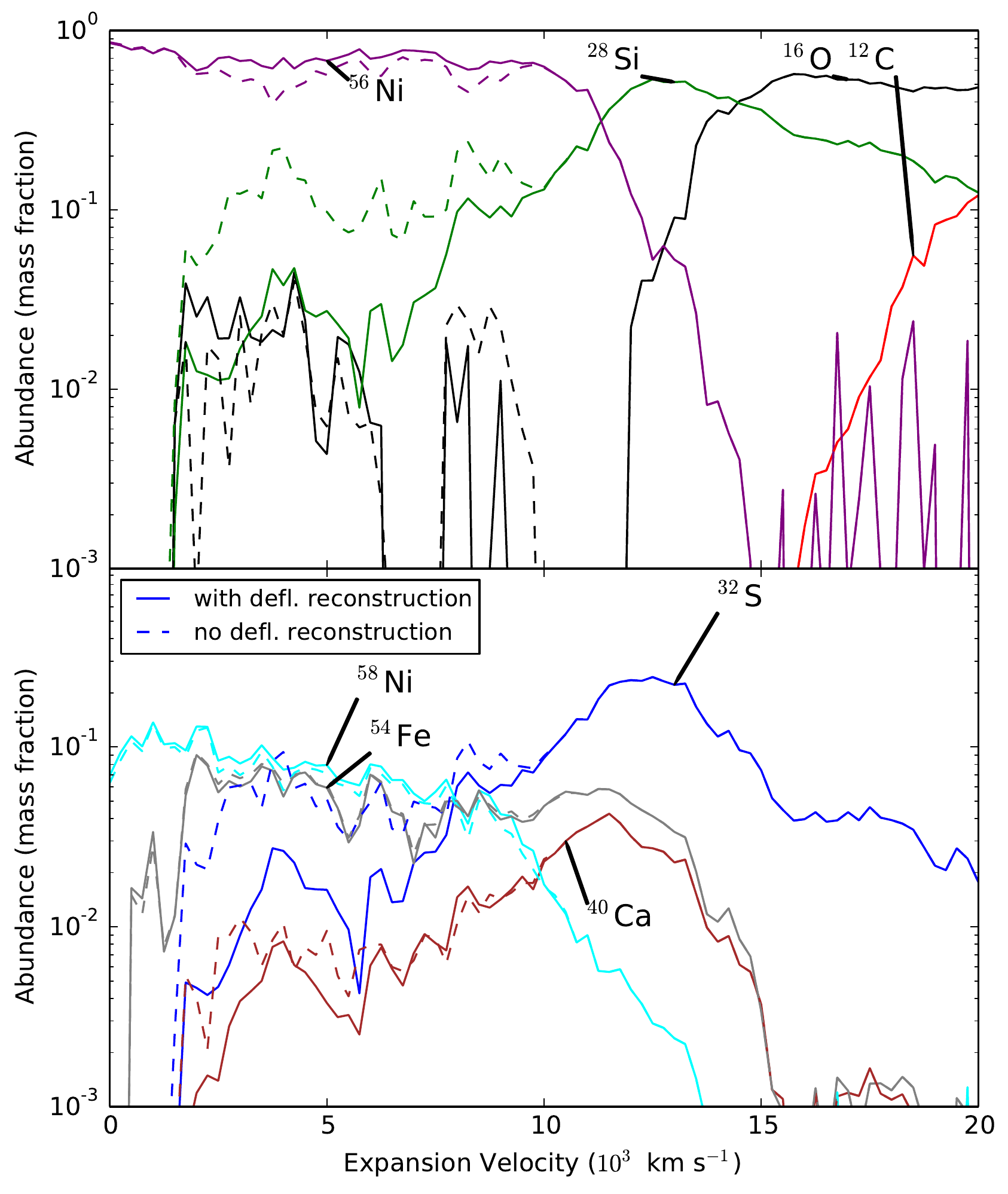}
\caption{\label{fig:ddt_yields}
Abundance profile of ejecta in velocity for 2D DDT simulation.
Upper and lower panels show different species from the same simulations.
Yields are averaged in spherical shell bins in velocity.  Cases are shown in
which the unresolved portion of the deflagration is explicitly reconstructed
(solid lines) and in which the temperature-density histories are directly
processed without reconstruction (dashed lines).  The main impact of
reconstruction is in capturing the peak temperature of the deflagration front,
giving more complete burning of Si- to Fe-group in the interior.
}
\end{figure}

Figure \ref{fig:ddt_yields} shows the nucleosynthetic yields for major
species from our 2D DDT
simulation with ignition distribution realization number 10 from
\citet{Kruegeretal2012} using the progenitor from that work with a central
density of $2\times 10^{-9}$~g~cm$^{-3}$.
The state shown is 4 seconds after ignition when the ejecta reaches approximate free expansion.
The Lagrangian tracer particles from the
simulation are binned based on their asymptotic radial velocity into bins of
250~km~s$^{-1}$ width.  For each bin, 100 randomly selected particles are
post-processed as described in previous sections.
See Appendix \ref{sec:samplinguncertainty} for a
discussion of the uncertainty arising from this choice of sampling.
For the purpose of comparison, we perform
nucleosynthetic post-processing both with and without the reconstruction of
the portion of deflagration histories within the artificially broadened
reaction front, as discussed in Section \ref{sec:def}.
Without this reconstruction, particle tracks are simply processed using their
$\rho(t)$, $T(t)$ history.

The abundance content of the ejecta from our 2D DDT simulations compare
fairly well with the general features seen in observations and the W7 profile
\citep{Stehleetal2005,Mazzalietal2008}.  Si group material is
fairly well-separated from the inner layers of Fe-group that is dominated by
$^{56}$Ni.
Reconstruction of deflagration tracks leads to more complete conversion of IME to IGE in the 2\,000-10\,000~km~s$^{-1}$ region due to the higher peak temperatures reached using reconstruction.
A notable difference from W7 is the absence, in our model, of a
contiguous region near the center that is depleted in $^{56}$Ni.  This loss
of such a core of stable
Fe-group material was seen also in our earlier work \citep{Kruegeretal2012},
and has since also been seen in 3D simulations as well
\citep{Seitenzahletal2013}.  It appears that without recourse to other mechanisms
of neutron enrichment in the core, the deflagration ash distribution produced
by multi-D DDT simulations does not in general produce an unmixed
core of stable Fe-group.

\begin{figure}
\includegraphics[width=\apjcolwidth]{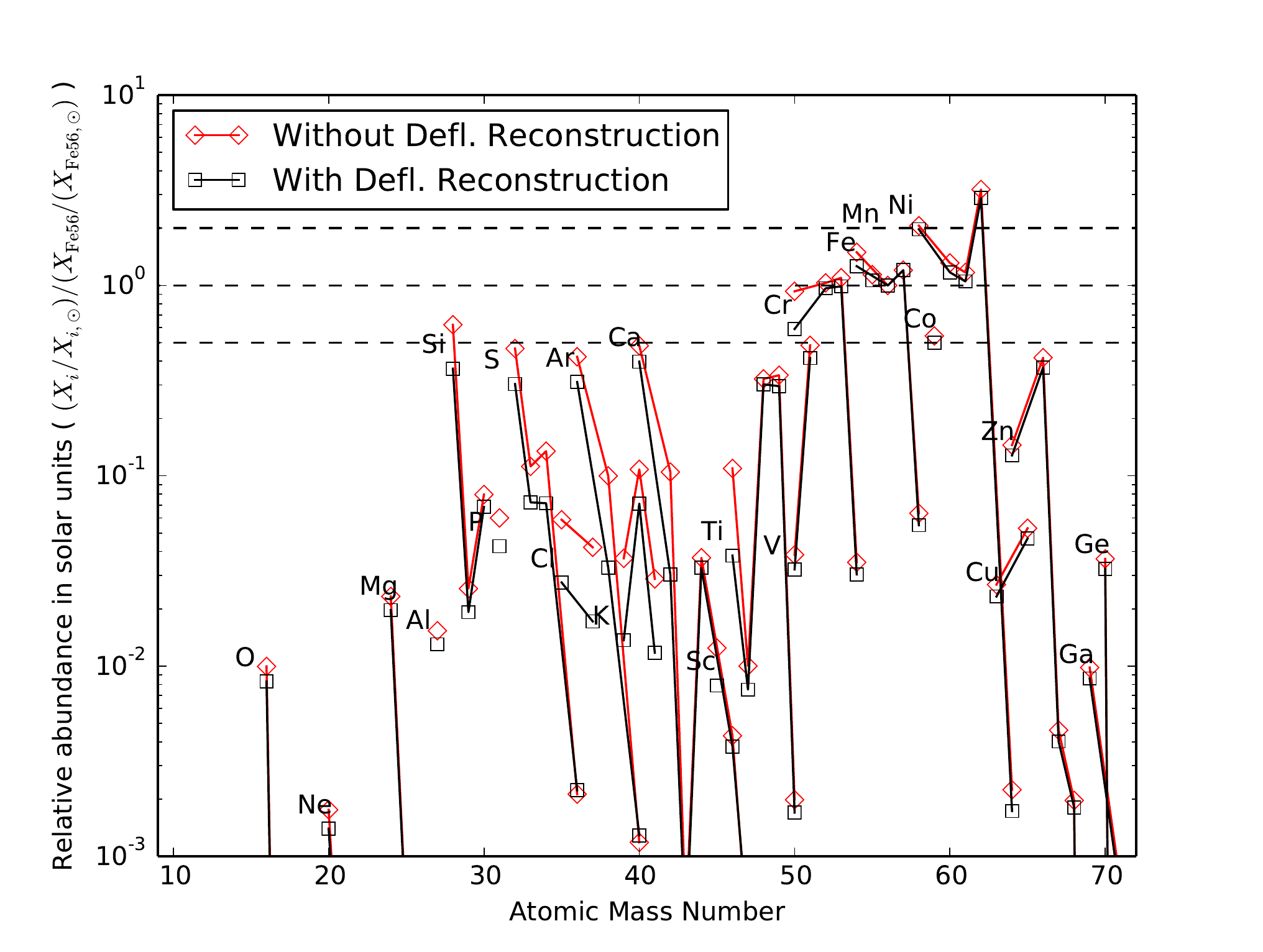}
\caption{\label{fig:isotopic_yields}
Isotopic yields in units of solar abundances and scaled to the $^{56}$Fe
yield.  Isotopes of a single element are connected by a line.  Results of
post processing particles with two methods are shown, in which the $rho$, $T$
history of the particle is used directly (without deflagration reconstruction,
red diamonds), and one in which the portion of the history within the
artificial reaction front is
reconstructed (with deflagration reconstruction, black squares).  Dashed
boundary lines at 0.5 and 2 are also shown for reference.  Differences in the
Fe group are mild, but the relative yield of IME is lower.
}
\end{figure}

The isotopic distribution in the overall yields after decay are shown in Figure
\ref{fig:isotopic_yields}, where integrated abundances are given in solar
units scaled by the Fe abundance.  The pattern observed is similar to that of
W7-like delayed detonation models \citep{Brachwitzetal2000}.
This simulation slightly underproduces the most neutron-rich isotope of several elements compared to solar abundances, as seen by \citet{Brachwitzetal2000} for a central density at ignition of $\rho_{\rm c,ign}=1.7\times 10^9$~g~cm$^{-3}$ (their ``C'' cases).
A slightly higher central density at ignition than that used for our progenitor, $\rho_{\rm c, ign}=2\times 10^9$~g~cm$^{-3}$, will give isotopic yields more similar to solar as in the cases of \citet{Brachwitzetal2000} with $\rho_{\rm c,ign}=2.1\times 10^9$ g~cm$^{-3}$ (their ``W'' cases).
We show separately
the yields obtained with and without reconstruction (black squares and red
diamonds respectively) of the deflagration history.
As is often found for delayed-detonation type models, there is an excess of
$^{62}$Ni.
The main difference with reconstruction is, as seen above, the higher peak
temperature obtained by using the reconstruction gives slightly more complete
burning for some tracks, leading to a lower relative fraction of Si-group
material in the case with reconstruction (black squares).

Yields of individual nuclides are tabulated in Appendix
\ref{sec:tabulatedyields}.  The total yield of $^{56}$Ni is $0.69M_\odot$
without reconstruction and $0.79M_\odot$ with deflagration reconstruction.
The total Fe-group yield, all elements with
$Z>22$, is $0.89M_\odot$ without reconstruction and $1.0M_\odot$
with reconstruction.
The $^{56}$Ni mass inferred from the burning model scalars on the
grid, as is done in \citet{Kruegeretal2010} is 0.70$M_\odot$, and the total
Fe-group mass inferred by integrating $\phi_{qn}\rho$ over the grid is
$0.86M_\odot$.  These values inferred from the progress variables are similar
to those obtained without reconstruction.
These differences reflect ambiguity introduced by material burned partially
by the artificial deflagration front, but then not fully
burned by the detonation in the hydrodynamics.  Generally this material has
$\phi_{\rm RD}>0.5$, and so is reconstructed in post-processing and ends up
fully burned,
but may remain incompletely burned in the burning model variables.
The second example track discussed in appendix \ref{sec:example_tracks} is of this type.
The discrepancy between the burning model and post-processing in final yields can be interpreted as an inconsistency of order 10\% between the ejected $^{56}$Ni mass and the ejection velocity.
The sense is that the ejection velocities are slightly lower than they should be if the burning were fully consistent.
This is the current level of uncertainty, and will vary somewhat for each simulation but can be estimated for a case by comparing these different yield estimates.

\section{Conclusions}
\label{sec:conclusions}

We have outlined methods for computing yields from multi-D simulations of thermonuclear supernovae and compared the accuracy of the results to benchmarks giving steady-state reaction front structures.
The model of burning presented here has been used in recent work on various aspects of SN~Ia systematic variation and physical assumptions
\citep{Jacksonetal2010,Kruegeretal2010,Kruegeretal2012,JacksonTownsleyCalder2014,Willcoxetal2016}.
The full post-processing method is used by \citet{Milesetal2016} to investigate possible spectral indicators of progenitor metallicity.
This paper accompanies the public release of our implementation, which will be integrated into the public release of Flash.

Our method uses a 3-stage model for carbon-oxygen fusion in hydrodynamics and Lagrangian fluid element histories that are recorded during the simulation and post-processed with a 225 nuclide nuclear reaction network.
Due to its necessarily limited spatial and time resolution compared to the reactions being modeled, reaction fronts are unresolved in the hydrodynamics.
In this work, for the first time, we attempt to
reconstruct the unresolved thermal structure of the reaction front in order
to obtain higher accuracy yields.  For verification, we compare the results of
hydrodynamic simulations to benchmark calculations performed using error-controlled methods and 200 nuclide reaction network.
These benchmarks give the reaction front structure in steady state for the detonation propagation mode.
Reproduction of benchmark detonation
structures required improvements to our previously used \citep{Calderetal07,Townsleyetal07,Townsleyetal09} parameterized model for carbon-oxygen fusion
in order to better characterize the conversion rate of Si- to Fe-group material.
We find that use of reconstruction for deflagrations increases the Fe-group yield by about 10\% over that inferred from the burning model alone due to improvement in representing the temperature peak in the deflagration front.
This implies a similar level of modest inconsistency between the $^{56}$Ni yield and the kinetic energy in our ejecta profiles as a current uncertainty in our simulation results.

The main remaining source of inconsistency arises for fluid elements which are processed by both the deflagration and detonation fronts in the simulation.
This leads to material that burns less completely in the hydrodynamic simulation than in post-processing when reconstruction is performed.
Future work may be able to improve the interaction between the detonation front and the thickened model flame front in order to improve this consistency.
We postpone a more thorough investigation until after we address unresolved structure in the detonation.

As an example, we computed yields for a 2D simulation of the deflagration-detonation transition scenario for a thermonuclear supernova.
The
resulting yields compare well to both previously successful 1D
delayed-detonation models of SNe~Ia, and the layered abundance structure
inferred from observations of normal SNe~Ia.  One significant difference,
however, is that the interior of the ejecta lacks a well-defined central
region that is depleted of Ni$^{56}$ via electron captures.  This is because
the material that undergoes strong electron capture during the
deflagration phase is mixed outward by buoyancy, and therefore is spread out
and diluted by surrounding material.  This is consistent with current
simulations of the multi-D DDT model
\citep{Kruegeretal2012,Seitenzahletal2011,Seitenzahletal2013}.

\acknowledgements

DMT acknowledges support from the Bart J. Bok fellowship at Steward Observatory,
The University of Arizona, during the early phases of this work.  ACC acknowledges 
support from the Department of Energy under grant DE-FG02-87ER40317.
We thank Aaron Jackson for his contributions to implementation and allowing his code to be released.
We thank Ivo
Seitenzahl for contributing his NSE tables to our released software.  The
software used in this work was in part developed by the DOE-supported
ASC/Alliances Center for Astrophysical Thermonuclear Flashes at the
University of Chicago. We thank Nathan Hearn for making his QuickFlash
analysis tools publicly available at http://quickflash.sourceforge.net.

\bibliography{master,timmes_master,townsley_master}

\begin{appendix}

\section{Propagation of spatially unresolved detonations with PPM}
\label{app:ppmdet}

As shown by the scales in Figure~\ref{fig:ltscales} and the benchmark reaction
structure in Figures~\ref{fig:space_compare_5e6} and
\ref{fig:space_compare_1e7},
our supernova simulations are performed on spatial grids which are very
coarse compared to the length scales involved in burning and with
hydrodynamic time steps many orders of magnitude larger than the timescales of
many of the principal energy-releasing reactions.  Since the simplified
burning kinetics includes the fastest burning step, carbon fusion, this
remains true in the simplified model as in the actual physics.  This brings
to light a verification problem: is our numerical treatment sufficient to
accurately capture salient features of the detonation and its products?
Here we will perform a verification that spatially unresolved hydrodynamic calculations give the same structure as that computed using the well-established ZND solution with explicit error control and the same reaction network (aprox13).

Of course an unresolved calculation
cannot accurately reproduce all aspects of the detonation dynamics, but it
may still be useful in some ways.  As an
example, in their study of the critical gradient necessary for detonation
ignition, \citet{Seitetal09} found that it was necessary to spatially resolve
the carbon burning length scale in order to obtain fully converged results
for the critical gradient.  However, they did find that unresolved
calculations were reasonably accurate, within an order of magnitude, compared
to the several orders of magnitude over which the size scale setting the
critical gradient varies across the densities and compositions of
interest.  Thus the unresolved calculations, though having known
deficiencies, were sufficiently accurate for the particular purpose.

In the present work we will be concerned with the steady-state
detonation structure.  The focus will therefore be on comparison with a
reference solution calculated from the ZND equations rather than on
comparison with a converged/resolved solution.  Notably, although
\citet{Seitetal09} found ``successful'' self-propagating detonations, they
did not confirm that the ZND structure was achieved.
It seems prudent to perform this verification before proceeding
further in our evaluation of our burning model.

We would like to demonstrate, as was done in \cite{Gamezoetal1999} for a
different hydrodynamics method than that in Flash, that when a portion of the detonation
structure is spatially unresolved on the grid, the thermal and compositional
structure of the resolved structures still match the ZND solution.
\citet{FryxMuelArne89} showed that Eulerian PPM with reactions disabled
within shocks produces the correct detonation speeds and
post-detonation state for a single step reaction.  Additionally, using an
alpha-chain network to study the initial stages of a detonation in carbon at
$\rho=10^9$~g~cm$^{-3}$, \citet{FryxMuelArne89} also saw good agreement among
resolutions at which the carbon reaction is resolved and those at which it
spatially is unresolved.  This indicates that the resolved stages did not
appear sensitive to lack of spatial resolution of the fastest stages.
Here, instead of comparing to a higher resolution, we will extend
verification to a comparison with the steady state reaction front structure
computed using the ZND equations.

Although the hydrodynamics method used in Flash is also Eulerian PPM, it
differs from the method used by \citet{FryxMuelArne89} in the way the
hydrodynamics and reactions are coupled.  Thus we cannot depend upon the
tests performed by \citet{FryxMuelArne89} as a verification of the method in
Flash.
The method described in \citet{FryxMuelArne89} uses the same
timestep for both hydrodynamics and nuclear reactions, limiting the changes
in any given species during one timestep to 5-10\%.  In contrast, for the
provided nuclear reaction networks, Flash uses a per-zone integration of the
reaction kinetics which is operator split from the hydrodynamics
\citep{Fryxelletal2000}.  This sub-hydro-step integration is performed with a
Bader-Deuflhard stiff ODE solver with an error-controlled adaptive timestep
\citep{Timm99,Presetal92}.  For the aprox13 network, during this integration
of the reaction kinetics, the temperature and density are taken to be
constant at the values given by the previous hydrodynamic time step.  Thus
while the variation of species abundance with time is always well-resolved,
due to being subject to error control, the spatial abundance and
thermodynamic structure as well as the time-history of the thermodynamics is
often severely under-resolved.

Flash does include the capability to limit the hydrodynamic timestep based on
energy release with a similar constraint on the change in species used by
\citet{FryxMuelArne89}.  However, this leads to a timestep so small
(nanoseconds) that it makes even 1D calculations intractable.
Therefore we
choose to keep the hydrodynamic timestep at that given by the standard CFL
limit.  For our typical 4~km resolution, this is about $10^{-4}$~s.

Finally, in order to execute our verification test, we must choose a regime
of parameter space in which to perform our comparison.  That is, we must
choose a fuel density and resolution for the hydrodynamic simulation.  Given
our application, we are lead to the natural choice of a density at which the
transition of abundances from Si-group to Fe-group is resolved on the 4~km
resolution grid that we typically use in SN~Ia calculations.  This is also
the critical process which will determine the amount of $^{56}$Ni produced in
the explosion.  Material that does not flash to NSE on short (unresolved)
time scales will have this burning stage quench as the star expands, freezing
in the final abundance structure.  A density of $10^{7}$~g~cm$^{-3}$ makes
about half of this burning stage resolved in a 4~km grid, as seen in
Figures~\ref{fig:ltscales} and \ref{fig:space_compare_1e7}.

\begin{figure}
\plotone{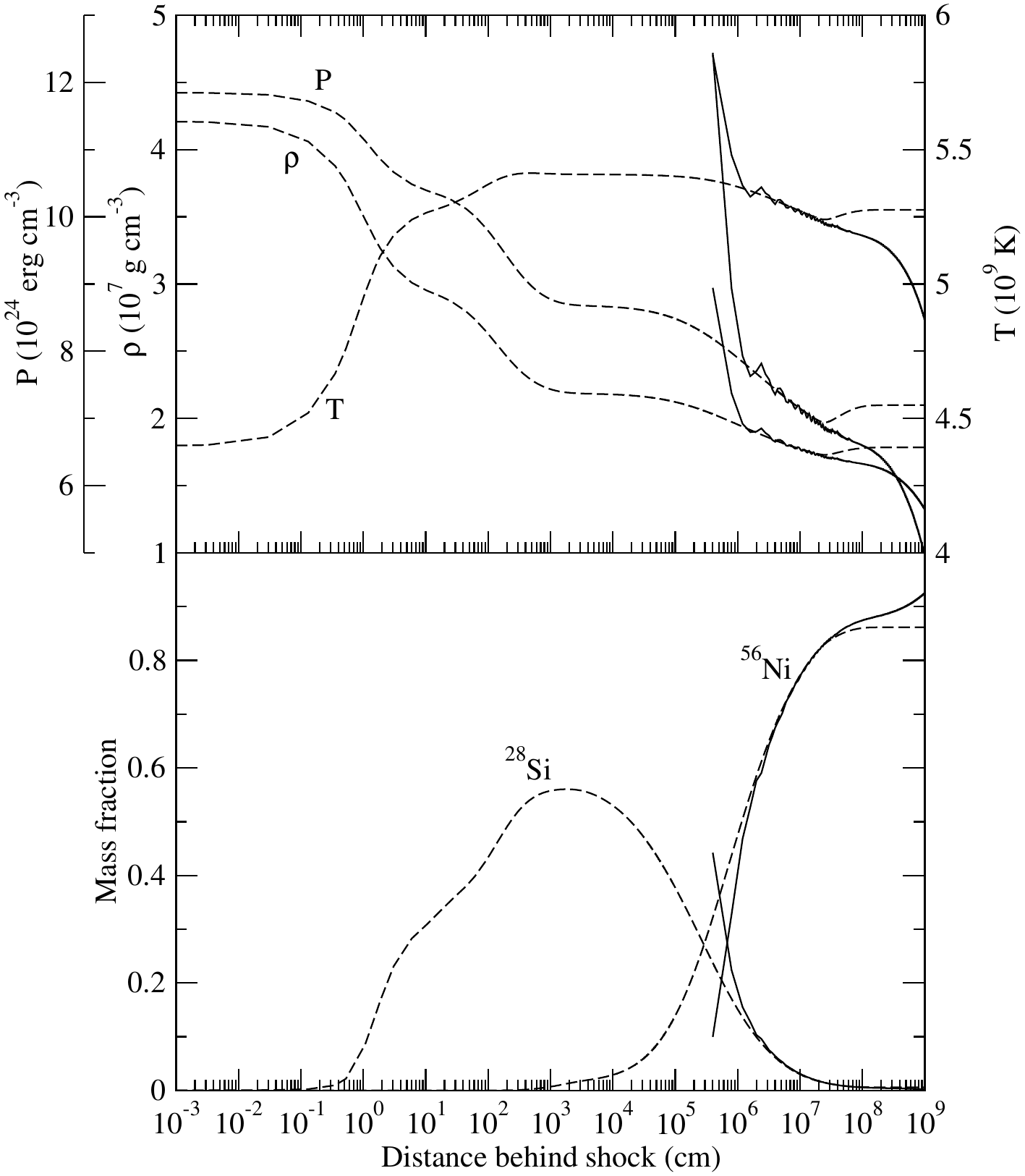}
\caption{\label{fig:PPM_check_1e7}
Comparison of detonation structure at $10^7$~g~cm$^{-3}$ calculated with the
ZND formalism (dashed) and simulated with the reactive hydrodynamics methods implemented in Flash (solid) in one dimension.  Both
methods use the aprox13 enhanced alpha-chain nuclear network.
}
\end{figure}

The initial condition for the hydrodynamic simulation has a constant density
($\rho=10^{7}$~g~cm$^{-3}$) mixture of 50/50 $^{12}$C and $^{16}$O at a
temperature of $4\times 10^8$~K away from the ignition point.  The simulation
is performed on a domain with a reflecting left boundary condition at $x=0$.
The right boundary condition is unimportant because the detonation wave
travels supersonically and the initial condition in the bulk material is in
equilibrium; a reflecting condition is used.  The detonation is ignited by
placing a linear temperature gradient which peaks at $1.8\times 10^9$~K at
$x=0$ and decreases to the background temperature at $x=128$~km.  This
configuration is only a very minor modification, for the ignition point, of
the ``Cellular'' Simulation setup included with the public Flash
distribution.  The standard adaptive refinement routines and setting were
used, which refine on pressure, density, and abundances of $^{28}$Si and
$^{12}$C.  Figure \ref{fig:PPM_check_1e7} compares the ZND structure
calculated with aprox13 (dashed lines) and the steady-state to which the
detonation asymptotes in the hydrodynamic simulation (solid lines).
A fairly large domain was necessary in order for the detonation
to come fully into steady state.  From Figure~\ref{fig:ltscales} the width to
completion of burning is nearly $10^9$~cm.  The domain used was $6.5\times
10^9$~cm, with a resolution of $4\times 10^5$~cm, and the
simulation was run for 5.4 seconds,
by which time the detonation nearly consumes the entire domain.  The
distance behind the shock in the hydrodynamic calculation is computed by
taking the distance from the first zone in which reactions are allowed by the
shock detection.

We find excellent agreement between the ZND solution and the result of the
hydrodynamic simulation despite the entire C and O burning stages being
unresolved.  After a slight overshoot in all $P$, $\rho$, and $T$ just behind
the shock front, the ZND solution is matched within better than 1\% out to
the pathological
point.  The hydrodynamic solution then extends smoothly to lower pressures as
expected for the unsupported solution.  Note that this solution, as expected
for an $\alpha$-chain network like
aprox13, is somewhat hotter than the more realistic detonation structure given
by a larger reaction set discussed in Section~\ref{sec:space_comparison}.

\section{Example Recorded and Reconstructed Lagrangian Histories}
\label{sec:example_tracks}

\begin{figure*}
\begin{tabular}{cc}
\includegraphics[width=\columnwidth]{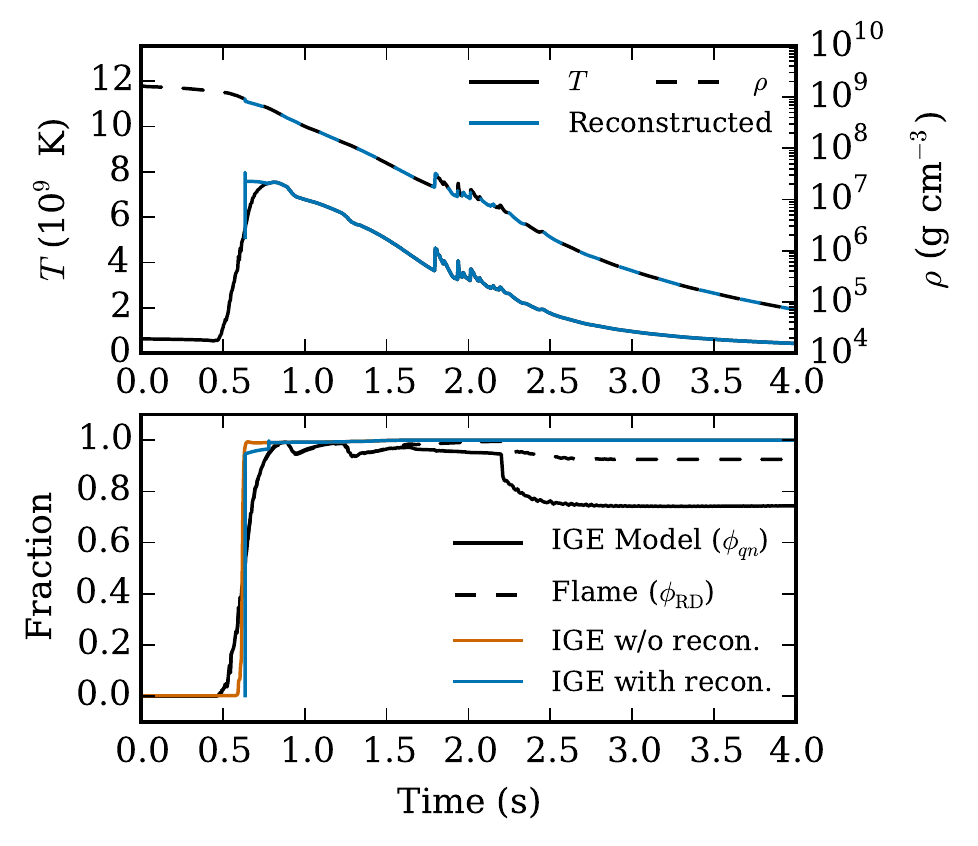}
&
\includegraphics[width=\columnwidth]{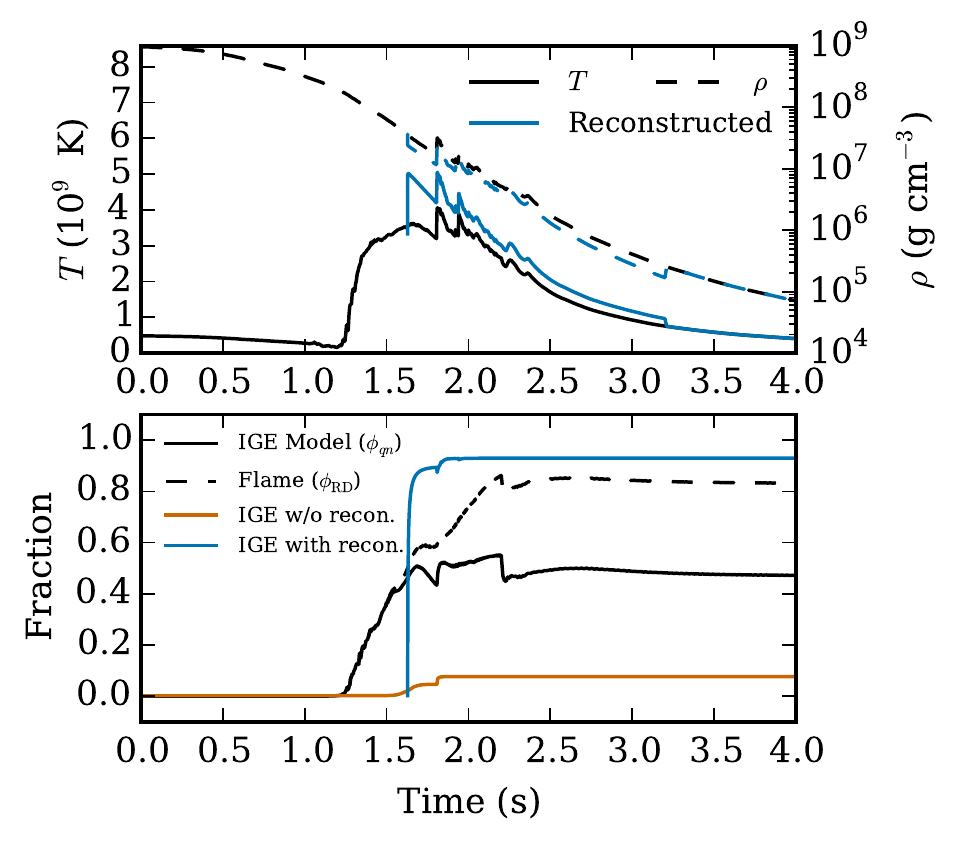}
\\
\includegraphics[width=\columnwidth]{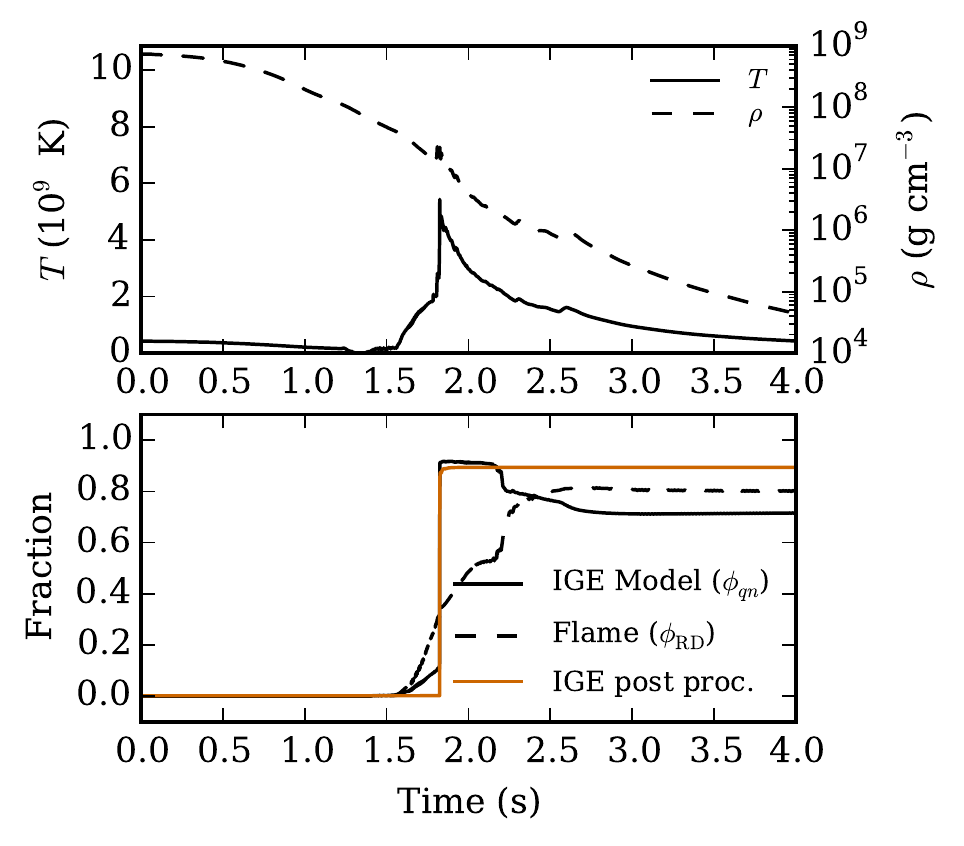}
&
\includegraphics[width=\columnwidth]{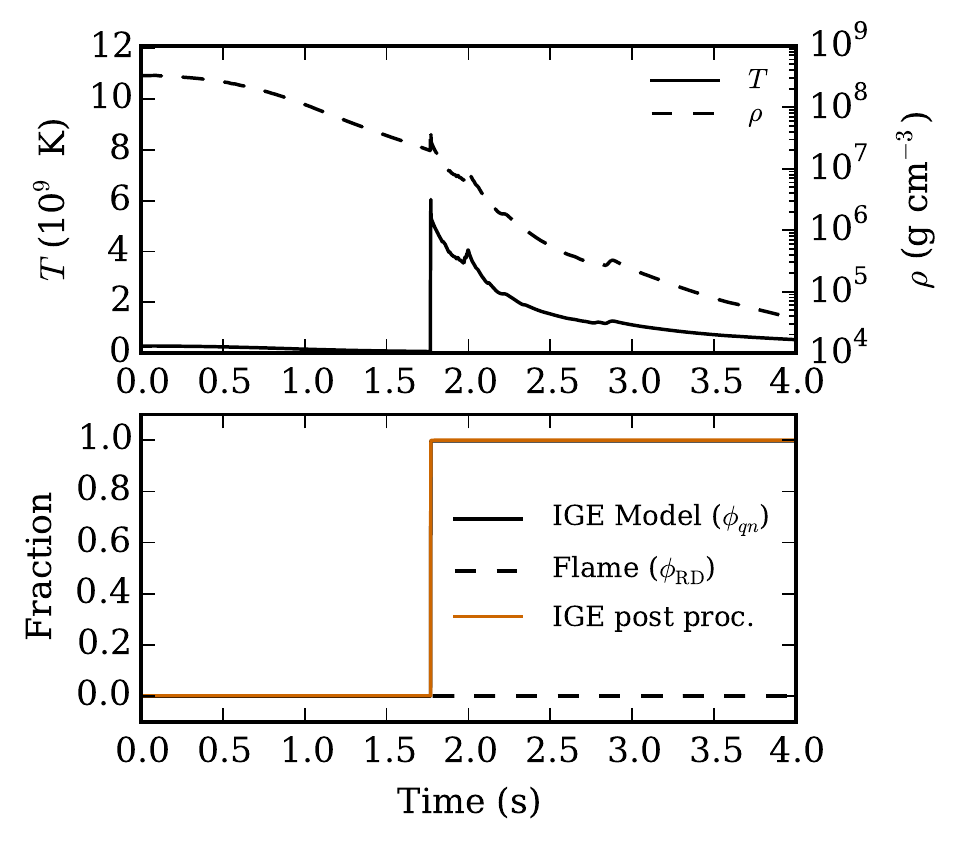}
\\
\includegraphics[width=\columnwidth]{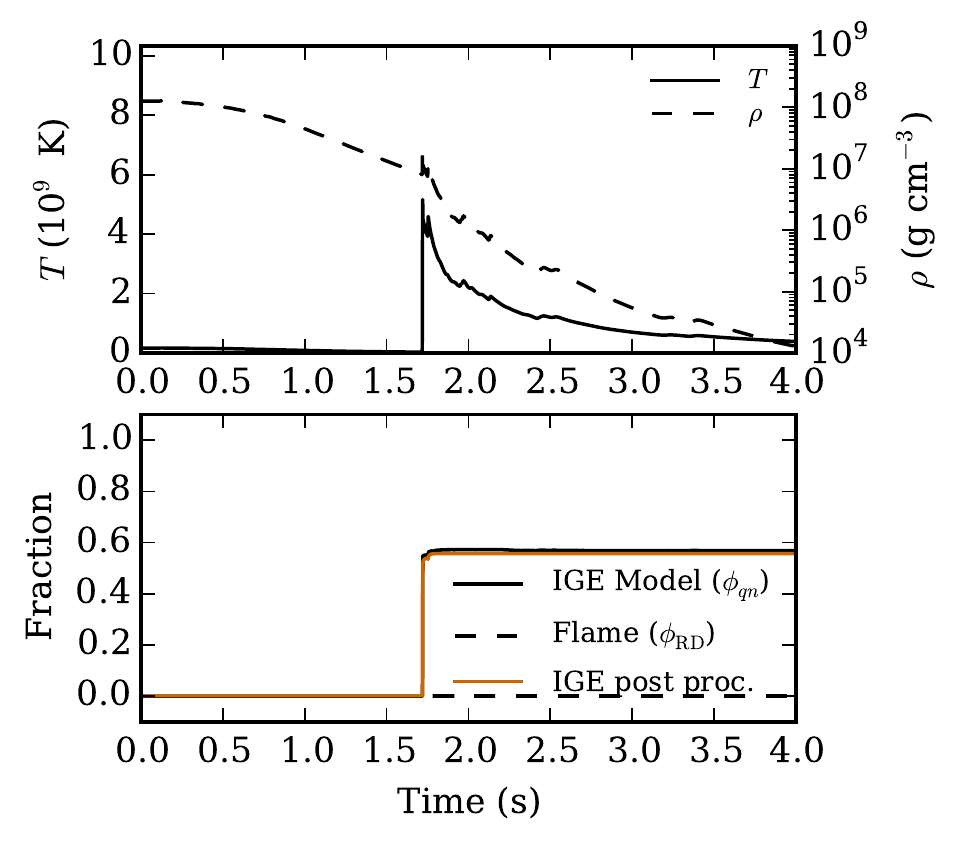}
&
\includegraphics[width=\columnwidth]{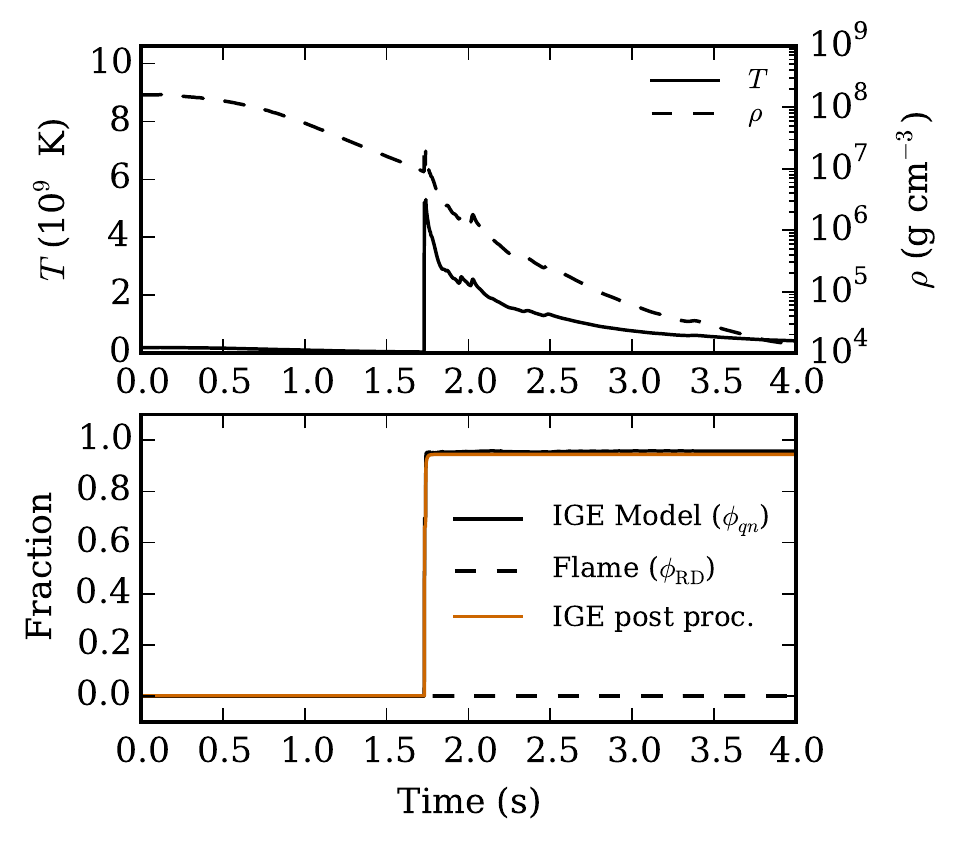}
\\
\end{tabular}
\caption{\label{fig:track_examples}
Example particle histories of various types, see text for individual descriptions.
The top panel of each pair shows the temperature (solid, left scale) and
density (dashed, right scale) recorded from the simulation (black) and, for deflagration tracks, the
reconstructed history (blue).  The bottom panel of each pair
shows the fraction of IGE, recorded from the burning model in hydrodynamics ($\phi_{qn}$, black) and determined in
post-processing with (blue) and without (red) reconstruction reconstruction.
Reconstruction is only performed for deflagration tracks.
Also shown is the progress variable for the artificial flame, $\phi_{\rm RD}$ (dashed).
}
\end{figure*}

In this appendix we present a range of example particle histories from the
simulation and the reconstruction obtained from the methods described in the
main text.  The distribution of tracks among the two burning modes,
deflagration and detonation, varies with position in the ejecta, with inner
layers having a large deflagration component and outer layers being mostly
dominated by detonation products.
Histories, both recorded and reconstructed,
of $\rho$ and $T$ as well as IGE fraction, which is represented by $\phi_{qn}$ in the burning model and \eqref{eq:phiqnabund} for the post-processed abundances, and $\phi_{\rm RD}$ are shown in
Figure \ref{fig:track_examples}.
These provide examples of the several broad classes of tracks produced by the simulation that we will now describe.

The top left panel in Figure \ref{fig:track_examples} shows a typical time history for a fluid element burned by a deflagration front.
The slow, several tenths of a second, rise to peak temperature is replaced in reconstruction by a quick rise followed by a steady decline as the density falls off.
The arrival of the detonation shock can be seen at around 1.8~s, and is relatively weak because this location is within the burned material so that the arriving shock is not an active detonation.
Some mixing artifacts, show by separation between the IGE fraction for the model and $\phi_{\rm RD}$, are apparent upon arrival of the detonation shock, and can be larger in other cases.
This is likely due to the proximity of slightly less burned material and may also indicate a mild mismatch between the advection of the particles and the hydrodynamics when a shock is present.
The grid in the simulation is forced to coarsen starting at 2.2~s, after burning has ceased.
The numerical mixing associated with the merging of cells can cause either a decrease, as seen here, or an increase of the IGE fraction recorded from the simulation.
The best time to compare the IGE fraction produced in post processing with that in the burning model is just before this coarsening.
As expected, we find a good but not precise match for deflagration tracks, within 10\% or so for this and other similar tracks, as the time the of deflagration is not precisely defined.

The top right panel in Figure \ref{fig:track_examples}
shows an example of cases that lead to the largest difference between the yields
estimated from the hydrodynamic burning model variables and the
post-processed yields.
In this deflagration track, when $\phi_{\rm RD}$
passes through 0.5, the density and temperature are still high enough for
fairly prompt full burning to Fe-group.  This is evidenced both in the
recorded $\phi_{qn}$ being similar to $\phi_{\rm RD}$ and the reconstructed
post-processing giving an IGE fraction that increases promptly to
close to unity.  However, as can be seen by the subsequent evolution of the
recorded history, $\phi_{qn}$ in the hydrodynamics does not continue to track
$\phi_{\rm RD}$.  As a result, the hydrodynamic progress variable does not
reach near unity as the fluid element passes the rest of the way through the
RD front, so that the processing of Si- to Fe-group is more complete in the
post-processing.
This is a result of the artificially thick and subsonic reaction front, creating an ambiguity in when the burning commences for this fluid element.
The fluid state can change
(expand) significantly while a particle is passing through the artificial
reaction front.
Note that when the detonation-produced shock arrives at about 1.8 seconds, it is too weak to cause much further progress in the production of IGE.
In some related cases, the shock is strong enough to further produce IGE.

A converse case in which the detonation arrives earlier in the process of artificially thick deflagration can be seen in the middle left panel in Figure \ref{fig:track_examples}.
Here a particle that has been partially burned by the flame is burned by the detonation.
Since the detonation front arrives just before $\phi_{\rm RD}=0.5$, the track is treated as a detonation with its $\rho$, $T$ history directly post-processed, and its IGE yield close to but not quite unity.
In this case the deflagration was taking place at low enough density that IGE production was reduced ($\phi_{qn}<\phi_{\rm RD}$), but the detonation created more complete burning.
The IGE abundance in the model and post-processing are fairly consistent just before the grid is coarsened at 2.2 seconds.

The right middle panel in Figure \ref{fig:track_examples} shows an example of a fairly clean detonation at higher density ($>10^7$~g~cm$^{-3}$).
At pre-detonation densities above $10^7$~g~cm$^{-3}$ burning proceeds fully to IGE in both the burning model and post-processing.
A large fraction of the IGE material is produced in this manner.

At lower densities, the burning in the detonation is less complete.
The material ejected at higher velocities above about 10\,000~km~s$^{-1}$ is almost all burned in the detonation mode to varying degrees of completeness, with the transition from complete to incomplete near a pre-detonation density of $10^7$ g~cm$^{-3}$.
While some of differences are attributable to density, even cases at very similar densities, like the two bottom panels in Figure \ref{fig:track_examples}, can lead to different IGE yields depending on the local strength of the detonation.
The weaker detonation shown in the left panel may be more curved \citep{DunkleySharpeFalle2013,MooreTownsleyBildsten2013} or less fully developed \citep{TownsleyMooreBildsten2012}.
Even for the stronger case, the Si burning is incomplete, giving an IGE fraction just short of unity.
Typically in these cases the post processing is quite consistent, within 5\% or so, of the IGE yield from the burning model.

\section{Sampling Uncertainty}
\label{sec:samplinguncertainty}

Computation of nucleosynthetic yields by post-processing of Lagrangian
histories introduces uncertainty due to the finite sampling of the overall
hydrodynamic solution.  It is useful to consider this uncertainty separately
from the uncertainty due to the finite resolution of the hydrodynamic
solution and any uncertainties introduced by assumptions in the models for
burning processes discussed in the main body of the paper.  Our Lagrangian
histories are placed in the hydrodynamic computation by initial position
randomly and evenly distributed in mass.  This makes the weighting for
computation of yields straightforward.  As discussed in section
\ref{sec:2dddtyields}, we additionally randomly sub-select up to 100 history
tracks for each 250~km~s$^{-1}$ bin in ejection velocity from those available
in that bin from the 100,000 tracks included in the hydrodynamic computation.

\begin{figure}
\includegraphics[width=\apjcolwidth]{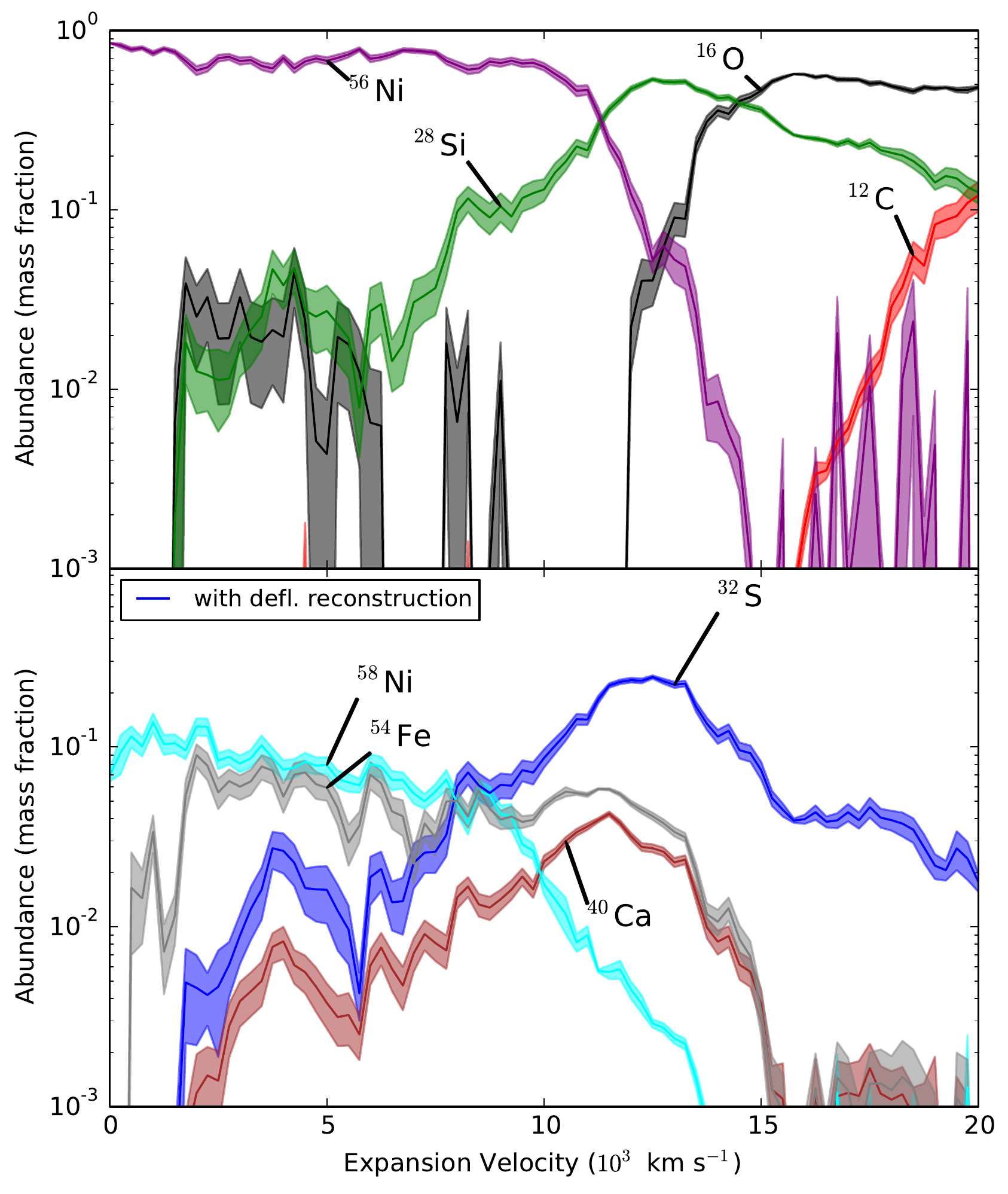}
\caption{\label{fig:profileuncertainty}
Profile computed using reconstruction of deflagration histories with colored
bands indicating uncertainty due to finite sampling by Lagrangian histories.
Uncertainty is taken as the standard deviation of the mean of the abundance
over the histories contributing to each velocity bin.
}
\end{figure}

To estimate the uncertainty due to the finite sampling represented by these
discrete tracks, we have computed the standard deviation of the mean for all
abundances in each ejection velocity bin.  The resulting uncertainty in the
major abundances for each velocity bin is shown in
Figure~\ref{fig:profileuncertainty}, intended to be compared directly with
the yield profiles shown in Figure~\ref{fig:ddt_yields}.  The major
abundances have uncertainties small enough for the comparisons made in this
manuscript, in which we are focusing on the major Fe-group and Si-group
yields.  For velocity bins between 1,000 and 18,000 km~s~$^{-1}$, 100 tracks
are processed, while for other velocity bins 100 are not available from the
100,000 included in the hydrodynamic computation.  The number of available
tracks falls to about 40 by 20,000 km~s$^{-1}$.

If smaller sampling uncertainty is desirable in work using the methods
described here, the number of tracks used or the choice of the initial
position distribution and weighting of the sampling can be modified to give
more samples in a particular portion of the ejecta
\citep[e.g.][]{Seitenzahletal2010}.  As long as the 100 samples in each bin
used here is sufficient to accurately characterize the variance of the
underlying distribution, the standard deviation of the mean should go as
$\propto N^{-1/2}$, where $N$ is the number of tracks.  For non-uniform mass
sampling, the simple standard deviation of the mean can no longer be used,
but it is straightforward to develop a similar measure of uncertainty by
estimating the variance of the distribution of yields using appropriate
weighting of the samples.

\begin{figure}
\includegraphics[width=\apjcolwidth]{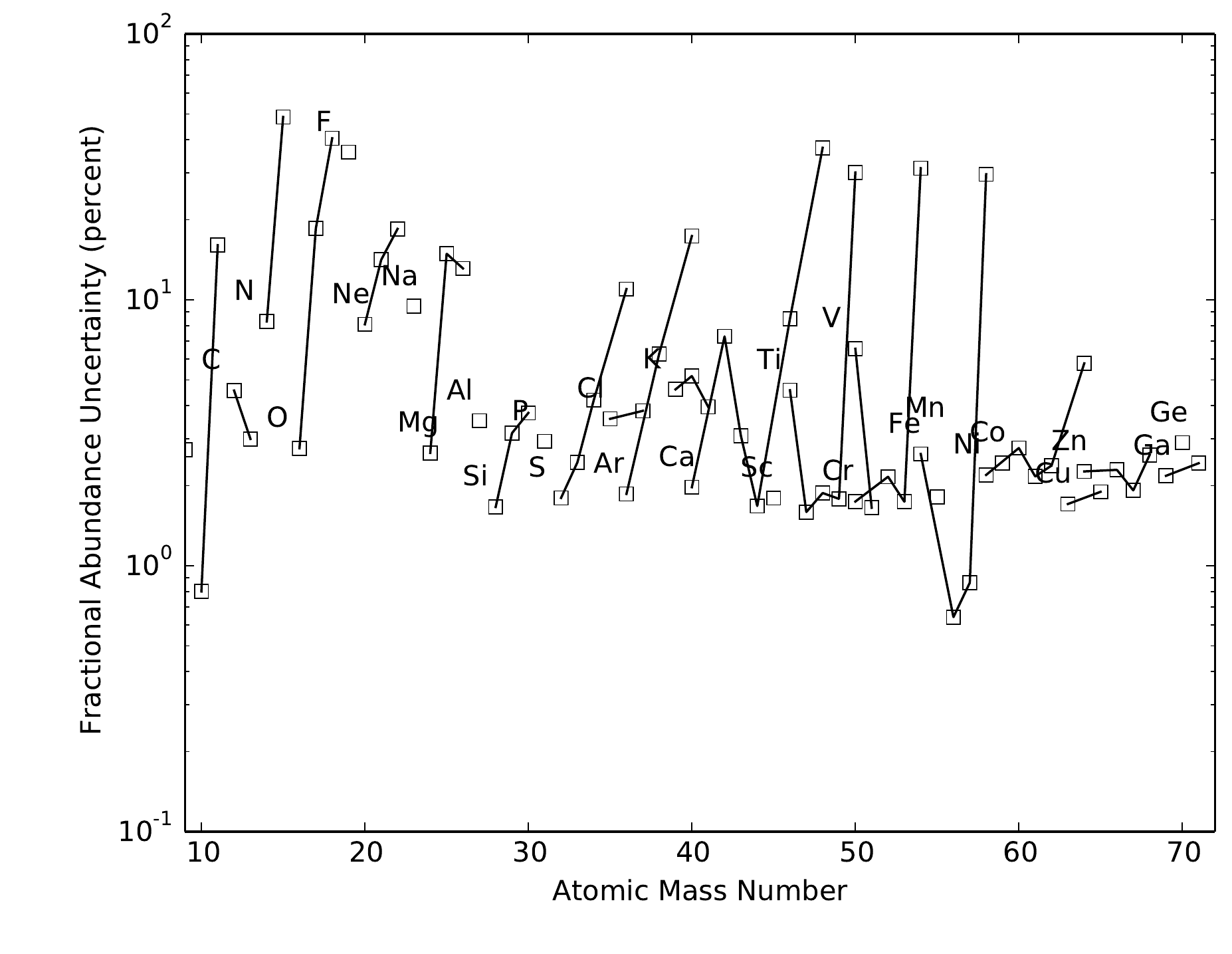}
\caption{\label{fig:isotopicuncertainty}
Uncertainty in total decayed yields due to finite sampling of Lagrangian
histories.
}
\end{figure}

The yield uncertainties can also be propagated in the usual way to
the computation of the total yields of all species when the sums over the
mass in each ejection velocity bin are performed.  The resulting
uncertainties are shown in Figure~\ref{fig:isotopicuncertainty} as a fraction
of each yield.  
Most of the uncertainties are in the 2 to 8 percent range, which is comparable
to or slightly better than our estimated uncertainty found by comparison to
steady-state detonation solutions in section \ref{sec:particle_verif}.  The
most neutron-rich isotopes have higher uncertainties because they are
produced in a relatively small amount of material, however even a 30\%
uncertainty is modest in a comparison like that shown in
Figure~\ref{fig:isotopic_yields}, which
spans 4 orders of magnitude in abundance.  If higher accuracy is desirable
for these isotopes in a particular study, more tracks can be included from
the regions producing them.

\section{Tabulated Yields}
\label{sec:tabulatedyields}

Table \ref{tab:yields1} lists the mass yields of all
nuclides with a mass of $10^{-9}M_\odot$ or more for our 2D DDT simulation.
Masses are listed at two times.  Masses in the first column are 4~s after the
beginning of the simulation, and those in the second column are after all
short-lived radionuclides, defined as those not present in solar abundances,
have decayed.
At each of these times we show the yields
obtained without any reconstruction of the thermal history
and with reconstruction of the thermal history near deflagration fronts as
described in section \ref{sec:def}.  The total Fe-group yield, all elements
with $Z>22$, is $0.89M_\odot$ without reconstruction and $1.0M_\odot$ with
deflagration reconstruction.

\section{Implementation of the burning model}
\label{sec:burning_implementation}

Here we mention several details about the implementation of the burning model
outlined in Section \ref{sec:burningmodel}.  By defining the burning model
principally by dynamical equations, Equations (\ref{eq:phfadot}),
(\ref{eq:phaqdot}), (\ref{eq:phqndot}), (\ref{eq:yedynamics}),
(\ref{eq:qbardynamics}), and (\ref{eq:yiondynamics}), we intend a
clear separation between the physical and numerical aspects of the
construction of the model.  These dynamical equations are summarized in Table
\ref{tab:oper_split}, where the operator splitting, discussed below, is also
indicated.

First we will address how the various dynamical variables are stored and
treated by the hydrodynamical evolution.  The fundamental thermodynamic and
hydrodynamic variables are the density field, $\rho(\vec x,t)$, and the
mass-specific energy, $\mathcal E(\vec x,t)$.  As given in \eqref{eq:rhodef},
this density is more clearly considered the local baryon density in a
particular choice of units.  Additional variables, used to describe the
two initial abundance fields, $X_{^{12}\rm C,i}(\vec x,t)$ and $X_{^{22}\rm
Ne,i}(\vec x,t)$, are stored but are not subject to any source terms.  This
$X_{^{22}\rm Ne,i}$ is used to represent the entire effective neutron excess
in the WD material, regardless of the actual nuclides that contribute to this
neutron excess.
The additional burning state
variables include the progress variables, $\phifa$, $\phiqn$,
$\phiaq$, the reaction-diffusion variable $\phi_{\rm RD}$, and the burning
state variables $\delta \bar q_{\rm qn}$, $\delta Y_{\rm ion,qn}$ and $Y_e$.
Care was taken in section \ref{sec:burningmodel} that all of these burning
variables are linear combinations of abundances, and therefore in the absence
of source terms evolve hydrodynamically as mass scalars.

Notably $Y_e$ is not stored as a partial like $\delta \bar q_{qn}$ or $\delta
Y_{\rm ion,qn}$.  From Equations (\ref{eq:yelong}-\ref{eq:qbarbreakout}), we
see that we basically have a choice for each of $Y_e$, $Y_{\rm ion}$, and
$\bar q$ whether to store the partial "$\delta$" value or the full value.
Either can be obtained from the other using the progress variables and
initial abundances.  In numerical tests we found that storing $\delta Y_e$
and constructing $Y_e$ when needed proved to not be well-behaved when solving
the hydrodynamic step.  We believe that this is related to the strong
dependence of the pressure on $Y_e$ in the highly degenerate material in the
interior of the WD.  This problem appears to have been wholly ameliorated by
using $Y_e$ as the advected mass scalar, deriving $\delta Y_e$ in order to
compute the time evolution given by the source term, and then recomputing
$Y_e$.

An important feature of our implementation of reactive hydrodynamics is the
splitting of the time evolution operator.  As described above in Appendix
\ref{app:ppmdet}, our reactive hydrodynamics code, Flash, is operator split
between hydrodynamics and energetic source terms.  We will also further split
our source terms to enable a high-efficiency sub-step integration.  In
Appendix \ref{app:ppmdet} the coupled reactions are integrated with a stiff
ODE solver that integrates through a hydrodynamical timestep by assuming a
constant $T$.  For our parameterized model of burning, we will assume that
the following quantities are constant during a hydrodynamic step:
$\langle\sigma v\rangle_{\rm C+C}$, $\tau_{\rm
NSQE}$, $\tau_{\rm NSE}$, $\bar q_{\rm NSE}$, $Y_{\rm ion,NSE}$ and $\dot
Y_{e,\rm NSE}$.  These are determined as described in Section
\ref{sec:sourceterms} depending on the proximity to the artificial flame.

Even with these values all
assumed to be constant, the burning model is still fairly tightly coupled.
In order to separate this coupling, as justified below, we will additionally operator split the
burning source terms as shown in Tables \ref{tab:oper_split} and \ref{tab:oper_split_2}.
The evolution represented by the Hydro column is computed first,
followed by the other columns in Table \ref{tab:oper_split} and then the other columns in Table \ref{tab:oper_split_2}.
The final results of each stage is used to compute the evolution of the next.
The important
aspect of this splitting is that each of the resulting source terms can be
analytically integrated through the hydrodynamic timestep, $\Delta t_{\rm
H}$.  As an example, the C-React operator update is performed as
\begin{equation}
\phi_{\rm fa, C+} =
1-\frac{(1-\phi_{\rm fa,C-})}{[1+r_{\rm CC}\,\Delta t(1-\phi_{\rm fa,C-})]}
\ ,
\end{equation}
where $\phi_{\rm fa,C-}$ is the value of $\phifa$ before the C-React
operator,and  $r_{\rm CC}=\rho X_{^{12}\rm C,f}N_{\rm A}\langle\sigma
v\rangle_{\rm C+C}$.  The other terms besides the flame are all exponential
relaxation and can therefore also be analytically integrated.  The evolution of
$\phi_{\rm RD}$ itself is not directly dependent on the other burning
variables.

This operator splitting is effective due to the separation of timescales
within the burning model.  Generally $\tau_{\rm CC}\ll \tau_{\rm
NSQE}\ll\tau_{\rm NSE}$, where each $\tau$ represents an approximate
timescale for C+C fusion, oxygen consumption / QSE adjustment, and completion
of Si burning.  Thus for a given time scale or time step, $\Delta t$,
generally only one variable is dynamically active and the others are either
nearly frozen out or tracking the dominant variable's behavior.

\end{appendix}

\onecolumngrid

\begin{deluxetable}{lllll|lllll}
\tablecolumns{10}
\tablecaption{Ejecta yields in $M_\odot$\label{tab:yields1}}
\tablehead{
& \multicolumn{2}{c}{at 4 seconds}
& \multicolumn{2}{c|}{decayed}
&& \multicolumn{2}{c}{at 4 seconds}
& \multicolumn{2}{c}{decayed}\\
Nuclide & w/o recon. & defl.\ recon.
            & w/o recon. & defl.\ recon.
&Nuclide & w/o recon. & defl.\ recon.
            & w/o recon. & defl.\ recon.
 }
\startdata
$^{4}$He & $8.8\times 10^{-3}$ & $9.0\times 10^{-3}$ & $8.8\times 10^{-3}$ & $9.0\times 10^{-3}$ 
& $^{39}$K & $7.7\times 10^{-5}$ & $3.3\times 10^{-5}$ & $7.7\times 10^{-5}$ & $3.3\times 10^{-5}$ \\
$^{12}$C & $1.4\times 10^{-3}$ & $1.5\times 10^{-3}$ & $1.4\times 10^{-3}$ & $1.5\times 10^{-3}$ 
& $^{40}$K & $2.9\times 10^{-8}$ & $2.2\times 10^{-8}$ & $2.9\times 10^{-8}$ & $2.2\times 10^{-8}$ \\
$^{14}$N & $2.0\times 10^{-9}$ & $1.9\times 10^{-9}$ & $3.0\times 10^{-9}$ & $2.8\times 10^{-9}$ 
& $^{41}$K & $1.2\times 10^{-8}$ & $1.8\times 10^{-8}$ & $4.5\times 10^{-6}$ & $2.1\times 10^{-6}$ \\
$^{16}$O & $5.7\times 10^{-2}$ & $5.5\times 10^{-2}$ & $5.7\times 10^{-2}$ & $5.5\times 10^{-2}$ 
& $^{42}$K & $1.2\times 10^{-8}$ & $1.5\times 10^{-8}$ & & \\
$^{19}$O & $6.5\times 10^{-9}$ & & & 
& $^{40}$Ca & $1.7\times 10^{-2}$ & $1.6\times 10^{-2}$ & $1.7\times 10^{-2}$ & $1.6\times 10^{-2}$ \\
$^{19}$F & & & $6.8\times 10^{-9}$ & 
& $^{41}$Ca & $4.5\times 10^{-6}$ & $2.1\times 10^{-6}$ & & \\
$^{20}$Ne & $1.7\times 10^{-3}$ & $1.6\times 10^{-3}$ & $1.7\times 10^{-3}$ & $1.6\times 10^{-3}$ 
& $^{42}$Ca & $2.6\times 10^{-5}$ & $8.7\times 10^{-6}$ & $2.6\times 10^{-5}$ & $8.8\times 10^{-6}$ \\
$^{21}$Ne & $1.8\times 10^{-7}$ & $2.0\times 10^{-7}$ & $1.8\times 10^{-7}$ & $2.0\times 10^{-7}$ 
& $^{43}$Ca & $5.3\times 10^{-8}$ & $3.2\times 10^{-8}$ & $2.5\times 10^{-7}$ & $2.3\times 10^{-7}$ \\
$^{22}$Ne & $4.5\times 10^{-6}$ & $4.5\times 10^{-6}$ & $4.5\times 10^{-6}$ & $4.5\times 10^{-6}$ 
& $^{44}$Ca & $5.9\times 10^{-8}$ & $4.7\times 10^{-8}$ & $3.1\times 10^{-5}$ & $3.2\times 10^{-5}$ \\
$^{23}$Ne & $8.7\times 10^{-9}$ & $8.7\times 10^{-9}$ & & 
& $^{45}$Ca & $5.2\times 10^{-9}$ & $4.7\times 10^{-9}$ & & \\
$^{22}$Na & $9.7\times 10^{-9}$ & $9.5\times 10^{-9}$ & & 
& $^{46}$Ca & $7.2\times 10^{-9}$ & $7.3\times 10^{-9}$ & $7.2\times 10^{-9}$ & $7.3\times 10^{-9}$ \\
$^{23}$Na & $1.2\times 10^{-5}$ & $1.3\times 10^{-5}$ & $1.3\times 10^{-5}$ & $1.4\times 10^{-5}$ 
& $^{47}$Ca & $9.2\times 10^{-9}$ & $1.3\times 10^{-8}$ & & \\
$^{24}$Na & $1.1\times 10^{-7}$ & $1.1\times 10^{-7}$ & & 
& $^{48}$Ca & $2.4\times 10^{-9}$ & $4.4\times 10^{-9}$ & $2.4\times 10^{-9}$ & $4.4\times 10^{-9}$ \\
$^{23}$Mg & $1.3\times 10^{-6}$ & $1.3\times 10^{-6}$ & & 
& $^{42}$Sc & $1.9\times 10^{-8}$ & $1.9\times 10^{-8}$ & & \\
$^{24}$Mg & $7.2\times 10^{-3}$ & $7.0\times 10^{-3}$ & $7.2\times 10^{-3}$ & $7.0\times 10^{-3}$ 
& $^{43}$Sc & $2.0\times 10^{-7}$ & $2.0\times 10^{-7}$ & & \\
$^{25}$Mg & $1.8\times 10^{-5}$ & $2.0\times 10^{-5}$ & $1.8\times 10^{-5}$ & $2.0\times 10^{-5}$ 
& $^{44}$Sc & $7.5\times 10^{-9}$ & $4.5\times 10^{-9}$ & & \\
$^{26}$Mg & $3.4\times 10^{-5}$ & $3.8\times 10^{-5}$ & $3.8\times 10^{-5}$ & $4.1\times 10^{-5}$ 
& $^{45}$Sc & $5.7\times 10^{-8}$ & $3.9\times 10^{-8}$ & $2.9\times 10^{-7}$ & $2.1\times 10^{-7}$ \\
$^{27}$Mg & $2.8\times 10^{-8}$ & $2.9\times 10^{-8}$ & & 
& $^{46}$Sc & $6.0\times 10^{-9}$ & $4.3\times 10^{-9}$ & & \\
$^{26}$Al & $3.5\times 10^{-6}$ & $3.2\times 10^{-6}$ & & 
& $^{47}$Sc & $8.4\times 10^{-9}$ & $8.5\times 10^{-9}$ & & \\
$^{27}$Al & $5.3\times 10^{-4}$ & $5.2\times 10^{-4}$ & $5.4\times 10^{-4}$ & $5.2\times 10^{-4}$ 
& $^{48}$Sc & $6.7\times 10^{-9}$ & $6.9\times 10^{-9}$ & & \\
$^{28}$Al & $1.3\times 10^{-7}$ & $1.4\times 10^{-7}$ & & 
& $^{49}$Sc & $2.4\times 10^{-9}$ & $2.2\times 10^{-9}$ & & \\
$^{27}$Si & $6.5\times 10^{-7}$ & $6.7\times 10^{-7}$ & & 
& $^{44}$Ti & $3.1\times 10^{-5}$ & $3.2\times 10^{-5}$ & & \\
$^{28}$Si & $2.4\times 10^{-1}$ & $1.6\times 10^{-1}$ & $2.4\times 10^{-1}$ & $1.6\times 10^{-1}$ 
& $^{45}$Ti & $2.3\times 10^{-7}$ & $1.7\times 10^{-7}$ & & \\
$^{29}$Si & $5.3\times 10^{-4}$ & $4.5\times 10^{-4}$ & $5.3\times 10^{-4}$ & $4.5\times 10^{-4}$ 
& $^{46}$Ti & $1.5\times 10^{-5}$ & $5.8\times 10^{-6}$ & $1.5\times 10^{-5}$ & $5.9\times 10^{-6}$ \\
$^{30}$Si & $1.1\times 10^{-3}$ & $1.1\times 10^{-3}$ & $1.1\times 10^{-3}$ & $1.1\times 10^{-3}$ 
& $^{47}$Ti & $3.3\times 10^{-7}$ & $1.8\times 10^{-7}$ & $1.3\times 10^{-6}$ & $1.1\times 10^{-6}$ \\
$^{31}$Si & $2.6\times 10^{-7}$ & $2.6\times 10^{-7}$ & & 
& $^{48}$Ti & $3.7\times 10^{-7}$ & $3.3\times 10^{-7}$ & $4.2\times 10^{-4}$ & $4.5\times 10^{-4}$ \\
$^{32}$Si & $1.7\times 10^{-8}$ & $1.7\times 10^{-8}$ & & 
& $^{49}$Ti & $2.3\times 10^{-8}$ & $1.9\times 10^{-8}$ & $3.3\times 10^{-5}$ & $3.3\times 10^{-5}$ \\
$^{30}$P & $4.3\times 10^{-6}$ & $3.7\times 10^{-6}$ & & 
& $^{50}$Ti & $2.0\times 10^{-7}$ & $1.9\times 10^{-7}$ & $2.0\times 10^{-7}$ & $1.9\times 10^{-7}$ \\
$^{31}$P & $2.9\times 10^{-4}$ & $2.4\times 10^{-4}$ & $2.9\times 10^{-4}$ & $2.4\times 10^{-4}$ 
& $^{51}$Ti & $2.7\times 10^{-9}$ & $3.1\times 10^{-9}$ & & \\
$^{32}$P & $2.1\times 10^{-7}$ & $2.1\times 10^{-7}$ & & 
& $^{52}$Ti & $2.9\times 10^{-9}$ & $3.5\times 10^{-9}$ & & \\
$^{33}$P & $1.7\times 10^{-7}$ & $2.0\times 10^{-7}$ & & 
& $^{46}$V & $9.0\times 10^{-8}$ & $9.1\times 10^{-8}$ & & \\
$^{34}$P & $1.4\times 10^{-9}$ & $1.4\times 10^{-9}$ & & 
& $^{47}$V & $9.0\times 10^{-7}$ & $8.8\times 10^{-7}$ & & \\
$^{31}$S & $4.9\times 10^{-7}$ & $4.6\times 10^{-7}$ & & 
& $^{48}$V & $8.7\times 10^{-8}$ & $5.7\times 10^{-8}$ & & \\
$^{32}$S & $1.1\times 10^{-1}$ & $8.3\times 10^{-2}$ & $1.1\times 10^{-1}$ & $8.3\times 10^{-2}$ 
& $^{49}$V & $2.8\times 10^{-7}$ & $2.1\times 10^{-7}$ & & \\
$^{33}$S & $2.2\times 10^{-4}$ & $1.6\times 10^{-4}$ & $2.2\times 10^{-4}$ & $1.6\times 10^{-4}$ 
& $^{50}$V & $2.1\times 10^{-8}$ & $2.1\times 10^{-8}$ & $2.1\times 10^{-8}$ & $2.1\times 10^{-8}$ \\
$^{34}$S & $1.5\times 10^{-3}$ & $9.3\times 10^{-4}$ & $1.5\times 10^{-3}$ & $9.3\times 10^{-4}$ 
& $^{51}$V & $5.7\times 10^{-7}$ & $5.7\times 10^{-7}$ & $1.1\times 10^{-4}$ & $1.1\times 10^{-4}$ \\
$^{35}$S & $1.0\times 10^{-7}$ & $9.1\times 10^{-8}$ & & 
& $^{52}$V & $4.9\times 10^{-9}$ & $4.0\times 10^{-9}$ & & \\
$^{36}$S & $1.2\times 10^{-7}$ & $1.4\times 10^{-7}$ & $1.2\times 10^{-7}$ & $1.4\times 10^{-7}$ 
& $^{53}$V & $2.3\times 10^{-9}$ & $1.7\times 10^{-9}$ & & \\
$^{34}$Cl & $1.1\times 10^{-7}$ & $9.5\times 10^{-8}$ & & 
& $^{48}$Cr & $4.2\times 10^{-4}$ & $4.5\times 10^{-4}$ & & \\
$^{35}$Cl & $8.9\times 10^{-5}$ & $4.8\times 10^{-5}$ & $8.9\times 10^{-5}$ & $4.8\times 10^{-5}$ 
& $^{49}$Cr & $3.3\times 10^{-5}$ & $3.3\times 10^{-5}$ & & \\
$^{36}$Cl & $2.9\times 10^{-7}$ & $2.7\times 10^{-7}$ & & 
& $^{50}$Cr & $4.2\times 10^{-4}$ & $3.0\times 10^{-4}$ & $4.2\times 10^{-4}$ & $3.0\times 10^{-4}$ \\
$^{37}$Cl & $4.1\times 10^{-7}$ & $4.3\times 10^{-7}$ & $2.2\times 10^{-5}$ & $1.0\times 10^{-5}$ 
& $^{51}$Cr & $4.5\times 10^{-6}$ & $2.5\times 10^{-6}$ & & \\
$^{38}$Cl & $1.1\times 10^{-8}$ & $1.3\times 10^{-8}$ & & 
& $^{52}$Cr & $4.8\times 10^{-4}$ & $4.5\times 10^{-4}$ & $9.2\times 10^{-3}$ & $9.9\times 10^{-3}$ \\
$^{36}$Ar & $2.0\times 10^{-2}$ & $1.7\times 10^{-2}$ & $2.0\times 10^{-2}$ & $1.7\times 10^{-2}$ 
& $^{53}$Cr & $4.1\times 10^{-6}$ & $4.1\times 10^{-6}$ & $1.1\times 10^{-3}$ & $1.2\times 10^{-3}$ \\
$^{37}$Ar & $2.1\times 10^{-5}$ & $9.7\times 10^{-6}$ & & 
& $^{54}$Cr & $9.2\times 10^{-6}$ & $9.1\times 10^{-6}$ & $9.2\times 10^{-6}$ & $9.1\times 10^{-6}$ \\
$^{38}$Ar & $9.2\times 10^{-4}$ & $3.5\times 10^{-4}$ & $9.2\times 10^{-4}$ & $3.5\times 10^{-4}$ 
& $^{55}$Cr & $1.1\times 10^{-8}$ & $1.1\times 10^{-8}$ & & \\
$^{39}$Ar & $1.9\times 10^{-8}$ & $2.0\times 10^{-8}$ & & 
& $^{56}$Cr & $2.8\times 10^{-8}$ & $3.2\times 10^{-8}$ & & \\
$^{40}$Ar & $1.9\times 10^{-8}$ & $2.3\times 10^{-8}$ & $1.9\times 10^{-8}$ & $2.3\times 10^{-8}$ 
& $^{57}$Cr & $1.3\times 10^{-9}$ & $1.4\times 10^{-9}$ & & \\
$^{41}$Ar & $6.0\times 10^{-9}$ & $8.8\times 10^{-9}$ & & 
& $^{58}$Cr & $3.8\times 10^{-9}$ & $5.3\times 10^{-9}$ & & \\
$^{38}$K & $3.1\times 10^{-7}$ & $3.1\times 10^{-7}$ & & 
& $^{51}$Mn & $1.0\times 10^{-4}$ & $1.1\times 10^{-4}$ & & 

\enddata
\end{deluxetable}

\renewcommand{\thetable}{1 (cont.)}

\begin{deluxetable}{lllll|lllll}
\tablecolumns{10}
\tablecaption{Ejecta yields in $M_\odot$ continued \label{tab:yields2}}
\tablehead{
& \multicolumn{2}{c}{at 4 seconds}
& \multicolumn{2}{c|}{decayed}
&& \multicolumn{2}{c}{at 4 seconds}
& \multicolumn{2}{c}{decayed}\\
Nuclide & w/o recon. & defl.\ recon.
            & w/o recon. & defl.\ recon.
&Nuclide & w/o recon. & defl.\ recon.
            & w/o recon. & defl.\ recon.
 }
\startdata
$^{52}$Mn & $2.3\times 10^{-6}$ & $1.9\times 10^{-6}$ & & 
& $^{63}$Zn & $2.1\times 10^{-6}$ & $2.2\times 10^{-6}$ & & \\
$^{53}$Mn & $1.1\times 10^{-4}$ & $9.9\times 10^{-5}$ & & 
& $^{64}$Zn & $8.9\times 10^{-6}$ & $9.1\times 10^{-6}$ & $8.6\times 10^{-5}$ & $8.8\times 10^{-5}$ \\
$^{54}$Mn & $2.3\times 10^{-6}$ & $2.3\times 10^{-6}$ & & 
& $^{65}$Zn & $1.1\times 10^{-6}$ & $1.1\times 10^{-6}$ & & \\
$^{55}$Mn & $2.6\times 10^{-5}$ & $2.6\times 10^{-5}$ & $9.1\times 10^{-3}$ & $9.8\times 10^{-3}$ 
& $^{66}$Zn & $1.1\times 10^{-5}$ & $1.2\times 10^{-5}$ & $1.5\times 10^{-4}$ & $1.5\times 10^{-4}$ \\
$^{56}$Mn & $3.0\times 10^{-8}$ & $2.7\times 10^{-8}$ & & 
& $^{67}$Zn & $4.2\times 10^{-8}$ & $4.1\times 10^{-8}$ & $2.4\times 10^{-7}$ & $2.4\times 10^{-7}$ \\
$^{57}$Mn & $2.2\times 10^{-8}$ & $1.5\times 10^{-8}$ & & 
& $^{68}$Zn & $3.2\times 10^{-7}$ & $3.4\times 10^{-7}$ & $4.8\times 10^{-7}$ & $5.1\times 10^{-7}$ \\
$^{58}$Mn & $4.3\times 10^{-9}$ & $4.3\times 10^{-9}$ & & 
& $^{69}$Zn & $6.1\times 10^{-9}$ & $6.9\times 10^{-9}$ & & \\
$^{59}$Mn & $2.2\times 10^{-8}$ & $2.1\times 10^{-8}$ & & 
& $^{63}$Ga & $2.9\times 10^{-6}$ & $2.9\times 10^{-6}$ & & \\
$^{52}$Fe & $8.7\times 10^{-3}$ & $9.5\times 10^{-3}$ & & 
& $^{64}$Ga & $1.4\times 10^{-6}$ & $1.4\times 10^{-6}$ & & \\
$^{53}$Fe & $1.0\times 10^{-3}$ & $1.1\times 10^{-3}$ & & 
& $^{65}$Ga & $3.9\times 10^{-7}$ & $4.0\times 10^{-7}$ & & \\
$^{54}$Fe & $6.4\times 10^{-2}$ & $6.2\times 10^{-2}$ & $6.4\times 10^{-2}$ & $6.2\times 10^{-2}$ 
& $^{66}$Ga & $2.3\times 10^{-8}$ & $2.3\times 10^{-8}$ & & \\
$^{55}$Fe & $1.1\times 10^{-3}$ & $1.1\times 10^{-3}$ & & 
& $^{67}$Ga & $6.4\times 10^{-8}$ & $6.4\times 10^{-8}$ & & \\
$^{56}$Fe & $1.7\times 10^{-2}$ & $1.7\times 10^{-2}$ & $7.0\times 10^{-1}$ & $8.1\times 10^{-1}$ 
& $^{68}$Ga & $1.6\times 10^{-8}$ & $1.6\times 10^{-8}$ & & \\
$^{57}$Fe & $4.0\times 10^{-5}$ & $4.0\times 10^{-5}$ & $2.1\times 10^{-2}$ & $2.4\times 10^{-2}$ 
& $^{69}$Ga & $2.1\times 10^{-7}$ & $2.1\times 10^{-7}$ & $2.3\times 10^{-7}$ & $2.4\times 10^{-7}$ \\
$^{58}$Fe & $1.4\times 10^{-4}$ & $1.4\times 10^{-4}$ & $1.4\times 10^{-4}$ & $1.4\times 10^{-4}$ 
& $^{70}$Ga & $7.2\times 10^{-9}$ & $6.9\times 10^{-9}$ & & \\
$^{59}$Fe & $3.0\times 10^{-7}$ & $3.0\times 10^{-7}$ & & 
& $^{71}$Ga & & & $1.1\times 10^{-8}$ & $1.1\times 10^{-8}$ \\
$^{60}$Fe & $4.3\times 10^{-6}$ & $3.9\times 10^{-6}$ & & 
& $^{64}$Ge & $7.6\times 10^{-5}$ & $7.7\times 10^{-5}$ & & \\
$^{61}$Fe & $7.7\times 10^{-7}$ & $9.3\times 10^{-7}$ & & 
& $^{65}$Ge & $5.5\times 10^{-6}$ & $5.6\times 10^{-6}$ & & \\
$^{55}$Co & $7.9\times 10^{-3}$ & $8.7\times 10^{-3}$ & & 
& $^{66}$Ge & $1.4\times 10^{-4}$ & $1.4\times 10^{-4}$ & & \\
$^{56}$Co & $6.3\times 10^{-5}$ & $6.4\times 10^{-5}$ & & 
& $^{67}$Ge & $1.4\times 10^{-7}$ & $1.4\times 10^{-7}$ & & \\
$^{57}$Co & $6.2\times 10^{-4}$ & $6.2\times 10^{-4}$ & & 
& $^{68}$Ge & $1.5\times 10^{-7}$ & $1.5\times 10^{-7}$ & & \\
$^{58}$Co & $3.9\times 10^{-6}$ & $4.0\times 10^{-6}$ & & 
& $^{69}$Ge & $1.4\times 10^{-8}$ & $1.4\times 10^{-8}$ & & \\
$^{59}$Co & $3.1\times 10^{-5}$ & $3.1\times 10^{-5}$ & $1.1\times 10^{-3}$ & $1.2\times 10^{-3}$ 
& $^{70}$Ge & $9.5\times 10^{-7}$ & $9.7\times 10^{-7}$ & $9.5\times 10^{-7}$ & $9.7\times 10^{-7}$ \\
$^{60}$Co & $2.0\times 10^{-6}$ & $1.5\times 10^{-6}$ & & 
& $^{71}$Ge & $1.1\times 10^{-8}$ & $1.1\times 10^{-8}$ & & \\
$^{61}$Co & $1.7\times 10^{-6}$ & $1.2\times 10^{-6}$ & & 
&  & & & & \\
$^{62}$Co & $1.3\times 10^{-7}$ & $1.1\times 10^{-7}$ & & 
&  & & & & \\
$^{63}$Co & $3.5\times 10^{-7}$ & $2.5\times 10^{-7}$ & & 
&  & & & & \\
$^{65}$Co & $4.5\times 10^{-8}$ & $4.0\times 10^{-8}$ & & 
&  & & & & \\
$^{56}$Ni & $6.9\times 10^{-1}$ & $7.9\times 10^{-1}$ & & 
&  & & & & \\
$^{57}$Ni & $2.0\times 10^{-2}$ & $2.3\times 10^{-2}$ & & 
&  & & & & \\
$^{58}$Ni & $6.1\times 10^{-2}$ & $6.7\times 10^{-2}$ & $6.1\times 10^{-2}$ & $6.7\times 10^{-2}$ 
&  & & & & \\
$^{59}$Ni & $2.9\times 10^{-4}$ & $3.0\times 10^{-4}$ & & 
&  & & & & \\
$^{60}$Ni & $4.4\times 10^{-3}$ & $4.5\times 10^{-3}$ & $1.5\times 10^{-2}$ & $1.6\times 10^{-2}$ 
&  & & & & \\
$^{61}$Ni & $1.6\times 10^{-5}$ & $1.6\times 10^{-5}$ & $6.0\times 10^{-4}$ & $6.2\times 10^{-4}$ 
&  & & & & \\
$^{62}$Ni & $2.7\times 10^{-4}$ & $2.7\times 10^{-4}$ & $5.3\times 10^{-3}$ & $5.5\times 10^{-3}$ 
&  & & & & \\
$^{63}$Ni & $3.8\times 10^{-7}$ & $3.4\times 10^{-7}$ & & 
&  & & & & \\
$^{64}$Ni & $9.8\times 10^{-7}$ & $8.7\times 10^{-7}$ & $9.8\times 10^{-7}$ & $8.7\times 10^{-7}$ 
&  & & & & \\
$^{65}$Ni & $2.7\times 10^{-7}$ & $2.9\times 10^{-7}$ & & 
&  & & & & \\
$^{58}$Cu & $1.2\times 10^{-6}$ & $1.3\times 10^{-6}$ & & 
&  & & & & \\
$^{59}$Cu & $7.7\times 10^{-4}$ & $8.3\times 10^{-4}$ & & 
&  & & & & \\
$^{60}$Cu & $3.9\times 10^{-5}$ & $4.1\times 10^{-5}$ & & 
&  & & & & \\
$^{61}$Cu & $6.4\times 10^{-6}$ & $6.6\times 10^{-6}$ & & 
&  & & & & \\
$^{62}$Cu & $5.0\times 10^{-7}$ & $5.1\times 10^{-7}$ & & 
&  & & & & \\
$^{63}$Cu & $3.5\times 10^{-6}$ & $3.5\times 10^{-6}$ & $9.3\times 10^{-6}$ & $9.2\times 10^{-6}$ 
&  & & & & \\
$^{64}$Cu & $4.6\times 10^{-7}$ & $4.4\times 10^{-7}$ & & 
&  & & & & \\
$^{65}$Cu & $1.2\times 10^{-6}$ & $1.2\times 10^{-6}$ & $8.4\times 10^{-6}$ & $8.6\times 10^{-6}$ 
&  & & & & \\
$^{66}$Cu & $5.5\times 10^{-8}$ & $5.1\times 10^{-8}$ & & 
&  & & & & \\
$^{60}$Zn & $1.1\times 10^{-2}$ & $1.1\times 10^{-2}$ & & 
&  & & & & \\
$^{61}$Zn & $5.8\times 10^{-4}$ & $6.0\times 10^{-4}$ & & 
&  & & & & \\
$^{62}$Zn & $5.1\times 10^{-3}$ & $5.3\times 10^{-3}$ & & 
&  & & & & 

\enddata
\end{deluxetable}

\renewcommand{\thetable}{\arabic{table}}
\setcounter{table}{1}

\clearpage
\begin{deluxetable}{rllll}
\tablecaption{Division of time evolution into operators\label{tab:oper_split}}
\tablehead{ & \colhead{Hydro}
& \colhead{Flame}
& \colhead{C-React$^{\tablenotemark{b}}$}
            & \colhead{O-consumption$^{\tablenotemark{b}}$} }
\startdata
$\displaystyle \frac{\partial \phi_{\rm RD}}{\partial t} =$
    & ${}-\vec v\cdot\nabla \phi_{\rm RD}$
    & ${}+\dot\phi_{\rm RD}$$^{\tablenotemark{a}}$
    &
    &
    \\\hline
$\displaystyle \frac{\partial\phi_{fa}}{\partial t} =$
    & ${}-\vec v\cdot\nabla \phi_{fa}$
    & ${}+\max\left[0,\dot\phi_{\rm RD}\right]$
    & ${}+\rho X_{^{12}{\rm C,f}}(1-\phifa)^2N_A\langle\sigma v\rangle/12$
    &
  \\
$\displaystyle \frac{\partial\phiaq}{\partial t} =$
    & ${}-\vec v\cdot\nabla \phiaq$
    &
    &
    & ${}+(\phifa-\phiaq)/\tau_{\rm NSQE}$
  \\
\enddata
\tablenotetext{a}{
    $\dot\phi_{\rm RD}=\kappa \nabla^2\phi_{\rm RD}+\frac{f}{4\tau}(\phi_{\rm
       RD}-\epsilon_0)(1-\phi_{\rm RD}+\epsilon_1)$}
\tablenotetext{b}{
    Analytically integrated over timestep.}
\end{deluxetable}

\begin{deluxetable}{rlll}
\tablecaption{Division of time evolution into operators continued\label{tab:oper_split_2}}
\tablehead{ & \colhead{Hydro}
	    & \colhead{Si-burning$^{\tablenotemark{b}}$}
	    &\colhead{Energy and Neutronization} }
\startdata
$\displaystyle \frac{\partial\phiqn}{\partial t} =$
    & ${}-\vec v\cdot\nabla \phiqn$
    & ${}+(\phiaq-\phiqn)^2/\tau_{\rm NSE}$
    &
  \\
$\displaystyle \frac{\partial(\delta Y_{e,n})}{\partial t} =$
    & ${}-\vec v\cdot\nabla (\delta Y_{e,n})$
    & ${}+\left[(\phiaq-\phiqn)^2/\tau_{\rm NSE}\right]Y_{e,0}$
    & ${}+\phiqn\dot Y_{e,\rm NSE}$
  \\
$\displaystyle \frac{\partial(\delta \bar q_{qn})}{\partial t} =$
    & ${}-\vec v\cdot\nabla (\delta \bar q_{qn})$
    & ${}+\left[(\phiaq-\phiqn)^2/\tau_{\rm NSE}\right]\bar q_{\rm QSE0}$
    & ${}+\left[(\phiaq-\phiqn)\bar q_{\rm QSE0}+\phiqn\bar q_{\rm NSE} -
      \delta \bar q_{qn}\right]/\tau_{\rm NSQE}$
      $^{\tablenotemark{b}}$
  \\
$\displaystyle \frac{\partial(\delta Y_{\rm ion,qn})}{\partial t} =$
    & ${}-\vec v\cdot\nabla (\delta Y_{\rm ion,qn})$
    & ${}+\left[(\phiaq-\phiqn)^2/\tau_{\rm NSE}\right]Y_{\rm ion,QSE0}$
    & ${}+\left[(\phiaq-\phiqn)Y_{\rm ion,QSE0}+\phiqn Y_{\rm ion,NSE} -
      \delta Y_{\rm ion,qn}\right]/\tau_{\rm NSQE}$
      $^{\tablenotemark{b}}$
  \\
\enddata
\tablenotetext{b}{
    Analytically integrated over timestep.}
\end{deluxetable}

\end{document}